\documentclass[
	reprint,
	superscriptaddress,
	showkeys,
	aps,
	pra,
	longbibliography,
	nofootinbib
]{revtex4-2}

\usepackage[utf8]{inputenc}

\usepackage{xcolor}
\definecolor{blau}{cmyk}{1.0,0.2,0.0,0.4}
\definecolor{rot}{cmyk}{0.04,1.0,0.8,0.07}
\definecolor{purple}{cmyk}{0.08,1.0,0.3,0.36}

\usepackage[
	colorlinks,
	pdfpagelabels,
	breaklinks,
	pdfstartview=FitH,
	bookmarksopen=true,
	bookmarksnumbered=true,
	bookmarksopenlevel=2,
	plainpages=false,
	hypertexnames=false,
	citecolor=rot,
	linkcolor=blau,
	urlcolor=purple,
	pdftitle={Quark-Diquark model},
	pdfauthor={Jonas Stoll, Niklas Zorbach, Jens Braun},
	pdfkeywords={diquark, FRG, numerical fluid dynamics}
]{hyperref}

\usepackage{bm,dsfont}
\usepackage{tensor}
\usepackage{amsmath,amssymb,amsthm}
\usepackage{physics}
\usepackage{upgreek}
\usepackage[bb=boondox]{mathalfa}
\usepackage{mathtools}
\usepackage{slashed}

\usepackage{bookmark}
\usepackage{graphicx}
\usepackage[caption=false]{subfig}
\usepackage{dcolumn}

\usepackage[utf8]{inputenc}
\usepackage{pgfplots}
\DeclareUnicodeCharacter{2212}{−}
\usepgfplotslibrary{groupplots,dateplot}
\usetikzlibrary{patterns,shapes.arrows}
\pgfplotsset{compat=newest}

\usepackage{ulem}

\usepackage{tikz}

\usepackage{orcidlink}

\usepackage[acronym]{glossaries-extra}
\glssetcategoryattribute{acronym}{nohyperfirst}{true}
\setabbreviationstyle[acronym]{long-short}

\def\GeV{\,\mathrm{GeV}}
\def\MeV{\,\mathrm{MeV}}

\newacronym{QCD}{QCD}{quantum chromodynamics}
\newacronym{QDM}{QDM}{quark-diquark model}
\newacronym{QMDM}{QMDM}{quark-meson-diquark model}
\newacronym{RBG}{RBG}{relativistic Bose gas}
\newacronym{fRG}{fRG}{functional renormalization group}
\newacronym{RG}{RG}{renormalization group}

\newacronym{UV}{UV}{ultraviolet}
\newacronym{IR}{IR}{infrared}

\newacronym{LPA}{LPA}{local potential approximation}
\newacronym{MFA}{MFA}{mean-field approximation}

\newacronym{BCS}{BCS}{Bardeen-Cooper-Schrieffer}
\newacronym{SSB}{SSB}{spontaneous symmetry breaking}

\newacronym{KT}{KT}{Kurganov-Tadmor}

\allowdisplaybreaks

\newcommand\numberthis{\addtocounter{equation}{1}\tag{\theequation}}

\def\d{\,\mathrm{d}}

\usepackage[capitalise,sort,compress]{cleveref}
\Crefname{section}{Sec.}{Sections}
\Crefname{table}{Tab.}{Tables}

\begin{document}

\title{Nonperturbative fluctuation effects of charged bosonic fields:\\ A quark-diquark model study at nonzero density}

\author{Jonas Stoll~\orcidlink{0000-0001-8204-9876}}
\email{jonas.stoll@tu-darmstadt.de}
\affiliation{
	Technische Universität Darmstadt, Department of Physics, Institut für Kernphysik, Theoriezentrum,\\
	Schlossgartenstraße 2, D-64289 Darmstadt, Germany
}

\author{Niklas Zorbach~\orcidlink{0000-0002-8434-5641}}
\email{niklas.zorbach@tu-darmstadt.de}
\affiliation{
	Technische Universität Darmstadt, Department of Physics, Institut für Kernphysik, Theoriezentrum,\\
	Schlossgartenstraße 2, D-64289 Darmstadt, Germany
}

\author{Jens Braun~\orcidlink{0000-0003-4655-9072}}
\email{jens.braun@tu-darmstadt.de}
\affiliation{
	Technische Universität Darmstadt, Department of Physics, Institut für Kernphysik, Theoriezentrum,\\
	Schlossgartenstraße 2, D-64289 Darmstadt, Germany
}
\affiliation{ExtreMe Matter Institute EMMI, GSI, Planckstra{\ss}e 1, D-64291 Darmstadt, Germany}

\begin{abstract}
	We study the renormalization group flow of the scale-dependent effective potential of a quark-diquark model with full field dependence at nonzero chemical potential.
	This includes a discussion of approximations in relation to complex bosonic fields and the Silver-Blaze property.
	The resulting flow equation for the scale-dependent effective potential can in principle be solved down to the infrared limit.
	For our quark-diquark model, which may serve as a low-energy model for dense strong-interaction matter, we find that a competition between the Bardeen-Cooper-Schrieffer singularity 
	and bosonic fluctuations can trigger a first-order phase transition at low temperatures that turns into a second-order phase transition at a tricritical point as the temperature increases.
\end{abstract}

\maketitle

\section{Introduction}\label{sec:introduction}
The phase diagram of the theory of the strong interaction, \gls{QCD}, at high baryon density remains a subject of ongoing research as an understanding of the details of the phase structure is ultimately required to better our understanding of, e.g., the physics underlying neutron stars. 
First-principles studies of various aspects of this region of the phase diagram, where \gls{QCD} is expected to be in a color superconducting state, are possible with functional methods (see, e.g., Refs.~\cite{Son:1998uk,Schafer:1999jg,Pisarski:1999bf,Pisarski:1999tv,Brown:1999aq,Evans:1999at,Hong:1999fh,Nickel:2006vf} for early ground-breaking studies and Refs.~\cite{Braun:2019aow,Leonhardt:2019fua,Braun:2021uua,Fukushima:2021ctq,Geissel:2024nmx,Fukushima:2024gmp,Geissel:2025vnp} for recent developments) but remain challenging. 
Even more, a systematic inclusion of interaction channels of the color-superconducting type may even be relevant for an accurate first-principles localization of the critical endpoint in the \gls{QCD} phase diagram~\cite{Fu:2019hdw}.
In light of the aforementioned challenges, the development of effective models for high-density \gls{QCD} remains of paramount importance in order to gain a more profound understanding of the mechanisms underlying dense, strong-interaction matter, see Refs.~\cite{Rischke:2000pv,Alford:2001dt,Buballa:2003qv,Shovkovy:2004me,Alford:2007xm,Anglani:2013gfu,Buballa:2014tba} for reviews. 
Progress in this respect has been recently made by a discussion of the renormalization of such models~\cite{Braun:2018svj,Braun:2022olp}, see Refs.~\cite{Andersen:2024qus,Gholami:2024diy,Gholami:2025afm,Andersen:2025uzh} for corresponding subsequent studies.

In our present work with a clear methodological focus, we discuss the spontaneous breakdown of a global $\text{SU}(3)$ color and $\text{U}(1)$ vector symmetry, $\text{SU}(3)\times\text{U}(1)_\mathrm{V}$, at nonzero temperature and large (quark) chemical potential in a \gls{QCD}-inspired \gls{QDM} with two massless quark flavors. 
In full \gls{QCD}, color symmetry is a gauge symmetry and, as such, cannot be broken spontaneously~\cite{Elitzur:1975im}. Nevertheless, our \gls{QDM} serves as a simple effective low-energy model for high-density \gls{QCD}, in much the same way that \gls{BCS} theory provides an effective description of superconductivity within electromagnetism~\cite{Bardeen:1957mv}. 
In this model, color superconductivity is then indicated by the formation of a nonzero diquark condensate, see, e.g., Refs.~\cite{Rischke:2000pv,Alford:2001dt,Buballa:2003qv,Shovkovy:2004me,Alford:2007xm,Anglani:2013gfu,Buballa:2014tba} for reviews. 
To investigate the dynamical formation of the diquark condensate, we apply the \gls{fRG} approach~\cite{Wetterich:1992yh} which allows us to study the effect of bosonic fluctuations beyond the \gls{MFA}.
To be more specific, by borrowing techniques from numerical fluid dynamics~\cite{Koenigstein:2021syz,Koenigstein:2021rxj,Steil:2021cbu,Zorbach:2024rre} (see also Ref.~\cite{Grossi:2019urj}), we can compute the full field-dependent effective potential from which the diquark condensate can be deduced. 
We provide a detailed discussion of the derivation of the corresponding flow equation, including the absence of spurious poles in the \gls{RG} flow and the role of the Silver-Blaze symmetry. 
For a recent overview of the range of applications of the \gls{fRG} approach, from statistical mechanics and quantum many-particle systems over high-energy physics to gravity, we refer the reader to Ref.~\cite{Dupuis:2020fhh}.

The present work is organized as follows: 
In \cref{sec:model}, we introduce a \gls{QDM} and discuss the relevant symmetries with a focus on the Silver-Blaze symmetry as it is relevant for the construction of truncations. 
After that, we briefly discuss the \gls{fRG} approach including a general discussion of the construction of truncations of the effective action in \cref{sec:framework}. 
In \cref{sec:mfa}, we then start with the \gls{MFA} which neglects fluctuation effects associated with the diquark fields. 
The construction of truncations including such effects at leading order in the derivative expansion is presented in \cref{sec:LPA}. 
Our results for the phase diagram of the \gls{QDM} can be found in \cref{sec:results}.
Further applications to other theories with complex scalar fields with nonvanishing chemical potentials are demonstrated in \cref{sec:applications}.
Our conclusions are presented in \cref{sec:conclusion}.

\section{Model}\label{sec:model}
In this section, we introduce the phenomenological model that we mostly use for an illustration of our considerations in the present work. 
This model is built up from quark and diquark fields.
We denote the quark fields by $\bar{\psi}$ and $\psi$ with $d_\gamma = 4$ Dirac, $N_\mathrm{f}=2$ flavor and $N_\mathrm{c}=3$ color degrees of freedom. 
The quarks are assumed to form diquark pairs of the two-flavor color-superconductor (2SC) type~\cite{Rapp:1997zu,Alford:1997zt,Berges:1998rc,Pisarski:1999tv,Pisarski:1999bf,Schafer:1999jg}. 
The three corresponding diquark fields $\Delta^*_c$ and $\Delta_c$ with~$c=1,2,3$ also carry an index referring to the fact that the diquarks carry a combination of two colors. 
In fact, these fields represent antisymmetric states in color space which result from the combination of two color triplets. 
In the following only the color index of the quark and diquark fields shall be given explicitly whereas other indices are suppressed for convenience. 

For our discussion it is convenient to introduce an additional (auxiliary) complex-valued field $\bm \mu$ with $\Re(\bm\mu)\geq0$, which is constant in spacetime. 
When evaluated at a real number, $\bm\mu$ reduces to the standard (quark) chemical potential, and can thus be viewed as its complex extension. 
In $d=4$ Euclidean spacetime dimensions, our \gls{QDM} is then defined by the following action:
\begin{align*}
	\numberthis\label{eq:action}
	S[\Phi] = \int_x \biggl[&\bar{\psi}_{c} \bigl( {\rm i} \slashed{\partial} - {\rm i} \bm\mu \gamma^0 \bigr) \psi_{c}  + \bar{\nu}^2 \Delta_{c}^* \Delta_{c} \nonumber \\
	&- \bar{\psi}_{a} \gamma_{5} C \tau^2 \Delta_{c}^* \frac{{\rm i}}{2} \epsilon_{a b c} \bar{\psi}_{b} \nonumber \\
	&+ \psi_{a} C \gamma_{5} \tau^2 \Delta_{c} \frac{{\rm i}}{2} \epsilon_{a b c} \psi_{b} \nonumber
	\biggr] \; ,
\end{align*}
where summation over repeated indices is implied, $C$ is the charge conjugation operator, $\tau^2$ is the second Pauli matrix living in flavor space, $\epsilon_{a b c}$ is the Levi-Civita symbol and $\bar{\nu}$ is a model parameter to be specified below. 
For convenience, we have introduced the super field~$\Phi$, 
\begin{align*}
	\numberthis\label{eq:super-field}
	\Phi = (\Delta_c, \Delta^*_c, \bar\psi_c, \psi_c, \bm\mu) 
\end{align*}
on the left-hand side of~\cref{eq:action}. 
The integral on the right-hand side is defined as 
\begin{align*}
	\numberthis\label{eq:integral}
	\int_{x} \equiv \int_{0}^{\beta} \d x_0 \int_{\mathds{R}^{d-1}} \d^{d-1}\vec{x} \, ,
\end{align*}
where $\beta = \frac{1}{T}$ is the inverse temperature and we impose periodic and antiperiodic boundary conditions for the bosonic and fermionic fields in the compactified temporal direction, respectively.

As can be seen in \cref{eq:action}, the field $\bm \mu$ is coupled to the quark bilinear $\bar{\psi}_{c} \gamma^0 \psi_{c}$. 
Assuming that this field is constant and real-valued, it can be considered a Lagrange multiplier which allows us to tune the net baryon number density. 
We add that the net baryon number is a conserved quantity which results from an invariance under $\text{U}(1)_\mathrm{V}$ transformations:
\begin{align*}
	\numberthis\label{eq:U(1)_V-transformation}
	\psi_c &\mapsto u \psi_c \, , \quad &\bar{\psi}_c &\mapsto \bar{\psi}_c u^{-1} \, , \\
	\Delta_c &\mapsto (u^{-1})^F \Delta_c \, , \quad &\Delta_c^* &\mapsto \Delta_c^* u^F \, ,
\end{align*}
where $u \in \text{U}(1)_\mathrm{V}$. Since diquarks are composites of two quarks, we have $F=2$.

At first glance, it may seem unnatural to introduce a complex-valued chemical potential $\bm \mu$, since we are often ultimately interested in real-valued chemical potentials.\footnote{
	Note that there are applications in which phenomenologically meaningful fields, such as the $\omega^0$ meson, enter a theory in the same way as the imaginary part of a complex-valued chemical potential see, e.g., Ref.~\cite{Haensch:2023sig}.}
However, promoting the chemical potential to a complex quantity can at least be useful for understanding certain peculiar properties of quantum field theories at low temperature.
One example is the invariance of the grand canonical partition function under shifts of the chemical potential, provided the latter remains below a critical value~\cite{Marko:2014hea,Khan:2015puu,Braun:2020bhy}. 
This invariance is known as the Silver-Blaze property~\cite{Cohen:2003kd}.
In practice, the promotion of the real-valued chemical potential in the standard bare action to a complex variable establishes an additional symmetry, described by the following set of transformations~\cite{Marko:2014hea,Khan:2015puu,Braun:2020bhy}:
\begin{align*}
	\numberthis\label{eq:sbs-transformation}
	\psi_c &\mapsto {\rm e}^{-{\rm i} \alpha x_0} \psi_c \, , \quad &\bar{\psi}_c &\mapsto \bar{\psi}_c {\rm e}^{{\rm i} \alpha x_0} \, , \\
	\Delta_c &\mapsto {\rm e}^{{\rm i} F \alpha x_0} \Delta_c \, , \quad &\Delta^*_c &\mapsto \Delta^*_c {\rm e}^{-{\rm i} F \alpha x_0} \, , \\
	\bm\mu &\mapsto \bm\mu - {\rm i}\alpha \, ,
\end{align*}
where $\alpha \in \mathds{R}$ and the associated symmetry group is $(\mathds{R},+)$. 
We shall refer to this symmetry as Silver-Blaze symmetry. 
At nonzero temperature, one additionally requires $\alpha=2 \uppi T n$ for $T > 0$ with $n \in \mathds{Z}$ to preserve the periodicity of bosonic fields and the antiperiodicity of fermionic fields along the temporal axis. 

The Silver-Blaze symmetry is a local symmetry which puts a constraint on derivatives of the fields in the temporal direction.
To be specific, temporal derivatives must always come with a suitably chosen term that depends on the chemical potential~\cite{Marko:2014hea,Khan:2015puu,Braun:2020bhy}. 
In terms of our auxiliary field~$\bm{\mu}$, we have $D_{\pm} \equiv \partial_0 \pm F \bm{\mu}$ with~$F=1$ for the quarks and $F=2$ for the diquarks.
Loosely speaking, the field $\bm \mu$ acts as a single-component gauge field similar to the covariant four-potential in, e.g., quantum electrodynamics.
However, the transformation of our ``gauge field" $\bm \mu$ does not depend on~$x_0$, see \cref{eq:sbs-transformation}. 
This is in contrast to the covariant four-potential which also transforms locally.
The aforementioned peculiarities of quantum field theories at low $T$ can be directly explained by the Silver-Blaze symmetry in combination with the requirement of complex analyticity in $\bm \mu$, see \cref{sec:Silver-Blaze}.

Last but not least, the action \eqref{eq:action} has a global $\text{SU}(N_\mathrm{c})$ color symmetry, 
\begin{align*}
	\numberthis\label{eq:color-transformation}
	\psi_b & \mapsto \left(  {\rm e}^{-{\rm i}\theta^a T^a}\right)_{bc} \psi_{c} \, , \quad &\bar{\psi}_{c} &\mapsto \bar{\psi}_{b} \left( {\rm e}^{{\rm i}\theta^a T^a}\right)_{bc} \, , \\
	\Delta_{b} &\mapsto \left( {\rm e}^{{\rm i}\theta^a T^a}\right)_{bc} \Delta_{c} \, , \quad &\Delta_{c}^* &\mapsto \Delta_{b}^*  \left(  {\rm e}^{-{\rm i}\theta^a T^a}\right)_{bc} \, ,
\end{align*}
where the $\theta^a$ are transformation angles and the~$T^a$ are the $\text{SU}(N_\mathrm{c})$ generators in the fundamental representation.

\section{Functional Renormalization Group}\label{sec:frg} 
\subsection{General Framework}\label{sec:framework}
In the \glsxtrlong{fRG} formalism one introduces an infrared regulator $R_k$ into the path integral.
This regulator is parametrized by the \gls{RG} scale $k$ that regularizes modes of the path integral with momenta $p^2 \lesssim k^2$, whereas modes with momenta $p^2 > k^2$ are not affected. 
From a Legendre transformation of this scale-dependent path integral with respect to the sources for the fields, one then obtains an \gls{RG} flow equation for the scale-dependent effective average action~$\Gamma_k$, the Wetterich equation \cite{Wetterich:1992yh}:
\begin{align*}
	\label{eq:wetterich-equation}\numberthis
	\partial_t \Gamma_k [ \Phi ] = \frac{1}{2}\,\mathrm{STr} \Big[  \big( \Gamma^{(2)}_k [ \Phi ] + R_k \big)^{-1} \cdot \partial_t R_k \Big] \, .
\end{align*}
Here, $\mathrm{STr}$ denotes the super trace which integrates over the temporal and spatial directions, sums over all internal indices and contributes an extra minus for fermionic degrees of freedom. Furthermore, $\partial_t = - k \partial_k$ is the \gls{RG} scale derivative. 
The solution of the Wetterich equation, the scale-dependent effective average action $\Gamma_k$, interpolates between a given initial condition~$\Gamma_{k=\Lambda}[\Phi]$ at the \gls{UV} scale~$\Lambda$ and the full quantum effective action $\Gamma[\Phi]$ in the \gls{IR} limit $k \to 0$, from which all physical observables can be obtained. 

The \gls{UV} scale~$\Lambda$ should be considered as an additional model parameter at which we fix the initial condition for~$\Gamma_k$ such that we recover the action \eqref{eq:action}.
This also implies that the regularization scheme belongs to the definition of the model.
In our present work, we choose standard {\it $3d$} spatial regulators for the quark and diquark fields (see, e.g., Refs.~\cite{Litim:2006ag,Blaizot:2006rj}) with corresponding Litim regulator shape functions~\cite{Litim:2000ci,Litim:2001up} and set~$\Lambda =1\GeV$.
Note that, while these regulators break Lorentz invariance, they do not break any of the symmetries discussed in \cref{sec:model}, see Refs.~\cite{Fu:2016tey,Braun:2017srn,Braun:2020bhy} and, in particular, Ref.~\cite{Braun:2022mgx} for a more detailed discussion of regulator-induced symmetry breaking.
Since \cref{eq:wetterich-equation} is a highly complicated functional differential equation and therefore in general not exactly solvable, it is required to consider truncations.
We define a truncation in terms of a map ${\mathds{T}}$ which is an endomorphism on the space of all actions.
This map is inserted on the right-hand side of the Wetterich equation \eqref{eq:wetterich-equation} before the field derivatives are performed, which leads to the following {\it truncated} Wetterich equation:
\begin{align*}
	\label{eq:truncated-wetterich-equation}\numberthis
	\partial_t \Gamma_k [ \Phi ] = \frac{1}{2}\,\mathrm{STr} \Big[  \big( ({\mathds{T}} \circ \Gamma_k)^{(2)} [ \Phi ] + R_k \big)^{-1} \cdot \partial_t R_k \Big] \, .
\end{align*}
When ${\mathds{T}}$ is the identity, we recover the exact Wetterich equation \eqref{eq:wetterich-equation}. 
We shall refer to the action ${\mathds{T}} \circ \Gamma_k$ as the {\it truncated} scale-dependent effective average action. 
We would like to emphasize that the notion of the truncation map is solely introduced to stress the difference between the solution of the truncated Wetterich equation \eqref{eq:truncated-wetterich-equation}, i.e., the scale-dependent effective average action $\Gamma_k$, and the truncated scale-dependent effective average action ${\mathds{T}} \circ \Gamma_k$.

Typically, the truncation map ${\mathds{T}}$ is used to reduce the complexity of a given action $\Gamma_k$ by extracting a set of scale-dependent couplings $\{\mathcal{O}_i(k)\}$. 
These couplings are then employed to construct a new (much simpler) truncated action ${\mathds{T}}\circ \Gamma_k$.
In other words, the truncation map ${\mathds{T}}$ reduces the information that enters the right-hand side of the Wetterich equation \eqref{eq:wetterich-equation}. 
As a consequence, the truncation map ${\mathds{T}}$ is inherently related to the definition of projection prescriptions $\mathrm{proj}_{\mathcal{O}_i}$ for each coupling ${\mathcal{O}_i}(k)$, i.e., $\mathcal{O}_i(k) := \mathrm{proj}_{\mathcal{O}_i}(\Gamma_k)$.
In general, the projections occurring in the truncation map should fulfill the consistency condition 
\begin{align*}
	\numberthis\label{eq:projections-prop}
	\mathrm{proj}_{\mathcal{O}_i}({\mathds{T}} \circ \Gamma_k) \stackrel{!}{=} \mathrm{proj}_{\mathcal{O}_i}(\Gamma_k) \, .
\end{align*}
Below, we shall discuss several truncations, such as \gls{MFA} and various versions of \gls{LPA} for which \cref{eq:projections-prop} is of high relevance. 
Note that \gls{LPA} is the lowest order in the derivative expansion and goes beyond \gls{MFA} as it includes bosonic fluctuation effects of the diquark fields. 

With these truncations, we investigate \gls{SSB} of the $\text{SU}(N_\mathrm{c}=3)\times \text{U}(1)_{V}$  symmetry of the \gls{QDM}, leading to the \gls{SSB} pattern
\begin{align*}
	\numberthis\label{eq:SSB-pattern}
	\text{SU}(3) \times \text{U}(1)_\mathrm{V} \overset{\mathrm{SSB}}{\longrightarrow} \text{SU}(2) \times \text{U}(1)\, ,
\end{align*}
i.e., five generators of $\text{SU}(3)\times \text{U}(1)_{V}$ are spontaneously broken. 
From a phenomenological standpoint, this amounts to study the formation of a nonzero diquark condensate associated with the simultaneous appearance of a gap in the quark excitation spectrum, see, e.g., Refs.~\cite{Rajagopal:2000wf,Alford:2007xm} for reviews. 
To this end, we shall compute the scale-dependent effective potential $U_k$ defined by the projection prescription
\begin{align*}
	\numberthis\label{eq:scale-dependent-effective-potential}
	U_k(\bar{\Delta}, \mu) = \mathrm{proj}_U(\Gamma_k) = \frac{1}{V_d} \Gamma_k[\Phi=\Phi_0] \, ,
\end{align*}
where $V_d$ is the $d$-dimensional spacetime volume and we evaluate the scale-dependent effective average action at the constant field configuration
\begin{align*}
	\numberthis\label{eq:super-field-evaluation}
	\Phi_0 = (&\Delta_c=\tfrac{1}{\sqrt{2}}\bar{\Delta}\,\delta_{c 3}, \Delta^*_c=\tfrac{1}{\sqrt{2}}\bar{\Delta}\,\delta_{c 3}, \\
	&\quad \bar\psi_c=0, \psi_c=0, \bm \mu = \mu) \, .
\end{align*}
Here, $\mu$ and $\bar{\Delta}$ are real-valued numbers. 
This choice implies $\Delta^*_c\Delta_c = \bar{\Delta}^2/2$. 
We refer to the (global) minimum $\bar{\Delta}_\mathrm{gs}$ of the effective potential $U = U_{k \to 0}$ as the diquark condensate, representing an order parameter for our studies of \gls{SSB} below. 

Let us close this subsection with a remark on the role of the Silver-Blaze symmetry in the construction of truncations. 
The Silver-Blaze symmetry of $\Gamma_k$ and its potential complex analyticity in $\bm \mu$ at zero temperature and low to intermediate real $\mu$ are of high phenomenological relevance, see also \cref{sec:model} and App.~\ref{sec:Silver-Blaze}.   
Therefore, neither of these properties of~$\Gamma_k$ should be explicitly violated by the truncation.
In practice, truncations for~$\Gamma_k$ are often formulated for real $\mu$. 
Therefore, they do not allow for an unambiguous assessment of the Silver-Blaze symmetry or complex analyticity in $\bm \mu$. 
However, if a generalization of such a truncation to complex $\bm \mu$ exists,\footnote{
	Here, we also assume that the generalized truncation reduces to the original truncation at real $\bm{\mu} = \mu$.} 
which does not explicitly destroy complex analyticity in~$\bm \mu$ and respects the Silver-Blaze symmetry, then the resulting~$\Gamma_k$ will itself obey the Silver-Blaze symmetry and may be analytic in some region of the complex $\bm \mu$ plane, depending on the model parameters.
Only with these generalized truncations can we eventually assess whether features like the Silver-Blaze property can appear in results at $\bm \mu=\mu$.
In the following, we will use a complex chemical potential to construct generalized truncations, which we then discuss with respect to analyticity and Silver-Blaze symmetry. For the actual derivation of the flow equation for the scale-dependent effective potential, however, we restrict ourselves to real chemical potentials $\bm{\mu}=\mu$.
Note that the dependence of the scale-dependent effective potential on $\Re(\bm\mu)$ can be studied without knowledge about its dependence on the imaginary part.
We rush to add a word of caution here. Even nonanalytic generalizations of truncations (e.g., truncations depending on the real and imaginary part of the chemical potential separately) can yield flow equations for the scale-dependent effective potential which agree identically with corresponding analytic generalizations at $\bm{\mu} = \mu$. Their difference only becomes apparent for complex values of~$\bm \mu$.
However, due to the lack of analyticity, these truncations cannot be used to assess whether features like the Silver-Blaze property are realized at real $\bm\mu=\mu$. 

\subsection{Mean-field approximation}\label{sec:mfa}
\begin{figure}[t]
	\includegraphics{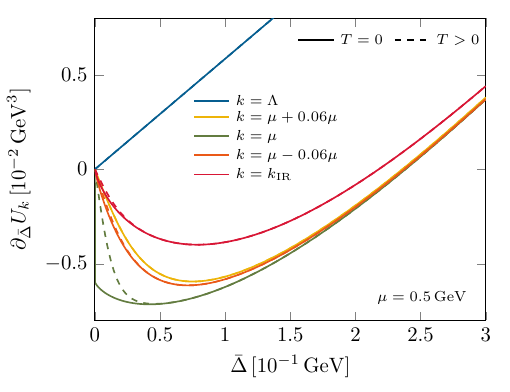}
	\caption{\label{fig:MF-BCS}Derivative of the scale-dependent effective potential in \gls{MFA} as a function of~$\bar{\Delta}$ for various \gls{RG} scales for $\mu = 0.5 \GeV$ and two values of the temperature,  $T = 0$ (solid lines) and $T = 5 \MeV$ (dashed lines).}
\end{figure}
Let us start our discussion of concrete truncations at the level of the \glsxtrlong{MFA}.
As discussed at the end of the previous subsection, we discuss a generalization of standard \gls{MFA} to complex-valued $\bm\mu$ defined by the truncation map ${\mathds{T}}^{\mathrm{MFA}}$: 
\begin{align*}
	\numberthis\label{eq:mf-truncation}
	{\mathds{T}}^{\mathrm{MFA}} \circ \Gamma_k[\Phi] &= \int_x \biggl[\bar{\psi}_{c} \bigl( {\rm i} \slashed{\partial} - {\rm i} \bm\mu \gamma^0 \bigr) \psi_{c} \nonumber \\
	&\qquad\quad - \bar{\psi}_{b} \gamma_5 C \tau^2 \Delta_{a}^* \frac{{\rm i}}{2} \epsilon_{a b c} \bar{\psi}_{c} \nonumber \\
	&\qquad\quad + \psi_{b} C \gamma_5 \tau^2 \Delta_{a} \frac{{\rm i}}{2} \epsilon_{a b c} \psi_{c} \nonumber
	\biggr]\,.
\end{align*}
This truncation map yields a $k$-independent functional where the structure is inherited from the \gls{QDM} action~\eqref{eq:action}. 
It thus fulfills all symmetries of our model including the Silver-Blaze symmetry and it does not explicitly violate complex analyticity in $\bm\mu$ since the real and imaginary parts do not occur separately.
However, this map removes the back coupling of the scale-dependent effective potential~$U_k$ on the \gls{RG} flow as well as the effect of kinetic terms of the diquark fields.
Still, among infinitely many other terms, a nontrivial scale-dependent effective potential as well as kinetic terms for the diquark fields are dynamically generated in the \gls{RG} flow.
For the flow equation of the scale-dependent effective potential we obtain\footnote{
	Note that the potential $U_k$ itself has an $\text{O}(2N_\mathrm{c})$ symmetry at $k=\Lambda$, see \cref{eq:action}, which is preserved during the \gls{RG} flow, because the right-hand side of \cref{eq:MF-medium} solely depends on the $\text{O}(2N_\mathrm{c})$ invariant $\Delta_{c}^* \Delta_{c}$.}
\begin{align*}
	\numberthis\label{eq:MF-medium}
	\partial_t U_k &= v_{d-1} N_\mathrm{f} d_\gamma \,k^d \sum_{\chi = \pm 1} \frac{k+\chi \mu}{2E^\mathrm{F}_\chi}\bigl(1-2 n_\mathrm{F}(\beta E^\mathrm{F}_\chi)\bigr) \, ,
\end{align*}
where $v_{d-1} = \mathrm{vol}({d-1})/(2\uppi)^{d-1}$, $\mathrm{vol}(d-1)$ is the volume of a ($d-1$)-dimensional unit ball, $n_{\mathrm{F}}(x) = 1/({\rm e}^x+1)$ is the Fermi-Dirac distribution function and
\begin{align*}
	\numberthis\label{eq:E-F}
	E^\mathrm{F}_\chi = \sqrt{\frac{\bar{\Delta}^2}{2} + (k+\chi\mu)^2} \, 
\end{align*}
are the quark energy functions.

The \gls{RG} flow of the scale-dependent effective potential is predominantly driven by the so-called \gls{BCS} singularity in the limit of vanishing temperature, see, e.g., Ref.~\cite{Braun:2020bhy} for a detailed discussion in the context of \gls{RG} flows. 
More precisely, the curvature of the scale-dependent effective potential is divergent at $\bar{\Delta}=0$ from $k=\mu$ down to the \gls{IR} limit associated with $k \to 0$.
This phenomenon is of high relevance for the numerical treatment of the \gls{QDM}. 
Therefore, we shall discuss it in some detail in the following. 

Starting with the zero-temperature limit of \cref{eq:MF-medium}, we find 
\begin{align*}
	\numberthis\label{eq:MF-T0}
	\partial_t U_k &= v_{d-1} N_\mathrm{f} d_\gamma\, k^d \sum_{\chi = \pm 1} \frac{k+\chi \mu}{2E^\mathrm{F}_\chi}\,.
\end{align*}
An integration with respect to the \gls{RG} scale~$k$ from $k^\prime$ to $\Lambda>\mu$ then yields
\begin{widetext}
	\begin{align*}
		\numberthis\label{eq:MF-pot}
		U_{k^\prime}
		&=\frac{1}{2}\bar{\nu}^2 \bar{\Delta}^2 + v_{d-1} N_\mathrm{f} d_\gamma \int^{\Lambda}_{k'} \frac{\d k}{k} \, k^{d} \sum_{\chi = \pm 1} \frac{k+\chi \mu}{2E^\mathrm{F}_\chi} \\
		&=\frac{1}{2}\bar{\nu}^2 \bar{\Delta}^2 - \frac{v_{d-1} N_\mathrm{f} d_\gamma}{64} 
		\biggl[ -2\mu(13 \bar{\Delta}^2 -4\mu^2-4k^2) \sum_{\chi = \pm 1} \chi \sqrt{\tfrac{\bar{\Delta}^2}{2} + (k+\chi\mu)^2} \\
		&\quad\quad+ 2k( -4k^2+3\bar{\Delta}^2-4\mu^2 ) \sum_{\chi = \pm 1} \sqrt{\tfrac{\bar{\Delta}^2}{2} + (k+\chi\mu)^2} 
		-3\bar{\Delta}^2(\bar{\Delta}^2-8\mu^2) \sum_{\chi = \pm 1} \mathrm{tanh}^{-1}\biggl\{ \frac{k+\chi \mu}{\sqrt{\tfrac{\bar{\Delta}^2}{2} + (k+\chi\mu)^2}} \biggr\}\biggr]_{k^\prime}^\Lambda\,,
	\end{align*}
\end{widetext}
where we have used the action \eqref{eq:action}.
Expanding this result for $k^\prime>\mu$ about $\bar{\Delta} =0$ reveals that it is analytic at this point and yields a convergent Taylor series.
In contrast, for $k^\prime \leq \mu$ the result is nonanalytic at $\bar{\Delta} =0$ due to logarithmic contributions originating from the inverse of the hyperbolic tangent, namely $\tfrac{3}{8}\mu^2\bar{\Delta}^2\ln \tfrac{\bar{\Delta}^2}{\Lambda^2}$ for $k^\prime<\mu$ and $\tfrac{3}{16}\mu^2\bar{\Delta}^2\ln \tfrac{\bar{\Delta}^2}{\Lambda^2}$ for $k^\prime=\mu$, see also, e.g., Ref.~\cite{Geissel:2024nmx}.
It directly follows that the curvature $\partial^2_{\bar{\Delta}} U_{k^\prime}$ for $k^\prime \leq \mu$ scales as $\sim\mu^2 \ln \tfrac{\bar{\Delta}^2}{\Lambda^2}$ for $(\bar{\Delta}/\Lambda)^2 \to 0$, i.e., it tends to $-\infty$ in this limit.\footnote{
	Since the nonanalytic behavior is encoded in Eq.~\eqref{eq:MF-pot} in terms which become independent of the regulator in the limit~\mbox{$k^\prime \to 0$}, it follows that an expansion of the scale-dependent effective potential about $\bar{\Delta}=0$ is ill-defined for~$k^\prime \leq \mu$ for any regulator from the class of standard $3d$ regulators.  
	This issue may be circumvented by integrating out fluctuations about the Fermi surface but this then requires the use of regulators that break the Silver-Blaze symmetry by construction~\cite{Braun:2020bhy}.
	Note that, as discussed in Ref.~\cite{Topfel:2024cll}, an interchange of the integration of the flow equation over the \gls{RG} scale~$k$ and an expansion in powers of a background field is in general delicate. However, this does not underlie the appearance of a divergent curvature of the effective potential at the origin in the present case.}
Consequently, for any~$\mu\neq 0$, the ground state of our model is governed by \gls{SSB}, i.e., a nonzero diquark condensate $\bar{\Delta}_\mathrm{gs}$ is necessarily generated for $k' \leq \mu$ and persists in the \gls{IR} limit, irrespective of our choice for the finite parameter $\bar{\nu}^2>0$ in the action~\eqref{eq:action}. 
Moreover, from an expansion of $\partial_{\bar{\Delta}} U$ in small $(\mu/\Lambda)^2$ and $(\bar{\Delta}/\Lambda)^2$, we can deduce the characteristic \gls{BCS}-like $\mu$-dependence of the condensate from the RG flow, see, e.g., Ref.~\cite{Braun:2020bhy}:
\begin{align*}
	\numberthis\label{eq:BCS-scaling}
	\bar{\Delta}_\mathrm{gs} \propto \displaystyle{\rm e}^{-c/\mu^2} \, ,
\end{align*}
where $c>0$ is a $\bar{\nu}^2$-dependent constant.

To illustrate the manifestation of the \gls{BCS} singularity in the \gls{RG} flow, we show the derivative of the scale-dependent effective potential $\partial_{\bar{\Delta}} U_k$ for $\mu = 0.5\GeV$ as a function of~$\bar{\Delta}$ for different values of $k$ in \cref{fig:MF-BCS}. 
For vanishing temperature, we find that $\partial_{\bar{\Delta}} U_k$ is discontinuous at $\bar{\Delta} = 0$ for $k = \mu$ and consequently the curvature diverges.
For $k<\mu$, the discontinuity disappears but the curvature at $\bar{\Delta} = 0$ still diverges.

At nonzero temperature, the \gls{BCS} singularity is screened. 
This is demonstrated in \cref{fig:MF-BCS} for $T=5\MeV$. 
To be specific, we observe that the discontinuity in $\partial_{\bar{\Delta}} U_k$ at~$\bar{\Delta}=0$ for $k = \mu$ disappears and the region around the origin turns into a ``region of steep descent", being a relic of the \gls{BCS} singularity. 
Outside of this region, we find that our results for $\partial_{\bar{\Delta}} U_k$ for $T = 0$ and $T=5\MeV$ agree almost identically in the \gls{IR} limit. 
Indeed, we observe pointwise convergence of our finite-temperature results for~$\partial_{\bar{\Delta}} U_k$ to the zero-temperature result as the temperature is decreased. 
In particular, the position of the nontrivial minimum of the potential~$U_k$ (zero of~$\partial_{\bar{\Delta}} U_k$) approaches smoothly the value obtained in the zero-temperature calculation. 

\subsection{Local potential approximation}\label{sec:LPA}
We now discuss the construction of truncations that incorporate bosonic fluctuations of the diquark fields at lowest order in the derivative expansion, commonly referred to as \glspl{LPA}.
Due to the complex nature of the diquark fields and their coupling to the ``gauge field" $\bm{\mu}$, this task is nontrivial.

In \gls{LPA} the change of the scale-dependent effective potential is taken into account in the truncation map.
To enable a discussion of the Silver-Blaze symmetry, the potential -- being explicitly dependent on the chemical potential -- must be consistently generalized to the complex-valued chemical potential~$\bm\mu$. 
Moreover, the truncation map should preserve complex analyticity in~$\bm\mu$.
A natural choice that does not break analyticity explicitly is
\begin{align}
	\numberthis\label{eq:new-potential}
	\bm U_k(\bar{\Delta},\bm \mu) = \mathrm{proj}_{\bm U}(\Gamma_k) = \frac{1}{V_d} \Gamma_k[\Phi=\tilde{\Phi}_0] \, ,
\end{align}
where the field configuration is now
\begin{align*}
	\numberthis\label{eq:new-super-field-evaluation}
	\tilde\Phi_0 = (&\Delta_c=\tfrac{1}{\sqrt{2}}\bar{\Delta}\,\delta_{c 3}, \Delta^*_c=\tfrac{1}{\sqrt{2}}\bar{\Delta}\,\delta_{c 3}, \\
	&\quad \bar\psi_c=0, \psi_c=0, \bm \mu) \,.
\end{align*}
Here, $\bm\mu$ is still a complex number and is not yet evaluated at a real number $\mu$ as done in \cref{eq:super-field-evaluation}.
The scale-dependent effective potential \eqref{eq:scale-dependent-effective-potential} and the generalized scale-dependent effective potential \eqref{eq:new-potential} are related~by 
\begin{align*}
	\numberthis\label{eq:relation-potentials}  
	U_k(\bar{\Delta},\mu)=\bm U_k(\bar{\Delta},\bm \mu = \mu) \, .
\end{align*}
Note that, in general, $\bm U_k(\bar{\Delta},\bm \mu)$ may depend on $\Re(\bm \mu)$ and $\Im(\bm \mu)$ separately, if it is nonanalytic in $\bm \mu$.

With~$\bm U_k$ at hand, we can now define a quite general abstract truncation map for \gls{LPA}:
\begin{align*}
	{\mathds{T}}^{\mathrm{LPA}} \circ \Gamma_k[\Phi] &= {\mathds{T}}^{\mathrm{MFA}} \circ \Gamma_k[\Phi] + \int_x \bm U_k(\sqrt{2\Delta^*_c \Delta_c}, \bm \mu) \\
	&\quad + \int_x \vec\nabla \Delta^*_c \vec\nabla \Delta_c + (\text{t.d.k.}) \, . 	\numberthis\label{eq:LPA-truncation-map}
\end{align*}
Here, ${\mathds{T}}^{\mathrm{MFA}} \circ \Gamma_k[\Phi]$ is the \gls{MFA} truncation map given in \cref{eq:mf-truncation} and ``t.d.k.'' stands for ``temporal diquark kinetic'' terms (i.e., contributions involving temporal derivatives of the diquark fields), which are assumed to be scale-independent within \gls{LPA}. 
Furthermore, we have promoted the field-argument $\bar{\Delta}$ of the generalized scale-dependent effective potential $\bm U_k$ to a position-dependent quantity, i.e., $\sqrt{2\Delta^*_c(x) \Delta_c(x)}$.
We will come back to the discussion of the Silver-Blaze symmetry in \gls{LPA} in a moment, but first let us discuss possible choices of temporal kinetic terms and their implications. 

As a consequence of our discussion in \cref{sec:model}, it may be tempting to employ the following ansatz as a first version for the temporal diquark kinetic term:
\begin{align*}
	\numberthis\label{eq:kinetic-v1}
	(\text{t.d.k.})_{\mathrm{1}} = \int_x D_{-} \Delta^*_c D_{+} \Delta_c\,.
\end{align*}
We shall refer to the truncation map with this temporal kinetic term as ${\mathds{T}}^{\mathrm{LPA}_1}$.
However, this choice leads to {\it double counting} in our calculations since the covariant temporal derivatives contribute a term $\sim \bm \mu^2 \Delta^*_c \Delta_c$ to the truncation map which is already included in corresponding second-order terms in the generalized scale-dependent effective potential~$\bm U_k$.
More explicitly, when evaluating \cref{eq:LPA-truncation-map} at $\tilde{\Phi}_0$ with \cref{eq:kinetic-v1} as the temporal kinetic term for the diquarks, we obtain 
\begin{align*}
	\numberthis\label{eq:double-counting}
	\frac{1}{V_d}({\mathds{T}}^{\mathrm{LPA}_1} \circ \Gamma_k)[\tilde{\Phi}_0]&=\bm U_k(\bar{\Delta}, \bm \mu) - \frac{1}{2}F^2 \bm \mu^2 \bar{\Delta}^2\\
	& \neq \frac{1}{V_d}\Gamma_k[\tilde{\Phi}_0] \stackrel{\eqref{eq:new-potential}}{=} \bm U_k(\bar{\Delta}, \bm \mu) \, .
\end{align*}
For nonzero $\bm \mu$, this is in contradiction to \cref{eq:projections-prop}; the latter requires to leave the generalized scale-dependent effective potential unchanged.
Below, in our discussion of the flow equations, which follow from our truncation maps, we shall see that this {\it double-counting} issue also leads to problems with respect to the phenomenological interpretation of the results. 

From these considerations we deduce that a proper version of \gls{LPA} must exclude potential-like contributions in the temporal diquark kinetic term. 
This issue may be solved by employing 
\begin{align*}
	\numberthis\label{eq:kinetic-v2}
	(\text{t.d.k.})_{\mathrm{2}} = \int_x \Big(&D_{-} \Delta^*_c D_{+} \Delta_c + F^2 \bm \mu^2 \Delta^*_c\Delta_c \Big) \,.
\end{align*}
In the following we shall refer to the truncation map with this temporal kinetic term as ${\mathds{T}}^{\mathrm{LPA}_2}$.
Note that, although not explicitly present in the truncation map, a
term $\sim \bm \mu^2 \Delta^*_c \Delta_c$ is dynamically generated in the \gls{RG} flow~\cite{Braun:2018svj} and eventually appears implicitly as part of the generalized scale-dependent effective potential $\bm U_k(\bar{\Delta}, \bm \mu)$ after performing the projection \eqref{eq:scale-dependent-effective-potential}. In our discussion below,
we shall carefully examine the consequences of these
two choices for the \gls{LPA} truncation map, also from a
phenomenological standpoint. 

Let us now discuss the Silver-Blaze symmetry and whether it is realized or not for our two versions of \gls{LPA}.
At first glance, by looking at the temporal diquark kinetic terms, we note that $(\text{t.d.k.})_{\mathrm{1}}$ is invariant under the Silver-Blaze transformations \eqref{eq:LPA-truncation-map} whereas $(\text{t.d.k.})_{\mathrm{2}}$ is not. 
However, this does not directly imply that ${\mathds{T}}^{\mathrm{LPA}_1}$ is a Silver-Blaze-invariant truncation, since the contribution from the generalized scale-dependent effective potential must also be taken into account.
In fact, both truncations are likely to break the Silver-Blaze symmetry. 
This is because the generalized scale-dependent effective potential in the truncation map \eqref{eq:LPA-truncation-map} can develop a highly nontrivial dependence on $\Im(\bm\mu)$ during the \gls{RG} flow, see also App.~\ref{sec:Silver-Blaze}.\footnote{
	Recall the transformation of the chemical potential in Eq.~\eqref{eq:sbs-transformation} and the fact that the generalized scale-dependent effective potential is in general built up from powers of the chemical potential up to arbitrarily high orders.}
Its transformation behavior under Silver-Blaze transformations can therefore, in general, not be compensated by the corresponding transformation of the simple temporal diquark kinetic terms $(\text{t.d.k.})_{\mathrm{1}}$ and $(\text{t.d.k.})_{\mathrm{2}}$, respectively.\footnote{
	Note that alternative generalizations of the scale-dependent effective potential with full $\mu$-dependence (e.g., involving only $\Re(\bm\mu$)) may also be invariant under Silver-Blaze transformations. However, unlike \cref{eq:new-potential}, they explicitly violate complex analyticity and consequently cannot be utilized to study features such as the Silver-Blaze property.}
Moreover, depending on the initial condition for the generalized scale-dependent effective potential, the Silver-Blaze symmetry may already be explicitly broken at the \gls{UV} scale.

In general, if we would like to construct a Silver-Blaze-symmetric version of \gls{LPA} and at the same time would like to use the full $\bm\mu$-dependence of the generalized scale-dependent effective potential as shown in \cref{eq:LPA-truncation-map}, then this would require a highly complicated and \gls{RG}-scale dependent temporal kinetic term for the diquark fields (in general involving derivatives of arbitrarily higher orders).
Therefore, with respect to studies in \gls{LPA}, we discard the Silver-Blaze symmetry and will use ${\mathds{T}}^{\mathrm{LPA}_2}$ for all numerical calculations presented below.

For completeness, let us now discuss the special case where $\bm U_k(\bar{\Delta},\bm \mu)$ is an analytic function in $\bm \mu$ in a neighborhood of $\bm \mu=0$. 
Complex analyticity implies that $\bm U_k(\bar{\Delta},\bm \mu)$ can be written in a Taylor series in $\bm \mu$ around $\bm \mu=0$ and, in particular, we are allowed to promote $\mu$ in $U_k(\bar{\Delta}, \mu)$ to a complex number, i.e., we have $\bm U_k(\bar{\Delta}, \bm\mu) \equiv U_k(\bar{\Delta}, \bm\mu)$.
Furthermore, in case of a complex analytic and Silver-Blaze-symmetric scale-dependent effective average action, any $\bm \mu$-dependence in the pure diquark sector has to be encoded in terms of the following building blocks:
\begin{align*}
	\numberthis\label{eq:building-block}
	D^n_{-} \Delta^*_c D^m_{+} \Delta_c \, ,
\end{align*}
where $m$ and~$n$ are integers with $m, n \geq 0$.
Consequently, in a Silver-Blaze-symmetric version of the truncation map \eqref{eq:LPA-truncation-map}, the pure diquark sector can be written~as 
\begin{align*}
	\numberthis\label{eq:SBS-improved-truncation}
	&\int_x \Big( \bm U_k(\sqrt{2\Delta^*_c \Delta_c}, \bm \mu) + (\text{t.d.k.}) \Big)\\
	&\quad \stackrel{!}{=} \int_x \Big(U_k(\sqrt{2\Delta^*_c \Delta_c}, \mu=0) + g_k(\{ D^n_{-} \Delta^*_c D^m_{+} \Delta_c\}) \Big)\, .
\end{align*}
Here, $U_k(\sqrt{2 \Delta^*_c \Delta_c}, \mu=0)$ is the scale-dependent effective potential evaluated at vanishing chemical potential. 
The \gls{RG}-scale dependent function $g_k$ does not carry an explicit dependence on~$\bm \mu$. 
It is solely constructed from the building blocks given in \cref{eq:building-block}. 
Apparently, the right-hand side of \cref{eq:SBS-improved-truncation} is invariant under the Silver-Blaze transformations in \cref{eq:sbs-transformation}.
However, similar to the previous discussion without limitation to complex analyticity, we can deduce from \cref{eq:SBS-improved-truncation} that the temporal diquark kinetic term is in general given by a highly nontrivial \gls{RG}-scale dependent function. 
Consequently, in the analytic case, any truncation that includes only a finite number of temporal derivative terms of the diquark fields but the {\it full} $\mu$-dependence of the scale-dependent effective potential as in \cref{eq:LPA-truncation-map} will in general break the Silver-Blaze symmetry.
However, if the truncation map includes only a Taylor expansion of the scale-dependent effective potential in $\bm \mu$ up to a finite order, i.e., if the full $\bm \mu$-dependence is not taken into account, it is still possible to preserve the Silver-Blaze symmetry.
Using \cref{eq:SBS-improved-truncation} as starting point for the construction of such a  truncation, it is reasonable to define the lowest order of the derivative expansion by the truncation map
\begin{align*}
	\numberthis\label{eq:LPA-SBS-truncation-map}	
	{\mathds{T}}^{\mathrm{LPA},\mathrm{SBS}} \circ \Gamma_k[\Phi] &= {\mathds{T}}^{\mathrm{MFA}} \circ \Gamma_k[\Phi] \\
	&\quad + \int_x U_k(\sqrt{2\Delta^*_c \Delta_c}, \mu=0) \\
	&\quad + \int_x \vec\nabla \Delta^*_c \vec\nabla \Delta_c + D_{-} \Delta^*_c D_{+} \Delta_c \, ,
\end{align*}
where SBS stands for Silver-Blaze symmetric. 
This map includes $\bm \mu$ dependence up to $\mathcal{O}(\bm\mu^2)$.
Note that, arbitrarily high orders of $\bm \mu$ and derivatives of diquark fields are still generated in the \gls{RG} flow.
The scale-dependent effective average action $\Gamma_k$ resulting from this map will be Silver-Blaze symmetric as well, provided that the regulator does not break the Silver-Blaze symmetry. 
However, since the complex analyticity of the scale-dependent effective average action cannot be guaranteed and the flow equation for this truncation exhibits systematic issues in the case of \gls{SSB}, as we shall discuss next in \cref{sec:flow-equation}, we do not employ it in our studies of specific models in Secs.~\ref{sec:results} and \ref{sec:applications}.  
Nevertheless, this truncation may still prove useful in applications where the diquark field is only used as an auxiliary field to resolve momentum dependences of four-quark correlators in phases without (color-)superconducting condensate, e.g., at high temperature and chemical potential in case of \gls{QCD}. 

\subsubsection{Flow equation}\label{sec:flow-equation}
\begin{figure}
	\includegraphics{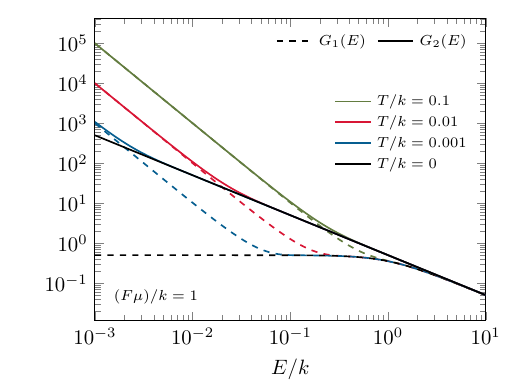}
	\caption{\label{fig:propagator-singularity}Diquark propagator functions $G_1$, see \cref{eq:bosonic-propagator}, and $G_2$, see \cref{eq:bosonic-propagator-2}, for $(F \mu)/k = 1$ and various temperatures with~$F=2$.  
	Note that, at~$T=0$, the propagator function $G_1$ remains finite in the limit~$E/k\to 0$, in contrast to the propagator function~$G_2$.}
\end{figure}
For all versions of \gls{LPA} discussed above, the flow equation for the scale-dependent effective potential assumes the same functional form:
\begin{align*}
	\numberthis\label{eq:LPA-flow-equation}
	\partial_t U_k &= -v_{d-1} k^{d} \, 2(N_\mathrm{c} - 1) G_1(E^{\mathrm{B}}_1) \\
	&\quad - v_{d-1}  k^{d}\sum_{\chi=\pm1}\Big(1 + \chi\frac{4 F^2 \mu^2}{\xi^2_+ - \xi^2_-}\Big) G_2(\xi_\chi)\\
	&\quad+ v_{d-1}  k^d N_\mathrm{f} d_\gamma \sum_{\chi = \pm 1} \frac{k+\chi \mu}{2E^\mathrm{F}_\chi}\bigl(1-2 n_\mathrm{F}(\beta E^\mathrm{F}_\chi)\bigr) \, ,
\end{align*}
where $E^\mathrm{F}_\chi$ is the quark energy function defined in \cref{eq:E-F} and
\begin{align*}
	\numberthis\label{eq:xis}
	\xi^2_{\pm} &= E_+^2 + 2 F^2\mu^2 \\
	&\quad\pm \sqrt{(E^2_+ + 2 F^2\mu^2)^2 - E^4_+ + E_-^4}
\end{align*}
with $E^2_\pm = \frac{1}{2}(E_2^{\mathrm{B}})^2 \pm \frac{1}{2}(E_1^{\mathrm{B}})^2$.
The diquark energy functions $E^{\mathrm{B}}_1$ and $E^{\mathrm{B}}_2$ differ for our different \gls{LPA} versions. 
We shall define them below.

In the flow equation \eqref{eq:LPA-flow-equation} two diquark propagator functions appear:
\begin{align*}
	\numberthis\label{eq:bosonic-propagator}
	G_1(E) &= \frac{k}{2\sqrt{F^2 \mu^2 + E^2}} \Big( 1 +\\
	&\quad + n_{\mathrm{B}}(\beta (\sqrt{F^2 \mu^2 + E^2}-F \mu)) \\
	&\quad + n_{\mathrm{B}}(\beta (\sqrt{F^2 \mu^2 + E^2}+F \mu))
	\Big) 
\end{align*}
and
\begin{align*}
	\numberthis\label{eq:bosonic-propagator-2}
	G_2(E) &= \frac{k}{2 E} \Big( 1 + 2n_{\mathrm{B}}(\beta E) \Big) \, ,
\end{align*}
where $n_\mathrm{B}(x)=1/({\rm e}^x - 1)$ is the Bose-Einstein distribution function.
Note that $0 \leq 4F^2\mu^2/(\xi^2_+ - \xi^2_-) \leq 1$ for all~$\mu$ and $(E^{\mathrm{B}}_1)^2>0, (E^{\mathrm{B}}_2)^2>0$.

In the limit of vanishing chemical potential the flow equation reduces to
\begin{align*}
	\partial_t U_k &= -v_{d-1} k^{d} (2N_\mathrm{c} - 1) G_2(E^{\mathrm{B}}_1) - v_{d-1} k^{d} G_2(E^{\mathrm{B}}_2)\\
	&\quad+ v_{d-1} k^{d+1} N_\mathrm{f} d_\gamma \frac{1-2 n_\mathrm{F}(\beta E^\mathrm{F}_0)}{E^\mathrm{F}_0} \, ,\numberthis\label{eq:LPA-flow-equation-mu=0}
\end{align*}
where we have used $\xi^2_{+}|_{\mu = 0} = (E^{\mathrm{B}}_2)^2$, $\xi^2_{-}|_{\mu = 0} = (E^{\mathrm{B}}_1)^2$, and the fact that $4F^2\mu^2/(\xi^2_+ - \xi^2_-)$ vanishes at least linearly in $\mu$.
The bosonic contribution can be split into two parts: one depending solely on $E^{\mathrm{B}}_1$ and the other depending solely on $E^{\mathrm{B}}_2$.
The factor $2 N_\mathrm{c} - 1 = 5$ in the $E^{\mathrm{B}}_1$ contribution is the number of broken generators as determined by the symmetry-breaking pattern~\eqref{eq:SSB-pattern}.
At nonzero chemical potential, see \cref{eq:LPA-flow-equation}, the bosonic contributions to the \gls{RG} flow of the scale-dependent effective potential cannot be split into the aforementioned two parts, see, e.g., Refs.~\cite{Nielsen:1975hm,Andersen:2005yk,Watanabe:2019xul} for a more detailed analysis of this fact. 
Note that the mixing of the $E^{\mathrm{B}}_1$ and $E^{\mathrm{B}}_2$ dependence in the bosonic contribution at nonzero $\mu$ renders the flow equation~\eqref{eq:LPA-flow-equation} numerically more challenging than, e.g., the \gls{LPA} flow equation of the quark-meson model where the corresponding energy functions associated with mesons do not mix, even at nonzero $\mu$, see App.~\ref{sec:numerical-treatment} for details of the numerical implementation.

Provided that the squared diquark energy functions are greater than zero, $(E_1^{\mathrm{B}})^2 > 0$ and $(E_2^{\mathrm{B}})^2 > 0$, we have $\xi^2_{\pm} > 0$ for all $\mu$, see \cref{eq:xis}.
If one of the (squared) energy functions approaches zero, we have $\xi_- \to 0$ while $\xi_+$ remains positive, $\xi_+ > 0$.
In this case, the diquark propagator function~$G_2(\xi_-)$ diverges in the \gls{RG} flow, see \cref{eq:LPA-flow-equation}. 
To be specific, we have $G_2(E) \sim k/(\beta E^2)$ for $E \to 0$. 
This behavior of the propagator prevents the \gls{RG} flow from actually reaching the singularity of $G_2(E)$ for~$k>0$. 
For the diquark energy functions, this implies that
\begin{align*}
	\numberthis\label{eq:constraints}
	(E_1^{\mathrm{B}})^2 \geq 0 \,, \quad 	(E_2^{\mathrm{B}})^2 \geq 0 \, .
\end{align*}
for any~$k\geq 0$. 
In fact, as we shall discuss below, this ``self-healing'' property of the flow equation eventually ensures that the scale-dependent effective potential becomes convex in the \gls{IR} limit, see Refs.~\cite{Litim:2006nn,Zorbach:2024zjx}.
Note that the singular behavior becomes weaker as the temperature is decreased. 
In fact, the propagator function only diverges as $G_2(E) \sim k / (2 E)$ at~$T=0$. 

The situation is different for the function $G_1$. 
While it also diverges as $G_1(E) \sim k/(\beta E^2)$ in the limit~$E\to 0$ for $T>0$, 
it converges pointwise to $1/(2 \sqrt{F^2 \mu^2 + E^2})$ for all $E > 0$  in the zero-temperature limit and approaches $1/(2 \sqrt{F^2 \mu^2})$ for $E \to 0$, which is finite.
Thus, $G_1$ does not exhibit a divergent behavior in the vicinity of $E = 0$ in the zero-temperature limit.\footnote{
	In contrast to neglecting the contributions associated with the function~$G_1$ in the \gls{RG} flow, neglecting the terms associated with~$G_2$ would imply that the \gls{RG} flow loses the aforementioned ``self-healing'' property \cite{Zorbach:2024zjx} in the limit~$T\to 0$, i.e., the system may potentially flow into a point with~$E=0$ for~$k>0$.}
For illustration purposes, we compare the functions~$G_1$ and~$G_2$ for various temperatures including the zero-temperature limit in \cref{fig:propagator-singularity}.

Let us start the discussion of the flow equation for the previously introduced Silver-Blaze-symmetric version of \gls{LPA}, see \cref{eq:LPA-SBS-truncation-map}. 
Its energy functions read 
\begin{subequations}
	\label{eq:bosonic-energy-function-sbs-sym}
	\begin{align}
		E^{\mathrm{B}}_1 &= \sqrt{k^2 + \tfrac{1}{\bar{\Delta}} \partial_{\bar{\Delta}} U_k(\bar{\Delta}, \mu=0) - F^2 \mu^2} \, , \\ E^{\mathrm{B}}_2 &= \sqrt{k^2 + \partial^2_{\bar{\Delta}} U_k(\bar{\Delta}, \mu=0) - F^2 \mu^2} \,.
	\end{align}
\end{subequations}
To obtain $U_k(\bar{\Delta}, \mu)$ for this truncation, we first compute $U_k(\bar{\Delta}, \mu=0)$ at vanishing chemical potential which can then be plugged into the right-hand side of the flow equation \eqref{eq:LPA-flow-equation} for nonzero chemical potential.
After that, the resulting equation can simply be integrated with respect to the \gls{RG} scale $k$.
Provided that the expressions under the square roots of the energy functions \eqref{eq:bosonic-energy-function-sbs-sym} are greater than zero, this procedure is in principle well-defined.
However, for a given solution $U_k(\bar{\Delta}, \mu=0)$, positions in $\bar{\Delta}$-space may exist (depending on $\mu$) for which the expressions under the square roots of the energy functions \eqref{eq:bosonic-energy-function-sbs-sym} become negative, especially in the case of \gls{SSB}, leading to an ill-defined $U_k(\bar{\Delta}, \mu)$. 
These points would then need to be excluded from the analysis, spoiling the consistency and applicability of this truncation.
Moreover, since the scale-dependent effective potential is not computed self-consistently with this procedure, it may in general not be convex. 

These issues can be resolved by using our \gls{LPA} truncations which take the $\mu$-dependence of the scale-dependent effective potential into account.
We start by examining the diquark energy functions for the truncation map~${\mathds{T}}^{\mathrm{LPA}_1}$ which suffers from the double-counting issue as discussed above. 
In this case, the diquark energy functions read
\begin{subequations}
	\label{eq:bosonic-energy-functions-v1}
	\begin{align}
		E^{\mathrm{B}}_1 &= \sqrt{k^2 + \tfrac{1}{\bar{\Delta}} \partial_{\bar{\Delta}} U_k - F^2 \mu^2} \, , \\
		E^{\mathrm{B}}_2 &= \sqrt{k^2 + \partial^2_{\bar{\Delta}} U_k- F^2 \mu^2} \,.
	\end{align}
\end{subequations}
These are similar to the ones in \cref{eq:bosonic-energy-function-sbs-sym}. 
However,  $U_k$ now denotes the full $\mu$-dependent scale-dependent effective potential $U_k(\bar{\Delta}, \mu)$.
Plugging these energy functions into the diquark propagator, we find that the presence of the aforementioned singular point in the flow equation results in the following constraints for $U_k$ in the \gls{IR} limit:
\begin{align*}
	\numberthis\label{eq:double-counting-2}
	\partial^2_{\bar{\Delta}} U \geq F^2 \mu^2 \, , \quad \frac{1}{\bar{\Delta}}\partial_{\bar{\Delta}} U \geq F^2 \mu^2 \, ,
\end{align*}
In words, the double-counting issue prevents the dynamical formation of a nontrivial minimum in the effective potential as associated with \gls{SSB}. 
Therefore, from here on, we shall discard this truncation.

Finally, let us discuss our third \gls{LPA} truncation, i.e., ${\mathds{T}}^{\mathrm{LPA}_2}$, which does not suffer from the double-counting issue.
Here, the diquark energy functions read
\begin{align*}
	\numberthis\label{eq:bosonic-energy-functions-v2}
	E^{\mathrm{B}}_1 = \sqrt{k^2 + \tfrac{1}{\bar{\Delta}} \partial_{\bar{\Delta}} U_k} \, , \quad E^{\mathrm{B}}_2 = \sqrt{k^2 + \partial^2_{\bar{\Delta}} U_k} \,.
\end{align*}
Plugging these energy functions into the diquark propagator, we now find that the singularity of the propagator~$G(E)$ entering the flow equation for the scale-dependent effective potential (see our discussion of \cref{eq:bosonic-propagator} above) yields the following constraints for~$U_k$ in the \gls{IR} limit:
\begin{align*}
	\numberthis\label{eq:convexity}
	\partial^2_{\bar{\Delta}} U \geq 0 \, , \quad \frac{1}{\bar{\Delta}}\partial_{\bar{\Delta}} U \geq 0\,.
\end{align*}
This implies that the scale-dependent effective potential becomes convex as~$k\to 0$, and that a nontrivial ground state as associated with~\gls{SSB} can be dynamically generated in the \gls{RG} flow. 
With this at hand, we shall discuss the phase structure of the \gls{QDM} model in \cref{sec:results}. 

\subsubsection{Ultraviolet potential}
Let us now discuss the initial condition for the scale-dependent effective potential at the initial \gls{RG} scale $\Lambda$.
For all discussed versions of \gls{LPA}, it has the form 
\begin{align*}
	\numberthis\label{eq:initial-condition-for-U}
	U_{\Lambda}(\bar{\Delta}, \mu) = \frac{1}{2}\bar\nu^2 \bar{\Delta}^2 \, ,
\end{align*}
which is $\mu$-independent.
Here, we would like to emphasize that the initial condition for the scale-dependent effective average action $\Gamma_{\Lambda} \equiv S$ should not be confused with the truncation ${\mathds{T}}\circ \Gamma_\Lambda$ at the cutoff scale $\Lambda$.
Since we discuss the lowest order in the derivative expansion, the truncation ${\mathds{T}}\circ \Gamma_k$ (even at $k = \Lambda$) comes with a kinetic term for the diquark fields, see \cref{eq:LPA-truncation-map}.
This is not the case for the action $\Gamma_{\Lambda} \equiv S$, see \cref{eq:action}.
Thus, we have $\Gamma_{\Lambda} \neq {\mathds{T}}\circ \Gamma_\Lambda$ in our present study.\footnote{
	However, we are still fulfilling the property \eqref{eq:projections-prop} for the only projection occurring in our truncation, i.e., the scale-dependent effective potential.}
This discrepancy can be lifted by going to a higher order in the derivative expansion by including the scale dependence of the wave-function renormalization of the diquark fields with initial condition~$Z_{k\to \Lambda}\to 0$.

If we had included a Silver-Blaze-symmetric kinetic term for the diquark fields in the action~\eqref{eq:action}, which would have corresponded to using $U_{\Lambda}(\bar{\Delta}, \mu) = \frac{1}{2}(\bar\nu^2 - F^2\mu^2) \bar{\Delta}^2$ as initial condition for~$U_k$ instead of \cref{eq:initial-condition-for-U}, the squared \gls{UV} mass $(\bar\nu^2 - F^2\mu^2)$ would decrease for increasing $\mu$, ``boosting" the formation of the diquark condensate.
Eventually, in the limit $F^2\mu^2 \to \bar\nu^2$ corresponding to~$U_{\Lambda}\equiv 0$, the diquark condensate would even diverge, which could, however, be cured by including higher order diquark self-interactions in the initial condition at the \gls{UV} scale~$\Lambda$. 
Instead, in our study, the quarks tend to drive the system dynamically into the regime associated with \gls{SSB}. 
We emphasize that this is different from models where the bosons are the fundamental degrees of freedom. 
For example, in purely bosonic models, a negative curvature of the scale-dependent effective potential at its origin is necessarily required at the initial \gls{RG} scale in order to encounter \gls{SSB} in the \gls{IR} limit, see also our discussion of Bose-Einstein condensation in the context of a \gls{RBG} in \cref{sec:rbg-at-finite-density}. 

Finally, we would like to emphasize that the aforementioned double-counting issue can {\it not} be resolved by solely adapting the initial condition.
However, it can be resolved by changing the definition of the scale-dependent effective potential in the flow equation.
This becomes apparent by introducing a new scale-dependent effective potential $V_k$ defined as $V_k = U_k + \tfrac{1}{2} F^2 \mu^2 \bar{\Delta}^2$.
Inserting it into the flow equation~\eqref{eq:LPA-flow-equation} with the energy functions~\eqref{eq:bosonic-energy-functions-v2}, we obtain the flow equation with the double-counting issue, where $U_k$ has been replaced by $V_k$.
Therefore, by initializing the ``double-counting flow equation" for $V_k$ with $V_\Lambda = \frac{1}{2}\bar\nu^2 \bar{\Delta}^2 + \tfrac{1}{2} F^2 \mu^2 \bar{\Delta}^2$, we obtain the same effective potential $U$ in the \gls{IR} limit as that obtained from the flow equation without double counting when we employ the relation $U_{k = 0} = V_{k = 0} - \tfrac{1}{2} F^2 \mu^2 \bar{\Delta}^2$.

\section{Phase structure of the quark-diquark model}\label{sec:results}
\begin{figure*}
	\subfloat[\label{fig:LPA-V1-phase-diagram-contour-full}%
	]{%
		\includegraphics{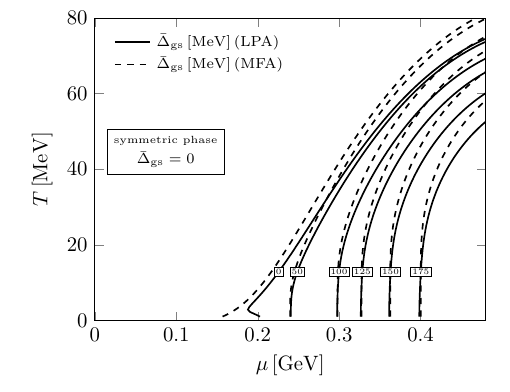}%
	}\hfill
	\subfloat[\label{fig:LPA-V1-phase-diagram-contour-first-order-regime}%
	]{%
		\includegraphics{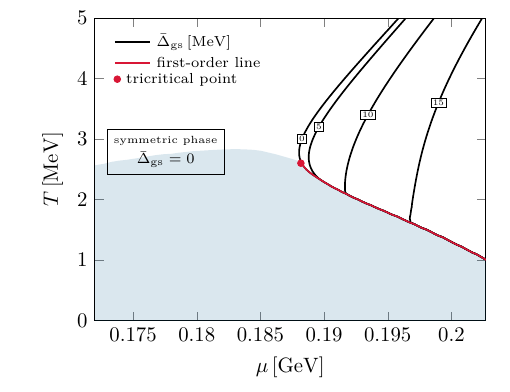}%
	}
	\caption{\label{fig:LPA-V1-phase-diagram-contour}Phase diagram of the \gls{QDM} for \gls{LPA} as defined by the truncation~${\mathds{T}}^{\mathrm{LPA}_2}$. 
	In the left panel, the contour lines are associated with the size of the condensate in \gls{LPA} (solid lines) and \gls{MFA} (dashed lines), respectively. 
	The right-hand panel shows a zoom into the first-order region depicted in the left panel. 
	This region opens up at low temperatures and moderate chemical potentials. 
	The red line depicts the first-order phase transition. 
	The red dot is the tricritical point located at $(\mu,T)\approx (188.2\MeV, 2.6\MeV)$. 
	In the shaded area, we find indications for the formation of a scale-dependent effective potential of the type associated with a first-order phase transition but the symmetry is eventually restored in the \gls{RG} flow, see, e.g., \cref{fig:LPA-V1-first-order-flow-mu-180}.}	
\end{figure*}
\begin{figure*}
	\subfloat[\label{fig:LPA-V1-first-order-flow-mu-180}%
	]{%
		\includegraphics{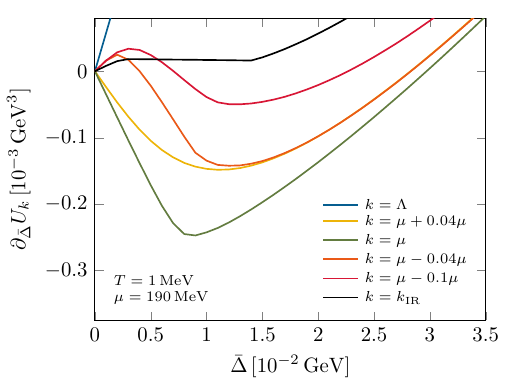}%
	}\hfill
	\subfloat[\label{fig:LPA-V1-first-order-flow-mu-200}%
	]{%
		\includegraphics{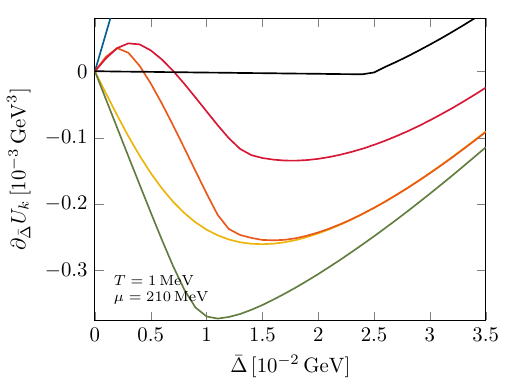}%
	}
	\caption{\label{fig:LPA-V1-first-order-flow}Derivative of the scale-dependent effective potential for various \gls{RG} scales (see left panel) at $T = 1 \MeV$ for $\mu=190\MeV$ (left panel, symmetric phase) and $\mu=210\MeV$ (right panel, phase associated with \gls{SSB}).}
\end{figure*}
\begin{figure}
	\includegraphics{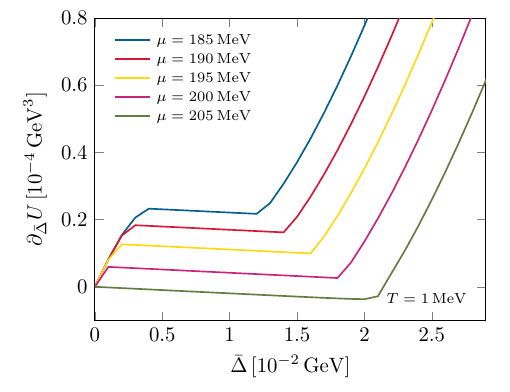}
	\caption{\label{fig:LPA-second-order}Derivative of the effective potential at $T = 1 \MeV$ for various values of $\mu$ near the first-order phase transition, see \cref{fig:LPA-V1-phase-diagram-contour}.}
\end{figure}
As discussed in detail in the previous section we shall restrict ourselves to the truncation \eqref{eq:LPA-truncation-map} with the temporal kinetic term for the diquark fields as defined in \cref{eq:kinetic-v2}, i.e., we choose ${\mathds{T}}^{\mathrm{LPA}_2}$ and compare the results to those from \gls{MFA}. 
To this end, we have to specify $\bar{\nu}^2$ in the action $\Gamma_\Lambda[\Phi] \equiv S[\Phi]$, see \cref{eq:action}. 
Inspired by \gls{QCD} phenomenology, see, e.g., Ref.~\cite{Morones-Ibarra:2017avu} and former studies on diquark condensation, see Refs.~\cite{Alford:2001dt,Alford:1997zt,Rapp:1997zu,Schafer:1999jg}, we choose $\bar{\nu}^2$ such that $\bar{\Delta}_\mathrm{gs} /\sqrt{2} \approx 100\MeV$ at a low temperature of $T=10\MeV$ and $\mu=350\MeV$, where the chiral symmetry is expected to be restored in \gls{QCD} with two massless quark flavors. 
For both \gls{MFA} and \gls{LPA}, we find $\bar{\nu}_\mathrm{MFA}^2=0.0585\GeV^2$ and $\bar{\nu}_\mathrm{LPA}^2=0.0575\GeV^2$, respectively. 
A complete list of the (numerical) parameters underlying all figures in the present work can be found in \cref{tab:numerical-data} in \cref{sec:numerical-treatment}.

Let us first consider the phase diagrams obtained from \gls{MFA} and \gls{LPA} in \cref{fig:LPA-V1-phase-diagram-contour-full} which are found to be quite similar.
Indeed, the contour lines associated with the size of the condensate agree very well at rather large~$\mu$ and small~$T$. 
While this observation is a trivial consequence of our parameter fixing procedure near $\mu \approx 350\MeV$, the good agreement over a wide range of chemical potentials is still remarkable and states that fluctuation effects are suppressed in this region of the phase diagram, at least with respect to the minimum of the effective potential. 
By increasing the temperature, we then find that the contour lines of the condensate start to deviate and bosonic fluctuations tend to boost the restoration of the symmetry. 

The most interesting difference between \gls{MFA} and \gls{LPA} occurs at the phase transition at low temperatures, $T\lesssim4\MeV$. 
While the phase boundary within \gls{MFA} is of second-order and bends towards lower $\mu$ for decreasing $T$,\footnote{
	In \gls{MFA}, we expect that the phase boundary tends to zero asymptotically for~$\mu\to 0$, since \gls{SSB} occurs at all $\mu$ for $T=0$ for our choice of model parameters, see \cref{sec:mfa}.} 
the phase boundary within \gls{LPA} has a turning point 
and bends towards higher $\mu$ for decreasing $T$ below that point.
This can be seen more clearly in \cref{fig:LPA-V1-phase-diagram-contour-first-order-regime}. 
From \cref{fig:LPA-V1-phase-diagram-contour-first-order-regime}, we also deduce that the order of the phase transition changes from second to first order slightly below the turning point, which is reflected by the fact that the contour lines merge at the phase boundary. 
This implies that a \textit{tricritical point} occurs at $(\mu,T)\approx (188.2\MeV, 2.6\MeV)$, where first- and second-order phase transition lines merge.
Since we are numerically restricted to computations with $T \gtrsim 1\MeV$, see \cref{sec:numerical-treatment}, we can not make definitive statements about the shape of the first-order phase transition at even lower temperatures. 
In any case, this phase structure indicates that the scaling behavior of the condensate \eqref{eq:BCS-scaling} in \gls{MFA} at $T=0$ is absent when fluctuation effects are taken into account. 

The first-order phase transition is generated by an interplay of the \gls{BCS} singularity\footnote{
	In the zero-temperature limit, the \gls{BCS} singularity manifests itself in a nonanalytic behavior of the scale-dependent effective potential associated with a singular behavior of its curvature at the origin. This singular behavior is removed at nonzero temperature. However, at sufficiently low temperatures, the physics is still dominated by the ``relic" of this singularity present in the zero-temperature limit. Here and in the following, we shall for simplicity also refer to this aspect as \gls{BCS} singularity.} 
in the quark sector and the singularity of the diquark propagator functions at low temperatures and moderate chemical potentials.
For low temperatures, the strength of the singularity of the diquark propagator functions is reduced (in the sense of the principle of strongest singularity, see Ref.~\cite{Zorbach:2024zjx}) and simultaneously the \gls{BCS} singularity controlling the dynamics in the quark sector becomes more pronounced. 
More precisely, the \gls{BCS} singularity generates steep negative slopes in $\partial_{\bar{\Delta}}U_k$ close to $\bar{\Delta}=0$ for \gls{RG} scales $k\lesssim \mu$, see also \cref{fig:MF-BCS}. 
The singularity in the diquark propagator functions is sensitive to the formation of this negative slope and causes $\partial_{\bar{\Delta}} U_k$ to exceed zero, resulting in the formation of a small bump in $\partial_{\bar{\Delta}} U_k$ close to $\bar{\Delta} = 0$ at $k \lesssim \mu$.
The blue shaded region in \cref{fig:LPA-V1-phase-diagram-contour-first-order-regime} shows where such a bump (i.e., a signal of a first-order phase transition) still occurs in the \gls{RG} flow for nonzero~$k$  in the symmetric phase. 
The boundary of this region intersects with the phase transition line at the tricritical point.
In the symmetric phase, close to the phase transition, the aforementioned bump remains present in form of a plateau in $\partial_{\bar{\Delta}} U_k$ down to the very small \gls{RG} scales, see \cref{fig:LPA-V1-first-order-flow-mu-180}, whereas it is fully removed in the broken phase, see \cref{fig:LPA-V1-first-order-flow-mu-200}.
More precisely, a comparison of \cref{fig:LPA-V1-first-order-flow-mu-180} and \cref{fig:LPA-V1-first-order-flow-mu-200} illustrates that the quark contribution becomes stronger with increasing $\mu$. 
This continuously lowers the plateau in $\partial_{\bar{\Delta}} U$ with increasing $\mu$, as depicted in \cref{fig:LPA-second-order}.
Therefore, the diquark condensate $\bar{\Delta}_\mathrm{gs}$ (as associated with the minimum of $U_k$ in the \gls{IR} limit) changes discontinuously at some critical value of $\mu$ where the height of the plateau becomes zero.

Finally, we would like to mention that a stronger singular behavior of the diquark propagator functions may prevent the formation of a bump associated with a first-order transition in the scale-dependent effective potential. 
This could potentially change the order of the phase transition in the corresponding regime of the phase diagram from first to second order.
However, we expect that a first-order phase transition should also be present for other regulator shape functions in the diquark sector since the Litim regulator already exhibits the strongest singular behavior within the standard class of regulator shape functions in \gls{LPA}, see Ref.~\cite{Zorbach:2024zjx}.

\section{Further applications}\label{sec:applications}
The considerations presented in \cref{sec:LPA} can also be applied to other models including complex scalar fields at nonzero chemical potential such as the \gls{RBG} and the \gls{QMDM}.

\subsection{Relativistic Bose gas at nonzero chemical potential}\label{sec:rbg-at-finite-density}
\begin{figure} 
	\includegraphics{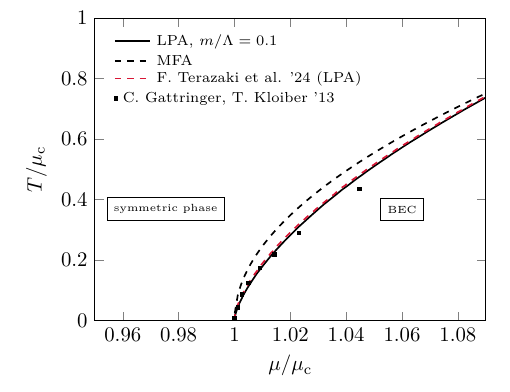}
	\caption{\label{fig:BEC}Phase boundary of the \gls{RBG} as a function of~$\mu/\mu_{\rm c}$ with~$\mu_{\rm c}$ being the critical chemical potential as obtained from our calculation with the truncation map ${\mathds{T}}^{\mathrm{LPA}_2}$ (solid line), \gls{MFA} (dashed black line), a previous \gls{fRG} study \cite{Terazaki:2024evv} (dashed red line), and a lattice study~\cite{Gattringer:2012df} (black squares).}
\end{figure}
\begin{figure}
	\includegraphics{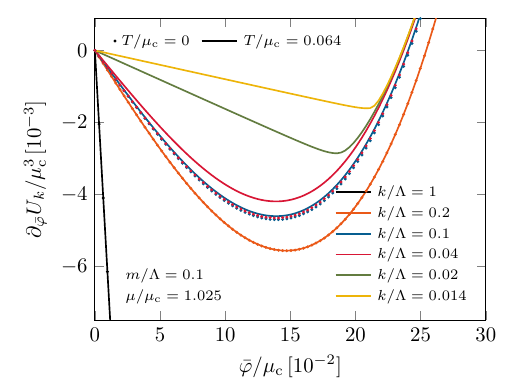}
	\caption{\label{fig:RBG-flow} Field derivative of the scale-dependent effective potential as a function of the field~$\bar{\varphi}$ in case of the \gls{RBG} for various \gls{RG} scales at $\mu/\mu_\mathrm{c}=1.024 $ for $T=0$ and $T/ \mu_\mathrm{c}=0.064$, respectively.}
\end{figure}
In the following, we consider a massive \gls{RBG}, see, e.g., Refs.~\cite{Svanes:2010we,Terazaki:2024evv,Andersen:2005yk,Marko:2014hea, Palhares:2012fv}, with a single complex scalar field~$\varphi$ at nonzero chemical potential and temperature. The super field $\Phi$ is now given by $\Phi^{\mathrm{RBG}} = (\varphi, \varphi^*, \bm\mu)$. 
The action reads
\begin{align*}
	\numberthis\label{eq:action-RBG}
	S[\Phi^{\mathrm{RBG}}] = \int_x \Big(&D_{-}\varphi^* D_{+}\varphi + \vec\nabla \varphi^* \vec\nabla \varphi \\
	&+ m^2 \varphi^* \varphi + \lambda (\varphi^* \varphi)^2\Big) \, ,
\end{align*}
where the covariant derivatives are $D_{\pm} = (\partial_0 \pm \bm\mu)$, $m$ is the mass parameter, and $\lambda$ is the four-boson coupling. 
The scale-dependent effective potential~$U_k$ is obtained from the projection rule~\eqref{eq:scale-dependent-effective-potential} with the homogeneous field configuration
\begin{align*}
	\numberthis\label{eq:scale-dependent-effective-potential-RBS}
	\Phi^{\mathrm{RBG}}_0 = (&\varphi=\tfrac{1}{\sqrt{2}}\bar{\varphi}, \varphi^*=\tfrac{1}{\sqrt{2}}\bar{\varphi}, \bm \mu = \mu) \, ,
\end{align*}
where $\bar{\varphi}$ and $\mu$ are real numbers. 
The flow equation for the scale-dependent effective potential reads
\begin{align*}
	\numberthis\label{eq:LPA-flow-equation-RBG}
	\partial_t U_k &= - v_{d-1} k^{d}\sum_{\chi=\pm1}\Big(1 + \chi\frac{4 \mu^2}{\xi^2_+ - \xi^2_-}\Big) G_2(\xi_\chi)\,.
\end{align*}
Here, $G_2(E)$ is given by \cref{eq:bosonic-propagator-2} with $F = 1$. 
The energy functions $E^{\mathrm{B}}_{1/2}$ entering in $\xi_{\pm}$ depend again on the variant of \gls{LPA} used in the concrete calculations, see \cref{sec:LPA}.
For the same reasons as discussed above for the \gls{QDM} study, we are solely considering ${\mathds{T}}^{\mathrm{LPA}_2}$ whose energy functions are given in \cref{eq:bosonic-energy-functions-v2} with $F = 1$.
The flow equation \eqref{eq:LPA-flow-equation-RBG} is initialized at $k = \Lambda$ with the potential 
\begin{align*}
	U_{\Lambda}(\bar{\varphi}, \mu) = \frac{1}{2}(m^2 - \mu^2) \bar{\varphi}^2 + \frac{1}{4}\lambda\bar{\varphi}^4 \, . 
\end{align*}
Note that this initial potential is $\mu$-dependent in contrast to the one chosen for the \gls{QDM}, see \cref{eq:initial-condition-for-U}. 
This $\mu$-dependence emerges from the temporal kinetic term for the bosons included in the action~$S$, see \cref{eq:action-RBG} and causes $U_{\Lambda}(\bar{\varphi}, \mu)$ to develop a nonzero minimum already at the scale $\Lambda$ for sufficiently large $\mu$.
It is precisely this $\mu$-dependence of the initial condition that can lead to a nontrivial ground state in the \gls{IR} limit, even for~$m^2>0$.
From a phenomenological standpoint, the presence of this nontrivial ground state is associated with \gls{SSB} and the formation of a Bose-Einstein condensate in this model, see, e.g., Refs.~\cite{Kapusta:1981aa,Andersen:2005yk}. 
Recall that bosonic fluctuations tend to restore the symmetry in the \gls{IR} limit. 

In \cref{fig:BEC}, we show our result for the phase boundary for the parameter choice~$m/\Lambda=0.1$ and $\lambda=1$.
We find $\mu_{\rm c}/\Lambda=0.1561$, which is the critical value of the chemical potential for Bose-Einstein condensation in the zero-temperature limit. 
For comparison, we included the results from a previous \gls{fRG} study~\cite{Terazaki:2024evv} a lattice calculation~\cite{Gattringer:2012df}, and the mean-field phase boundary~\cite{Kapusta:1981aa} given by  $T_\mathrm{MFA}(\mu)=\sqrt{3 (\mu^2-m^2)/\lambda}$ for $\mu \geq m$. 
Note that the temperature as well as the chemical potential are given in units of~$\mu_{\rm c}$. 
In all these studies, the phase transition is found to be of second-order and the corresponding critical temperature increases with the chemical potential. 
Moreover, we find very good agreement with the previous \gls{fRG} study. 
However, our results deviate significantly from \gls{MFA} results. 
Notably, our results are in good agreement with those from the lattice calculations at low temperatures, indicating that the truncation map ${\mathds{T}}^{\mathrm{LPA}_2}$ indeed provides us with a good truncation although it explicitly breaks the Silver-Blaze symmetry. 
For increasing chemical potential, we then find that our results for the critical temperature start to deviate from the lattice results which may at least partially be traced back to cutoff artifacts.

Finally, we present two examples of \gls{RG} flows in the Bose-Einstein condensation phase in \cref{fig:RBG-flow}: one corresponding to $T/\mu_\mathrm{c}=0.064 $ and the other one to the zero-temperature limit. 
Both \gls{RG} flows break down at (different) nonzero \gls{RG} scales due to the limitation of the numerical precision in our calculations.
For $T=0$, the \gls{RG} flow breaks down at higher \gls{RG} scales than for $T/\mu_\mathrm{c}=0.064$. 
This can be traced back to the strength of the singularity of the propagator function $G_2$ which becomes stronger when the temperature is increased, see our discussion below \cref{eq:constraints} and also \cref{fig:propagator-singularity}.
This also implies that the scale-dependent effective potential becomes convex faster in the \gls{RG} flow which is indicated by the typical flattening of $\partial_{\bar{\varphi}} U_k$ as illustrated in \cref{fig:RBG-flow}.
At the respective (numerical) breakdown scales, the position of the minimum of $U_k$ (zero of $\partial_{\bar{\varphi}}U_k$) is already almost converged in both cases shown in \cref{fig:RBG-flow}. 
This suggests that our results do not suffer from the numerical instabilities in this respect. 

\subsection{Quark-Meson-Diquark model}\label{sec:QMDM}
In this subsection we add meson fields $\phi_i$ (with $i = 1,2,3,4$) to our \gls{QDM} action in \cref{eq:action} which leads us to a \gls{QMDM}.  
Models of this type have previously been studied with quarks coming in two colors, see, e.g., Refs.~\cite{Strodthoff:2011tz,Cichutek:2020bli,Khan:2015etz,Khan:2015puu}, and three colors, see, e.g., Refs.~\cite{Lakaschus:2020caq,Lakaschus:2021ewd,Khan:2015etz, Gholami:2024diy,Yuan:2023dco,Braun:2018svj}.

In the following we consider the action 
\begin{align*}
	\numberthis\label{eq:action-QMD}
	S[\Phi^{\mathrm{QMD}}] = \int_x \biggl[&\bar{\psi}_{c} \bigl( {\rm i} \slashed{\partial} - {\rm i} \bm\mu \gamma^0 \bigr) \psi_{c}  + \bar{\nu}^2 \Delta_{c}^* \Delta_{c} \nonumber \\
	&- h_\Delta \bar{\psi}_{c_1} \gamma^\mathrm{ch} C \tau^2 \Delta_{c_0}^* \frac{{\rm i}}{2} \epsilon_{c_0 c_1 c_2} \bar{\psi}_{c_2} \nonumber \\
	&+ h_\Delta \psi_{c_1} C \gamma^\mathrm{ch} \tau^2 \Delta_{c_0} \frac{{\rm i}}{2} \epsilon_{c_0 c_1 c_2} \psi_{c_2} \nonumber\\
	&+h_\sigma \bar\psi_c {\rm i} ([\phi_1 + {\rm i} \sum_{i=2}^{4}\tau^{i-1} \phi_i \gamma^5]\big) \psi_c \nonumber\\
	&+\tfrac{1}{2} \partial_\nu\phi_i\partial_\nu\phi_i + \frac{1}{2} m^2 \phi_i \phi_i + \frac{1}{4} \lambda (\phi_i \phi_i )^2
	\biggr] \; ,
\end{align*}
where $\tau^{i}$ with $i = 1,2,3$ are the three Pauli matrices, $m$ and $\lambda$ are the mass and quartic coupling associated with the meson fields, respectively. The Yukawa couplings~$h_\Delta$ and $h_\sigma$ are associated with the diquark and meson fields, respectively.
The super field is defined as  
\begin{align*}
	\numberthis\label{eq:super-field-QMD}
	\Phi^{\mathrm{QMD}} = (\Delta_c, \Delta^*_c, \bar\psi_c, \psi_c, \phi_i, \bm\mu) \, .
\end{align*}
A numerical study of this model is beyond the scope of the present work. 
Here, we shall only show its \gls{LPA} flow equation as obtained from a straightforward but proper generalization of our truncation map ${\mathds{T}}^{\mathrm{LPA}_2}$ underlying the \gls{QDM} study. 
To this end, we define the projection rule for the scale-dependent effective potential as follows:
\begin{align*}
	\numberthis\label{eq:scale-dependent-effective-potential-QMD}
	U_k(\sigma, \bar{\Delta}, \mu) = \frac{1}{V_d} \Gamma_k[\Phi=\Phi^{\mathrm{QMD}}_0] \, ,
\end{align*}
where
\begin{align*}
	\numberthis\label{eq:super-field-evaluation-QMD}
	\Phi^{\mathrm{QMD}}_0 = (&\Delta_c=\tfrac{1}{\sqrt{2}}\bar{\Delta}\,\delta_{c 3}, \Delta^*_c=\tfrac{1}{\sqrt{2}}\bar{\Delta}\,\delta_{c 3}, \\
	&\quad \bar\psi_c=0, \psi_c=0, \phi_i=\sigma\,\delta_{i 1}, \bm \mu = \mu) \, .
\end{align*}
Note that the scale-dependent effective potential depends on one more field direction than the \gls{QDM} model, requiring the use of a suitable framework in numerical studies, see, e.g.,  Ref.~\cite{Zorbach:2024rre}.

Applying ${\mathds{T}}^{\mathrm{LPA}_2}$ to the \gls{QMDM}, we eventually obtain the flow equation for the scale-dependent effective potential:
\begin{widetext}
	\begin{align*}
		\numberthis\label{eq:LPA-flow-equation-QMDM}
		\partial_t U_k = &-v_{d-1} k^d \bigg(2(N_\mathrm{c} - 1) \, G_1(E_{\mathrm{G},\Delta}) + (N_\mathrm{f}^2 - 1) \, G_2(E_{\mathrm{G},\sigma}) + G_3(E_{\mathrm{G},\Delta}, E_{\mathrm{R},\Delta}, E_{\mathrm{R},\sigma}, \partial_{\bar{\Delta}} \partial_\sigma U_k)\bigg) \\
		&+ v_{d-1} k^d N_\mathrm{f} d_\gamma \bigg(\sum_{\chi = \pm 1} \frac{k}{E^{\mathrm{F}}} \frac{E^{\mathrm{F}}+\chi \mu}{2E^\mathrm{F}_\chi}\bigl(1-2 n_\mathrm{F}(\beta E^\mathrm{F}_\chi)\bigr) + \frac{k}{2E^\mathrm{F}} \Bigl(1- n_\mathrm{F}\big(\beta(E^\mathrm{F}+\mu)\big)- n_\mathrm{F}\big(\beta(E^\mathrm{F}-\mu)\big)\Bigr)\bigg)\, ,
	\end{align*}
\end{widetext}
where 
\begin{subequations}
	\label{eq:energy-functions-QMDM}
	\begin{align}
		E_{\mathrm{G},\Delta}&=\sqrt{k^2 + \frac{\partial_{\bar{\Delta}} U_k}{\bar{\Delta}}}\, , \\
		E_{\mathrm{R},\Delta}&=\sqrt{k^2 + \partial^2_{\bar{\Delta}} U_k}\, ,\\
		E_{\mathrm{G},\sigma}&=\sqrt{k^2 + \frac{\partial_\sigma U_k}{\sigma}}\, , \\ 
		E_{\mathrm{R},\sigma}&=\sqrt{k^2 + \partial^2_\sigma U_k}\, , \\ 
		E^\mathrm{F}_\chi&=\sqrt{\frac{1}{2} h_\Delta^2 \bar{\Delta}^2 + (E^{\mathrm{F}}+\chi\mu)^2}\, , \\ 
		E^{\mathrm{F}}&=\sqrt{k^2+h_\sigma^2\sigma^2} \,
	\end{align}
\end{subequations}
are the bosonic and quark energy functions and $G_1(E)$ is given in \cref{eq:bosonic-propagator}, $G_2(E)$ in \cref{eq:bosonic-propagator-2} and 
\begin{align*}
	\numberthis \label{eq:new-propa-QMDM}
	&G_3(E_{\mathrm{G},\Delta}, E_{\mathrm{R},\Delta}, E_{\mathrm{R},\sigma}, \partial_{\bar{\Delta}} \partial_\sigma U_k) \\
	& \qquad =k \sum_{j=1}^{3} \frac{\mathcal{N}(r_j)}{\sqrt{r_j} \prod\limits_{l=1, l\neq j}^{3}(r_j-r_l)} \Bigl[ \frac{1}{2} + n_\mathrm{B}(\beta \sqrt{r_j}) \Bigr] \, .
\end{align*}
Here, we have introduced the auxiliary function~$\mathcal{N}(r)$,
\begin{align*}
	\numberthis
	&\mathcal{N}(r) = (-r+E_{\mathrm{R},\sigma}^2)\bigl( (-r+E_{\mathrm{R},\Delta}^2) + (-r+E_{\mathrm{G},\Delta}^2) \bigr) \\
	&- (\partial_{\bar{\Delta}} \partial_\sigma U_k)^2 + (-r+E_{\mathrm{R},\Delta}^2) (-r+E_{\mathrm{G},\Delta}^2) + 4 F^2\mu^2 (-r) \, .
\end{align*}
In \cref{eq:new-propa-QMDM}, $r_j$ are the roots of 
\begin{align*}
	\numberthis
	&\mathcal{D}(r)= - (\partial_{\bar{\Delta}} \partial_\sigma U_k)^2 (-r+E_{\mathrm{G},\Delta}^2) \\
	&+(-r+E_{\mathrm{R},\sigma}^2)\bigl( (-r+E_{\mathrm{R},\Delta}^2) (-r+E_{\mathrm{G},\Delta}^2) + 4 F^2\mu^2 (-r) \bigr) \, .
\end{align*}
As in the case of the \gls{QDM}, \gls{SSB} at large chemical potential is controlled by an interplay of the \gls{BCS} singularity in the quark sector and bosonic fluctuations. 
In the \gls{IR} limit, the singularity in the bosonic sector then ensures convexity of the scale-dependent effective potential.

\section{Conclusions}\label{sec:conclusion}
In the present work, we constructed various truncation maps to define \glsxtrlongpl{LPA} for theories which involve complex scalar fields coupled to a chemical potential. 
We find that the Silver-Blaze-symmetric version of \gls{LPA} is not suitable for investigating \gls{SSB}. 
One of the Silver-Blaze-violating maps, i.e., the one defined by \cref{eq:LPA-truncation-map} with \cref{eq:kinetic-v2}, turns out to be best suited for an investigation of \gls{SSB} in these type of theories.
Our discussion of Silver-Blaze-symmetric truncation maps remains a well-defined starting point for a systematic construction of truncations for studies of theories at small chemical potentials at higher order in the derivative expansions.

In \cref{sec:results} and \cref{sec:rbg-at-finite-density}, we have applied the truncation map~\eqref{eq:LPA-truncation-map} to a \gls{QDM} and to a \gls{RBG}, respectively. 
For the \gls{QDM}, we find a first-order phase transition at low temperatures which becomes a second-order phase transition at a tricritical point. 
The emergence of a first-order phase transition is the result of a nontrivial competition of the \gls{BCS} singularity in the quark sector and the singular behavior of the diquark propagator functions at small temperatures. 
Interestingly, our results indicate the existence of a first-order phase transition from a trivial ground state to a color superconductor at a nonzero critical chemical potential at zero temperature. 
This suggests that the typical \gls{BCS}-type exponential scaling behavior~\eqref{eq:BCS-scaling} of the gap usually observed in \gls{MFA} studies of this type of models is absent when fluctuations of the order-parameter field are taken into account as done in \gls{LPA}. 
For the \gls{RBG}, we find a second-order phase transition for all considered values of the chemical potential. 
Notably, our results are found to agree well with those from previous \gls{fRG} and lattice studies. 

Finally, we briefly discussed how a \glsxtrlong{QMDM} can be constructed from our \gls{QDM} by suitably including mesonic degrees of freedom and provided a corresponding truncation map, see \cref{sec:QMDM}. 
The resulting flow equation for the scale-dependent effective potential may serve as a starting point of future numerical studies of this model which may provide us with a fresh insight into the competition of chiral symmetry breaking and the formation of a color superconductor in dense two-flavor \gls{QCD} matter.

\acknowledgements
We thank A.~Gei\ss el, P.~Guttandin, A.~Koenigstein, J.~M.~Pawlowski, F.~Rennecke, B.~Schallmo, S.~T\"opfel, M.~Thies, and A.~Zorbach for discussions and comments on the manuscript. 
As members of the fQCD collaboration~\cite{fQCD}, the authors also would like to thank the members of this collaboration for discussions.
This work is supported in part by the Deutsche Forschungsgemeinschaft (DFG, German Research Foundation) through the Collaborative Research Center CRC 1245 ``Nuclei: From Fundamental Interactions to Structure and Stars" -- project number 279384907 -- SFB 1245, the CRC-TR 211 ``Strong-interaction matter under extreme conditions" -- project number 315477589 -- TRR 211, and the State of Hesse within the Research Cluster ELEMENTS (Project No. 500/10.006). 

\appendix

\section{Silver-Blaze symmetry and analytic properties of the effective potential}\label{sec:Silver-Blaze}
This appendix considers the dependence of the scale-dependent effective potential and thermodynamic observables on the chemical potential in theories with Silver-Blaze symmetry at zero temperature on more general grounds. 
To this end, we use the notation for the generalized scale-dependent effective potential as introduced in \cref{sec:LPA}. 

In theories where bosons do not couple to the chemical potential (e.g., the multi-component meson field~$\phi$ in a two-flavor quark-meson model), Silver-Blaze symmetry implies, at the level of the scale-dependent effective average action, that
\begin{align*}
	\numberthis
	\Gamma_k[\bar{\psi}_c,\psi_c,\phi](\bm \mu) = \Gamma_k[\bar{\psi}_c {\rm e}^{{\rm i} \alpha \cdot},  {\rm e}^{-{\rm i} \alpha \cdot} \psi_c, \phi](\bm \mu - {\rm i}\alpha)\,,
\end{align*}
where the symbol $\cdot$ represents a placeholder for the argument of the transformed fields. 
Note that we have used \cref{eq:sbs-transformation} and that uncharged scalar fields transform trivially under Silver-Blaze transformations. 
By evaluating both sides of this relation on a constant meson background field configuration~$\phi=\sigma$  and setting~$\bar{\psi}_c=\psi_c=0$, we arrive at the following relation for the generalized scale-dependent effective potential:
\begin{align*}
	\numberthis \label{eq:Ukrelation}
	{\bm U}_k(\sigma,\bm\mu) = {\bm U}_k(\sigma,\bm\mu -{\rm i}\alpha)\,,
\end{align*}
where we have used the definition 
\begin{align*}
	\numberthis
	{\bm U}_k(\sigma,\bm \mu) = \frac{1}{V_d}\Gamma_k[0,0,\sigma](\bm \mu)\,.
\end{align*}
When considering the zero-temperature limit, $\alpha$ is a continuous variable. 
From \cref{eq:Ukrelation} we then deduce that~${\bm U}_k$ does not depend on the imaginary part of the chemical potential.
Moreover, if the generalized scale-dependent effective potential~${\bm U}_k(\sigma, \bm \mu)$ is complex analytic in $\bm \mu$ for a given tuple $(k, \sigma)$, it must also be independent of the real part of $\bm \mu$ at that tuple.
Looking now at the Wetterich equation, we note that the \gls{RG} flow of the generalized scale-dependent effective potential is determined by the full scale-dependent two-point function $\Gamma^{(2)}_k$ evaluated on a homogeneous background field $\sigma$. 
In practice, the $k$- and $\sigma$-dependent poles of this quantity therefore restrict the analyticity of $\bm U_k$ at most to a strip in the complex plane defined by $0 \leq \Re(\bm\mu) < \tilde\mu(k,\sigma)$ and we will assume the existence of such a strip in the following.
Assuming that~$\tilde{\mu}(k,\sigma)$ is a strictly increasing and continuous function in $\sigma$ for any $k$, the equation $\tilde{\mu}=\tilde{\mu}(k,\sigma)$ can be uniquely solved for~$\sigma$ which yields a critical value~$\sigma_{\rm c}(k,\tilde{\mu})$. 
For~$\sigma > \sigma_{\rm c}(k,\tilde{\mu})$, it then follows that the generalized scale-dependent effective potential is complex analytic in the domain $0 \leq \Re(\bm\mu) < \tilde{\mu}$ and, together with \cref{eq:Ukrelation}, it follows that 
\begin{align*}
	\numberthis \label{eq:Uk_indep_mu}
	{\bm U}_{k}(\sigma,\bm\mu)= {\bm U}_{k}(\sigma,\bm\mu=0) \, .
\end{align*}
Let us now turn to the \gls{IR} limit $k \to 0$.
Provided that~$\sigma_{\rm gs}({\bm\mu}=0) > \sigma_{\rm c}(k=0,\tilde{\mu})$, where $\sigma_{\rm gs}({\bm\mu}=0)$ is the position of the ground state of the generalized effective potential in the vacuum, \cref{eq:Uk_indep_mu} implies that $\sigma_{\rm gs}({\bm\mu})$ does not depend on $\bm\mu$ in the corresponding chemical potential domain. 
Conversely, the equation $\mu_c=\tilde{\mu}(k\!=\! 0,\sigma_{\rm gs}({\bm\mu}\!=\!0))$ defines a critical value of the chemical potential below which the ground state and thus also the pressure $p=-{\bm U}_{k\to 0}(\sigma_\mathrm{gs}(\bm \mu),{\bm \mu})$ as well as all thermodynamic observables derived from it remain independent of~$\bm\mu$ 
(which is dubbed Silver-Blaze phenomenon/property~\cite{Cohen:2003kd} in the case of real $\bm\mu$).\footnote{
	The Silver-Blaze phenomenon/property should not be confused with the Silver-Blaze symmetry of the effective action. The Silver-Blaze phenomenon/property refers to the fact that observables do not depend on the chemical potential below some critical value. This phenomenon  necessarily requires the Silver-Blaze symmetry of the effective action. However, the symmetry of the effective action under Silver-Blaze transformations does not imply the existence of a finite Silver-Blaze region where thermodynamic observables are independent of the chemical potential.}

For example, this is a well-known zero-temperature feature of models in which bosons do not couple to the chemical potential, such as the two-flavor quark-meson model and the Gross-Neveu model in the zero-temperature limit.\footnote{
	For instance, in the Gross-Neveu model using a spatial Litim regulator in \gls{MFA}, $\bm U_k(\sigma, \bm\mu)$ remains $\bm\mu$-independent, provided that $k^2 + h_\sigma^2 \sigma^2 > \Re(\bm\mu)^2$, see, e.g., Refs.~\cite{Stoll:2021ori,Topfel:2024cll}.
	The resulting critical value is $\mu_{\mathrm{c}} = h_\sigma \sigma_{\mathrm{gs}}(\bm \mu=0)$.
	However, note that $\bm U_{k \to 0}(\sigma, \bm \mu)$ may still exhibit a $\bm \mu$-dependence for $\sigma < \sigma_{\mathrm{gs}}(\bm \mu=0)$, even within $0 \leq \Re(\bm\mu) < \mu_c$.}

For a theory where bosons couple to the chemical potential (as it is case for the \gls{QDM} and the \gls{RBG}), the situation is different. 
On the level of the scale-dependent effective average action, Silver-Blaze symmetry implies in this case that
\begin{align*}
	\numberthis
	& \Gamma_k[\bar{\psi}_c,\psi_c,\Delta_c^{\ast},\Delta_c](\bm \mu) \\ 
	& \quad = \Gamma_k[
 	\bar{\psi}_c {\rm e}^{{\rm i} \alpha \cdot},
 	{\rm e}^{-{\rm i} \alpha \cdot} \psi_c,
  	\Delta^*_c {\rm e}^{-{\rm i} F \alpha \cdot},
   	{\rm e}^{{\rm i} F \alpha \cdot} \Delta_c](\bm \mu - {\rm i}\alpha)\,,
\end{align*}
where we have again used~\cref{eq:sbs-transformation}.
However, by evaluating now both sides on a constant diquark background field configuration~$\Delta^{\ast}_c=\Delta_c=\bar{\Delta}/\sqrt{2}$ and setting~$\bar{\psi}_c=\psi_c=0$, we arrive at
\begin{align*}
	\numberthis \label{eq:Ukcharged}
	& {\bm U}_k(\bar{\Delta},\bm\mu) \\
	& 
	\quad = \frac{1}{V_d} \Gamma_k[0, 0, (\bar{\Delta}/\sqrt{2})  {\rm e}^{-{\rm i} F \alpha \cdot},
   	{\rm e}^{{\rm i} F \alpha \cdot} (\bar{\Delta}/\sqrt{2}) ](\bm \mu - {\rm i}\alpha) 
\end{align*}
with
\begin{align*}
	\numberthis
 	{\bm U}_k(\bar{\Delta},\bm\mu) = \frac{1}{V_d}  \Gamma_k[0, 0, \bar{\Delta}/\sqrt{2} , \bar{\Delta}/\sqrt{2} ](\bm \mu)\,.
\end{align*}
Relation~\eqref{eq:Ukcharged} corresponds to relation \eqref{eq:Ukrelation}, which is derived for a theory where the bosons do {\it not} couple to the chemical potential. 
In contradistinction to theories of the latter type, however, it is not possible to draw any conclusions from \cref{eq:Ukcharged} regarding the dependence of the generalized scale-dependent effective potential on the chemical potential for our theory with charged bosons.
In this case, the generalized scale-dependent effective potential may therefore carry a dependence on the chemical potential at any \gls{RG} scale~$k$, even if the underlying scale-dependent effective average action is invariant under Silver-Blaze transformations. 
For example, this is the case for the generalized scale-dependent effective potential of the \gls{QDM} in \gls{MFA}, see Eq.~\eqref{eq:MF-pot}. 
In particular, the generalized scale-dependent effective potential may exhibit a nontrivial dependence on $\text{Im}({\bm \mu})$. 
However, in the absence of \gls{SSB},\footnote{
	We also tacitly assume the absence of any explicit symmetry breaking in terms of, e.g., external sources.} 
where the ground state is associated with~$\bar{\Delta}_\mathrm{gs}=0$, we can deduce from \cref{eq:Ukcharged} that
\begin{align*}
	\numberthis
	{\bm U}_k(0,\bm\mu) = {\bm U}_k(0,\bm\mu -{\rm i}\alpha)\,.
\end{align*}
Thus, in the absence of \gls{SSB} at $T=0$, ${\bm U}_k(0,\bm\mu)$ does not depend on the imaginary part of the chemical potential.  
If, additionally, ${\bm U}_k(0,\bm\mu)$ is complex analytic in $\bm \mu$ for some $k$, then ${\bm U}_k(0,\bm\mu)$ is even independent of the real part of the chemical potential. 
If this holds in the limit $k\to 0$ for~$0\leq \Re({\bm \mu}) < \mu_{\rm c}$, where $\mu_{\rm c}$ again denotes the extent of the complex analytic domain in the direction of~$\Re({\bm \mu})$, we conclude that thermodynamic observables in the absence of \gls{SSB} are independent of the chemical potential below a critical value~$\mu_c$.  

Let us finally comment on the existence of a nonzero critical value~$\mu_c$ of the chemical potential. 
In our discussion above, we assumed complex analyticity in the domain~$\Re(\bm \mu)< \mu_{\rm c}$.
From a thermodynamic standpoint, it is reasonable to assume the existence of such a domain. 
Indeed, the chemical potential is the change in free energy when a charged particle (i.e., a particle that couples to a given chemical potential, such as a quark or diquark in case of a quark chemical potential) is added to or removed from the system. 
Therefore, the density of the charged particles can become finite only for $\mu \geq \mu_{\rm c}$, where the critical value~$\mu_{\rm c}$ is set by the vacuum pole mass of the lightest charged particle in the spectrum of the theory,\footnote{
	Strictly speaking, in the relativistic systems considered in this work, it is not the density but the difference in the densities of charged particles and their antiparticles.} 
see Ref.~\cite{Braun:2020bhy} for a detailed discussion. 
For~$\mu < \mu_{\rm c}$, the density remains zero and is independent of the chemical potential.  
Since the density is directly related to the first derivative of the grand canonical partition function with respect to chemical potential, the latter (and therefore also the pressure) does not depend on the chemical potential for~$\mu < \mu_{\rm c}$. 
As discussed above, this invariance of the grand canonical partition function under a shift of the chemical potential below $\mu_{\rm c}$ can be understood as a consequence of the Silver-Blaze symmetry of the effective action and complex analyticity of the ground-state value of the generalized effective potential in $\bm\mu$. 

\section{Numerical implementation of the flow equations}\label{sec:numerical-treatment}
The flow equation \eqref{eq:LPA-flow-equation} is a partial differential equation that can be recast into a continuity equation by performing a field-derivative on both sides, see Refs.~\cite{Grossi:2019urj,Koenigstein:2021syz}. 
In this continuity formulation, the \gls{MFA} contribution is a source term, the pure $E^{\mathrm{B}}_1$-dependent bosonic contribution is an advection term and the mixed $E^{\mathrm{B}}_1$ and $E^{\mathrm{B}}_2$-dependent bosonic contribution
is an advection-diffusion term. 
This formulation enables the use of the well-established \gls{KT} scheme \cite{Kurganov:2000ovy} to solve the flow equation \eqref{eq:LPA-flow-equation}.
Therefore, we always worked on the level of the field derivative of the scale-dependent effective potential rather than the potential itself.
As a side effect, this formulation establishes a well-defined boundary condition $\partial_{\bar{\Delta}} U_k(\bar{\Delta}=0, \mu)=0$ at $\bar{\Delta}=0$ for $T>0$. 
At low $T$, this is where the ``relic'' of the \gls{BCS} singularity determines the slope of $\partial_{\bar{\Delta}} U_k$ and where it is therefore crucial to control the dynamics very well.
\begin{table}
	\begin{ruledtabular}
		\setlength\extrarowheight{8pt}
		\begin{tabular}{c c c c c c c}
			Figure & Resolution & Range & \gls{UV} parameter & $k_{\mathrm{IR}}$ & $tol$ &\\
			\hline
			\cref{fig:MF-BCS} & $0.1$ & $300$ & $\bar{\nu}^2=58.5\times10^3$ & $5$ & $10^{-14}$ &\\
			\cref{fig:LPA-V1-phase-diagram-contour-full} & $2$ & $2\times10^3$ & $\bar{\nu}_\mathrm{LPA}^2=57.5\times10^3$ & $75$ & $10^{-10}$ &\\
			\cref{fig:LPA-V1-phase-diagram-contour-full} & $2$ & $2\times10^3$ & $\bar{\nu}_\mathrm{MFA}^2=58.5\times10^3$ & $75$ & $10^{-10}$ &\\
			\cref{fig:LPA-V1-phase-diagram-contour-first-order-regime} & $1$ & $10^3$ & $\bar{\nu}^2=57.5\times10^3$ & $13$ & $10^{-10}$ &\\
			\cref{fig:LPA-V1-first-order-flow} & $1$ & $10^3$ & $\bar{\nu}^2=57.5\times10^3$ & $14$ & $10^{-14}$ &\\
			\cref{fig:LPA-second-order} & $1$ & $10^3$ & $\bar{\nu}^2=57.5\times10^3$ & $14$ & $10^{-14}$ &\\
			\cref{fig:BEC} & $0.5$ & $10^3$ & $m^2=10^4, \lambda=1$ & $1$ & $10^{-10}$ & \\
			\cref{fig:RBG-flow} & $0.5$ & $10^3$ & $m^2=10^4, \lambda=1$ & $14$ & $10^{-12}$
		\end{tabular}
	\end{ruledtabular}
	
	\caption{\label{tab:numerical-data}List of model and numerical parameters for all plots presented in this work. 
	All dimensionful variables are given in powers of $\MeV$ according to their mass dimension. 
	The \gls{UV} scale has always been chosen to be $\Lambda=1000\MeV$ and $rtol=atol=tol$. 
	``Resolution" and ``Range" refer to the grid in field space. 
	To resolve the phase boundary (``zero height contour line") in \gls{LPA} more accurately, see \cref{fig:LPA-V1-phase-diagram-contour-full}, we employed a range of $10^3$ and a resolution of $1$ in field space with $k_\mathrm{IR}=20$.}
\end{table}

We implemented the \gls{KT} scheme in a semi-discrete way, i.e., we discretized the field space, but used a continuous \gls{RG} scale, which leads to a coupled system of ordinary differential equations \cite{Stoll:2021ori}. 
We treated the mixed advection-diffusion term as advective in the sense of the \gls{KT} scheme.
We then used Python's \textit{solve\_ivp} \cite{2020SciPy-NMeth} in \textit{Python 3}~\cite{10.5555/1593511} (using various libraries~\cite{2020SciPy-NMeth,Hunter:2007,harris2020array}) to solve this system with ``LSODA'' as numerical time stepper with the parameters $rtol$ and $atol$ for its relative and absolute error.
Some of the results have been cross-checked with \texttt{Mathematica}~\cite{Mathematica:12.1}.
A plain list of model parameters and numerical parameters for all figures shown in this work can be found in \cref{tab:numerical-data}.
Note that in some domains associated with \gls{SSB}, as shown in \cref{fig:LPA-V1-phase-diagram-contour-full}, but far from the phase boundary, we stopped the RG flow at comparatively large values of $k_{\text{IR}}$. This is justified because the \gls{RG} flow close to the minimum of the scale-dependent effective potential ``freezes out'' early in the \gls{RG} flow anyhow, due to strong \gls{SSB} in these domains.

However, dealing with mixed advection-diffusion terms is challenging also for the \gls{KT} scheme.
In regions of the phase diagram, where the quark contribution creates a deep well, i.e., at low temperatures and large chemical potentials, we find that numerical noise occurs when we consider very high resolutions.
More precisely, the solution $\partial_{\bar{\Delta}} U_k(\bar{\Delta}, \mu)$ develops spurious oscillations around the minimum of this well, i.e., where the mixed contribution changes from being diffusion dominated to advection dominated. 
However, the influence of these artifacts on the results presented in our present work appear to be negligible. 
In fact, the results appear to be properly converged coming from lower grid resolutions.

\vfill

\bibliography{bib}

\begin{thebibliography}{80}%
\makeatletter
\providecommand \@ifxundefined [1]{%
 \@ifx{#1\undefined}
}%
\providecommand \@ifnum [1]{%
 \ifnum #1\expandafter \@firstoftwo
 \else \expandafter \@secondoftwo
 \fi
}%
\providecommand \@ifx [1]{%
 \ifx #1\expandafter \@firstoftwo
 \else \expandafter \@secondoftwo
 \fi
}%
\providecommand \natexlab [1]{#1}%
\providecommand \enquote  [1]{``#1''}%
\providecommand \bibnamefont  [1]{#1}%
\providecommand \bibfnamefont [1]{#1}%
\providecommand \citenamefont [1]{#1}%
\providecommand \href@noop [0]{\@secondoftwo}%
\providecommand \href [0]{\begingroup \@sanitize@url \@href}%
\providecommand \@href[1]{\@@startlink{#1}\@@href}%
\providecommand \@@href[1]{\endgroup#1\@@endlink}%
\providecommand \@sanitize@url [0]{\catcode `\\12\catcode `\$12\catcode
  `\&12\catcode `\#12\catcode `\^12\catcode `\_12\catcode `\%12\relax}%
\providecommand \@@startlink[1]{}%
\providecommand \@@endlink[0]{}%
\providecommand \url  [0]{\begingroup\@sanitize@url \@url }%
\providecommand \@url [1]{\endgroup\@href {#1}{\urlprefix }}%
\providecommand \urlprefix  [0]{URL }%
\providecommand \Eprint [0]{\href }%
\providecommand \doibase [0]{https://doi.org/}%
\providecommand \selectlanguage [0]{\@gobble}%
\providecommand \bibinfo  [0]{\@secondoftwo}%
\providecommand \bibfield  [0]{\@secondoftwo}%
\providecommand \translation [1]{[#1]}%
\providecommand \BibitemOpen [0]{}%
\providecommand \bibitemStop [0]{}%
\providecommand \bibitemNoStop [0]{.\EOS\space}%
\providecommand \EOS [0]{\spacefactor3000\relax}%
\providecommand \BibitemShut  [1]{\csname bibitem#1\endcsname}%
\let\auto@bib@innerbib\@empty
\bibitem [{\citenamefont {Son}(1999)}]{Son:1998uk}%
  \BibitemOpen
  \bibfield  {author} {\bibinfo {author} {\bibfnamefont {D.~T.}\ \bibnamefont
  {Son}},\ }\bibfield  {title} {\bibinfo {title} {{Superconductivity by long
  range color magnetic interaction in high density quark matter}},\ }\href
  {https://doi.org/10.1103/PhysRevD.59.094019} {\bibfield  {journal} {\bibinfo
  {journal} {Phys. Rev.}\ }\textbf {\bibinfo {volume} {D59}},\ \bibinfo {pages}
  {094019} (\bibinfo {year} {1999})},\ \Eprint
  {https://arxiv.org/abs/hep-ph/9812287} {arXiv:hep-ph/9812287 [hep-ph]}
  \BibitemShut {NoStop}%
\bibitem [{\citenamefont {Sch{\"a}fer}\ and\ \citenamefont
  {Wilczek}(1999)}]{Schafer:1999jg}%
  \BibitemOpen
  \bibfield  {author} {\bibinfo {author} {\bibfnamefont {T.}~\bibnamefont
  {Sch{\"a}fer}}\ and\ \bibinfo {author} {\bibfnamefont {F.}~\bibnamefont
  {Wilczek}},\ }\bibfield  {title} {\bibinfo {title} {{Superconductivity from
  perturbative one gluon exchange in high density quark matter}},\ }\href
  {https://doi.org/10.1103/PhysRevD.60.114033} {\bibfield  {journal} {\bibinfo
  {journal} {Phys. Rev.}\ }\textbf {\bibinfo {volume} {D60}},\ \bibinfo {pages}
  {114033} (\bibinfo {year} {1999})},\ \Eprint
  {https://arxiv.org/abs/hep-ph/9906512} {arXiv:hep-ph/9906512 [hep-ph]}
  \BibitemShut {NoStop}%
\bibitem [{\citenamefont {Pisarski}\ and\ \citenamefont
  {Rischke}(2000{\natexlab{a}})}]{Pisarski:1999bf}%
  \BibitemOpen
  \bibfield  {author} {\bibinfo {author} {\bibfnamefont {R.~D.}\ \bibnamefont
  {Pisarski}}\ and\ \bibinfo {author} {\bibfnamefont {D.~H.}\ \bibnamefont
  {Rischke}},\ }\bibfield  {title} {\bibinfo {title} {{Gaps and critical
  temperature for color superconductivity}},\ }\href
  {https://doi.org/10.1103/PhysRevD.61.051501} {\bibfield  {journal} {\bibinfo
  {journal} {Phys. Rev.}\ }\textbf {\bibinfo {volume} {D61}},\ \bibinfo {pages}
  {051501} (\bibinfo {year} {2000}{\natexlab{a}})},\ \Eprint
  {https://arxiv.org/abs/nucl-th/9907041} {arXiv:nucl-th/9907041 [nucl-th]}
  \BibitemShut {NoStop}%
\bibitem [{\citenamefont {Pisarski}\ and\ \citenamefont
  {Rischke}(2000{\natexlab{b}})}]{Pisarski:1999tv}%
  \BibitemOpen
  \bibfield  {author} {\bibinfo {author} {\bibfnamefont {R.~D.}\ \bibnamefont
  {Pisarski}}\ and\ \bibinfo {author} {\bibfnamefont {D.~H.}\ \bibnamefont
  {Rischke}},\ }\bibfield  {title} {\bibinfo {title} {{Color superconductivity
  in weak coupling}},\ }\href {https://doi.org/10.1103/PhysRevD.61.074017}
  {\bibfield  {journal} {\bibinfo  {journal} {Phys. Rev.}\ }\textbf {\bibinfo
  {volume} {D61}},\ \bibinfo {pages} {074017} (\bibinfo {year}
  {2000}{\natexlab{b}})},\ \Eprint {https://arxiv.org/abs/nucl-th/9910056}
  {arXiv:nucl-th/9910056 [nucl-th]} \BibitemShut {NoStop}%
\bibitem [{\citenamefont {Brown}\ \emph {et~al.}(2000)\citenamefont {Brown},
  \citenamefont {Liu},\ and\ \citenamefont {Ren}}]{Brown:1999aq}%
  \BibitemOpen
  \bibfield  {author} {\bibinfo {author} {\bibfnamefont {W.~E.}\ \bibnamefont
  {Brown}}, \bibinfo {author} {\bibfnamefont {J.~T.}\ \bibnamefont {Liu}},\
  and\ \bibinfo {author} {\bibfnamefont {H.-c.}\ \bibnamefont {Ren}},\
  }\bibfield  {title} {\bibinfo {title} {{On the perturbative nature of color
  superconductivity}},\ }\href {https://doi.org/10.1103/PhysRevD.61.114012}
  {\bibfield  {journal} {\bibinfo  {journal} {Phys. Rev.}\ }\textbf {\bibinfo
  {volume} {D61}},\ \bibinfo {pages} {114012} (\bibinfo {year} {2000})},\
  \Eprint {https://arxiv.org/abs/hep-ph/9908248} {arXiv:hep-ph/9908248
  [hep-ph]} \BibitemShut {NoStop}%
\bibitem [{\citenamefont {Evans}\ \emph {et~al.}(2000)\citenamefont {Evans},
  \citenamefont {Hormuzdiar}, \citenamefont {Hsu},\ and\ \citenamefont
  {Schwetz}}]{Evans:1999at}%
  \BibitemOpen
  \bibfield  {author} {\bibinfo {author} {\bibfnamefont {N.~J.}\ \bibnamefont
  {Evans}}, \bibinfo {author} {\bibfnamefont {J.}~\bibnamefont {Hormuzdiar}},
  \bibinfo {author} {\bibfnamefont {S.~D.~H.}\ \bibnamefont {Hsu}},\ and\
  \bibinfo {author} {\bibfnamefont {M.}~\bibnamefont {Schwetz}},\ }\bibfield
  {title} {\bibinfo {title} {{On the QCD ground state at high density}},\
  }\href {https://doi.org/10.1016/S0550-3213(00)00253-4} {\bibfield  {journal}
  {\bibinfo  {journal} {Nucl. Phys.}\ }\textbf {\bibinfo {volume} {B581}},\
  \bibinfo {pages} {391} (\bibinfo {year} {2000})},\ \Eprint
  {https://arxiv.org/abs/hep-ph/9910313} {arXiv:hep-ph/9910313 [hep-ph]}
  \BibitemShut {NoStop}%
\bibitem [{\citenamefont {Hong}\ \emph {et~al.}(2000)\citenamefont {Hong},
  \citenamefont {Miransky}, \citenamefont {Shovkovy},\ and\ \citenamefont
  {Wijewardhana}}]{Hong:1999fh}%
  \BibitemOpen
  \bibfield  {author} {\bibinfo {author} {\bibfnamefont {D.~K.}\ \bibnamefont
  {Hong}}, \bibinfo {author} {\bibfnamefont {V.~A.}\ \bibnamefont {Miransky}},
  \bibinfo {author} {\bibfnamefont {I.~A.}\ \bibnamefont {Shovkovy}},\ and\
  \bibinfo {author} {\bibfnamefont {L.~C.~R.}\ \bibnamefont {Wijewardhana}},\
  }\bibfield  {title} {\bibinfo {title} {{Schwinger-Dyson approach to color
  superconductivity in dense QCD}},\ }\href
  {https://doi.org/10.1103/PhysRevD.61.056001, 10.1103/PhysRevD.62.059903}
  {\bibfield  {journal} {\bibinfo  {journal} {Phys. Rev.}\ }\textbf {\bibinfo
  {volume} {D61}},\ \bibinfo {pages} {056001} (\bibinfo {year} {2000})},\
  \bibinfo {note} {[Erratum: Phys. Rev. {\bf D62}, 059903 (2000)]},\ \Eprint
  {https://arxiv.org/abs/hep-ph/9906478} {arXiv:hep-ph/9906478 [hep-ph]}
  \BibitemShut {NoStop}%
\bibitem [{\citenamefont {Nickel}\ \emph {et~al.}(2006)\citenamefont {Nickel},
  \citenamefont {Wambach},\ and\ \citenamefont {Alkofer}}]{Nickel:2006vf}%
  \BibitemOpen
  \bibfield  {author} {\bibinfo {author} {\bibfnamefont {D.}~\bibnamefont
  {Nickel}}, \bibinfo {author} {\bibfnamefont {J.}~\bibnamefont {Wambach}},\
  and\ \bibinfo {author} {\bibfnamefont {R.}~\bibnamefont {Alkofer}},\
  }\bibfield  {title} {\bibinfo {title} {{Color-superconductivity in the
  strong-coupling regime of Landau gauge QCD}},\ }\href
  {https://doi.org/10.1103/PhysRevD.73.114028} {\bibfield  {journal} {\bibinfo
  {journal} {Phys. Rev. D}\ }\textbf {\bibinfo {volume} {73}},\ \bibinfo
  {pages} {114028} (\bibinfo {year} {2006})},\ \Eprint
  {https://arxiv.org/abs/hep-ph/0603163} {arXiv:hep-ph/0603163} \BibitemShut
  {NoStop}%
\bibitem [{\citenamefont {Braun}\ \emph {et~al.}(2020)\citenamefont {Braun},
  \citenamefont {Leonhardt},\ and\ \citenamefont {Pospiech}}]{Braun:2019aow}%
  \BibitemOpen
  \bibfield  {author} {\bibinfo {author} {\bibfnamefont {J.}~\bibnamefont
  {Braun}}, \bibinfo {author} {\bibfnamefont {M.}~\bibnamefont {Leonhardt}},\
  and\ \bibinfo {author} {\bibfnamefont {M.}~\bibnamefont {Pospiech}},\
  }\bibfield  {title} {\bibinfo {title} {{Fierz-complete NJL model study III:
  Emergence from quark-gluon dynamics}},\ }\href
  {https://doi.org/10.1103/PhysRevD.101.036004} {\bibfield  {journal} {\bibinfo
   {journal} {Phys. Rev. D}\ }\textbf {\bibinfo {volume} {101}},\ \bibinfo
  {pages} {036004} (\bibinfo {year} {2020})},\ \Eprint
  {https://arxiv.org/abs/1909.06298} {arXiv:1909.06298 [hep-ph]} \BibitemShut
  {NoStop}%
\bibitem [{\citenamefont {Leonhardt}\ \emph {et~al.}(2020)\citenamefont
  {Leonhardt}, \citenamefont {Pospiech}, \citenamefont {Schallmo},
  \citenamefont {Braun}, \citenamefont {Drischler}, \citenamefont {Hebeler},\
  and\ \citenamefont {Schwenk}}]{Leonhardt:2019fua}%
  \BibitemOpen
  \bibfield  {author} {\bibinfo {author} {\bibfnamefont {M.}~\bibnamefont
  {Leonhardt}}, \bibinfo {author} {\bibfnamefont {M.}~\bibnamefont {Pospiech}},
  \bibinfo {author} {\bibfnamefont {B.}~\bibnamefont {Schallmo}}, \bibinfo
  {author} {\bibfnamefont {J.}~\bibnamefont {Braun}}, \bibinfo {author}
  {\bibfnamefont {C.}~\bibnamefont {Drischler}}, \bibinfo {author}
  {\bibfnamefont {K.}~\bibnamefont {Hebeler}},\ and\ \bibinfo {author}
  {\bibfnamefont {A.}~\bibnamefont {Schwenk}},\ }\bibfield  {title} {\bibinfo
  {title} {{Symmetric nuclear matter from the strong interaction}},\ }\href
  {https://doi.org/10.1103/PhysRevLett.125.142502} {\bibfield  {journal}
  {\bibinfo  {journal} {Phys. Rev. Lett.}\ }\textbf {\bibinfo {volume} {125}},\
  \bibinfo {pages} {142502} (\bibinfo {year} {2020})},\ \Eprint
  {https://arxiv.org/abs/1907.05814} {arXiv:1907.05814 [nucl-th]} \BibitemShut
  {NoStop}%
\bibitem [{\citenamefont {Braun}\ and\ \citenamefont
  {Schallmo}(2022{\natexlab{a}})}]{Braun:2021uua}%
  \BibitemOpen
  \bibfield  {author} {\bibinfo {author} {\bibfnamefont {J.}~\bibnamefont
  {Braun}}\ and\ \bibinfo {author} {\bibfnamefont {B.}~\bibnamefont
  {Schallmo}},\ }\bibfield  {title} {\bibinfo {title} {{From quarks and gluons
  to color superconductivity at supranuclear densities}},\ }\href
  {https://doi.org/10.1103/PhysRevD.105.036003} {\bibfield  {journal} {\bibinfo
   {journal} {Phys. Rev. D}\ }\textbf {\bibinfo {volume} {105}},\ \bibinfo
  {pages} {036003} (\bibinfo {year} {2022}{\natexlab{a}})},\ \Eprint
  {https://arxiv.org/abs/2106.04198} {arXiv:2106.04198 [hep-ph]} \BibitemShut
  {NoStop}%
\bibitem [{\citenamefont {Fukushima}\ \emph {et~al.}(2022)\citenamefont
  {Fukushima}, \citenamefont {Pawlowski},\ and\ \citenamefont
  {Strodthoff}}]{Fukushima:2021ctq}%
  \BibitemOpen
  \bibfield  {author} {\bibinfo {author} {\bibfnamefont {K.}~\bibnamefont
  {Fukushima}}, \bibinfo {author} {\bibfnamefont {J.~M.}\ \bibnamefont
  {Pawlowski}},\ and\ \bibinfo {author} {\bibfnamefont {N.}~\bibnamefont
  {Strodthoff}},\ }\bibfield  {title} {\bibinfo {title} {{Emergent hadrons and
  diquarks}},\ }\href {https://doi.org/10.1016/j.aop.2022.169106} {\bibfield
  {journal} {\bibinfo  {journal} {Annals Phys.}\ }\textbf {\bibinfo {volume}
  {446}},\ \bibinfo {pages} {169106} (\bibinfo {year} {2022})},\ \Eprint
  {https://arxiv.org/abs/2103.01129} {arXiv:2103.01129 [hep-ph]} \BibitemShut
  {NoStop}%
\bibitem [{\citenamefont {Gei{\ss}el}\ \emph {et~al.}(2024)\citenamefont
  {Gei{\ss}el}, \citenamefont {Gorda},\ and\ \citenamefont
  {Braun}}]{Geissel:2024nmx}%
  \BibitemOpen
  \bibfield  {author} {\bibinfo {author} {\bibfnamefont {A.}~\bibnamefont
  {Gei{\ss}el}}, \bibinfo {author} {\bibfnamefont {T.}~\bibnamefont {Gorda}},\
  and\ \bibinfo {author} {\bibfnamefont {J.}~\bibnamefont {Braun}},\ }\bibfield
   {title} {\bibinfo {title} {{Pressure and speed of sound in two-flavor
  color-superconducting quark matter at next-to-leading order}},\ }\href
  {https://doi.org/10.1103/PhysRevD.110.014034} {\bibfield  {journal} {\bibinfo
   {journal} {Phys. Rev. D}\ }\textbf {\bibinfo {volume} {110}},\ \bibinfo
  {pages} {014034} (\bibinfo {year} {2024})},\ \Eprint
  {https://arxiv.org/abs/2403.18010} {arXiv:2403.18010 [hep-ph]} \BibitemShut
  {NoStop}%
\bibitem [{\citenamefont {Fukushima}\ and\ \citenamefont
  {Minato}(2025)}]{Fukushima:2024gmp}%
  \BibitemOpen
  \bibfield  {author} {\bibinfo {author} {\bibfnamefont {K.}~\bibnamefont
  {Fukushima}}\ and\ \bibinfo {author} {\bibfnamefont {S.}~\bibnamefont
  {Minato}},\ }\bibfield  {title} {\bibinfo {title} {{Speed of sound and trace
  anomaly in a unified treatment of the two-color diquark superfluid, the
  pion-condensed high-isospin matter, and the 2SC quark matter}},\ }\href
  {https://doi.org/10.1103/PhysRevD.111.094006} {\bibfield  {journal} {\bibinfo
   {journal} {Phys. Rev. D}\ }\textbf {\bibinfo {volume} {111}},\ \bibinfo
  {pages} {094006} (\bibinfo {year} {2025})},\ \Eprint
  {https://arxiv.org/abs/2411.03781} {arXiv:2411.03781 [hep-ph]} \BibitemShut
  {NoStop}%
\bibitem [{\citenamefont {Gei{\ss}el}\ \emph {et~al.}(2025)\citenamefont
  {Gei{\ss}el}, \citenamefont {Gorda},\ and\ \citenamefont
  {Braun}}]{Geissel:2025vnp}%
  \BibitemOpen
  \bibfield  {author} {\bibinfo {author} {\bibfnamefont {A.}~\bibnamefont
  {Gei{\ss}el}}, \bibinfo {author} {\bibfnamefont {T.}~\bibnamefont {Gorda}},\
  and\ \bibinfo {author} {\bibfnamefont {J.}~\bibnamefont {Braun}},\ }\bibfield
   {title} {\bibinfo {title} {{Color superconductivity under neutron-star
  conditions at next-to-leading order}},\ }\Eprint
  {https://arxiv.org/abs/2504.03834} {arXiv:2504.03834 [hep-ph]}  (\bibinfo
  {year} {2025})\BibitemShut {NoStop}%
\bibitem [{\citenamefont {Fu}\ \emph {et~al.}(2020)\citenamefont {Fu},
  \citenamefont {Pawlowski},\ and\ \citenamefont {Rennecke}}]{Fu:2019hdw}%
  \BibitemOpen
  \bibfield  {author} {\bibinfo {author} {\bibfnamefont {W.-j.}\ \bibnamefont
  {Fu}}, \bibinfo {author} {\bibfnamefont {J.~M.}\ \bibnamefont {Pawlowski}},\
  and\ \bibinfo {author} {\bibfnamefont {F.}~\bibnamefont {Rennecke}},\
  }\bibfield  {title} {\bibinfo {title} {{QCD phase structure at finite
  temperature and density}},\ }\href
  {https://doi.org/10.1103/PhysRevD.101.054032} {\bibfield  {journal} {\bibinfo
   {journal} {Phys. Rev. D}\ }\textbf {\bibinfo {volume} {101}},\ \bibinfo
  {pages} {054032} (\bibinfo {year} {2020})},\ \Eprint
  {https://arxiv.org/abs/1909.02991} {arXiv:1909.02991 [hep-ph]} \BibitemShut
  {NoStop}%
\bibitem [{\citenamefont {Rischke}\ and\ \citenamefont
  {Pisarski}(2000)}]{Rischke:2000pv}%
  \BibitemOpen
  \bibfield  {author} {\bibinfo {author} {\bibfnamefont {D.~H.}\ \bibnamefont
  {Rischke}}\ and\ \bibinfo {author} {\bibfnamefont {R.~D.}\ \bibnamefont
  {Pisarski}},\ }\bibfield  {title} {\bibinfo {title} {{Color superconductivity
  in cold, dense quark matter}},\ }in\ \href@noop {} {\emph {\bibinfo
  {booktitle} {{5th Workshop on QCD (QCD 2000)}}}}\ (\bibinfo {year} {2000})\
  pp.\ \bibinfo {pages} {220--230},\ \Eprint
  {https://arxiv.org/abs/nucl-th/0004016} {arXiv:nucl-th/0004016} \BibitemShut
  {NoStop}%
\bibitem [{\citenamefont {Alford}(2001)}]{Alford:2001dt}%
  \BibitemOpen
  \bibfield  {author} {\bibinfo {author} {\bibfnamefont {M.~G.}\ \bibnamefont
  {Alford}},\ }\bibfield  {title} {\bibinfo {title} {{Color superconducting
  quark matter}},\ }\href
  {https://doi.org/10.1146/annurev.nucl.51.101701.132449} {\bibfield  {journal}
  {\bibinfo  {journal} {Ann. Rev. Nucl. Part. Sci.}\ }\textbf {\bibinfo
  {volume} {51}},\ \bibinfo {pages} {131} (\bibinfo {year} {2001})},\ \Eprint
  {https://arxiv.org/abs/hep-ph/0102047} {arXiv:hep-ph/0102047} \BibitemShut
  {NoStop}%
\bibitem [{\citenamefont {Buballa}(2005)}]{Buballa:2003qv}%
  \BibitemOpen
  \bibfield  {author} {\bibinfo {author} {\bibfnamefont {M.}~\bibnamefont
  {Buballa}},\ }\bibfield  {title} {\bibinfo {title} {{NJL model analysis of
  quark matter at large density}},\ }\href
  {https://doi.org/10.1016/j.physrep.2004.11.004} {\bibfield  {journal}
  {\bibinfo  {journal} {Phys. Rept.}\ }\textbf {\bibinfo {volume} {407}},\
  \bibinfo {pages} {205} (\bibinfo {year} {2005})},\ \Eprint
  {https://arxiv.org/abs/hep-ph/0402234} {arXiv:hep-ph/0402234} \BibitemShut
  {NoStop}%
\bibitem [{\citenamefont {Shovkovy}(2005)}]{Shovkovy:2004me}%
  \BibitemOpen
  \bibfield  {author} {\bibinfo {author} {\bibfnamefont {I.~A.}\ \bibnamefont
  {Shovkovy}},\ }\bibfield  {title} {\bibinfo {title} {{Two lectures on color
  superconductivity}},\ }\href {https://doi.org/10.1007/s10701-005-6440-x}
  {\bibfield  {journal} {\bibinfo  {journal} {Found. Phys.}\ }\textbf {\bibinfo
  {volume} {35}},\ \bibinfo {pages} {1309} (\bibinfo {year} {2005})},\ \Eprint
  {https://arxiv.org/abs/nucl-th/0410091} {arXiv:nucl-th/0410091} \BibitemShut
  {NoStop}%
\bibitem [{\citenamefont {Alford}\ \emph {et~al.}(2008)\citenamefont {Alford},
  \citenamefont {Schmitt}, \citenamefont {Rajagopal},\ and\ \citenamefont
  {Sch\"afer}}]{Alford:2007xm}%
  \BibitemOpen
  \bibfield  {author} {\bibinfo {author} {\bibfnamefont {M.~G.}\ \bibnamefont
  {Alford}}, \bibinfo {author} {\bibfnamefont {A.}~\bibnamefont {Schmitt}},
  \bibinfo {author} {\bibfnamefont {K.}~\bibnamefont {Rajagopal}},\ and\
  \bibinfo {author} {\bibfnamefont {T.}~\bibnamefont {Sch\"afer}},\ }\bibfield
  {title} {\bibinfo {title} {{Color superconductivity in dense quark matter}},\
  }\href {https://doi.org/10.1103/RevModPhys.80.1455} {\bibfield  {journal}
  {\bibinfo  {journal} {Rev. Mod. Phys.}\ }\textbf {\bibinfo {volume} {80}},\
  \bibinfo {pages} {1455} (\bibinfo {year} {2008})},\ \Eprint
  {https://arxiv.org/abs/0709.4635} {arXiv:0709.4635 [hep-ph]} \BibitemShut
  {NoStop}%
\bibitem [{\citenamefont {Anglani}\ \emph {et~al.}(2014)\citenamefont
  {Anglani}, \citenamefont {Casalbuoni}, \citenamefont {Ciminale},
  \citenamefont {Ippolito}, \citenamefont {Gatto}, \citenamefont {Mannarelli},\
  and\ \citenamefont {Ruggieri}}]{Anglani:2013gfu}%
  \BibitemOpen
  \bibfield  {author} {\bibinfo {author} {\bibfnamefont {R.}~\bibnamefont
  {Anglani}}, \bibinfo {author} {\bibfnamefont {R.}~\bibnamefont {Casalbuoni}},
  \bibinfo {author} {\bibfnamefont {M.}~\bibnamefont {Ciminale}}, \bibinfo
  {author} {\bibfnamefont {N.}~\bibnamefont {Ippolito}}, \bibinfo {author}
  {\bibfnamefont {R.}~\bibnamefont {Gatto}}, \bibinfo {author} {\bibfnamefont
  {M.}~\bibnamefont {Mannarelli}},\ and\ \bibinfo {author} {\bibfnamefont
  {M.}~\bibnamefont {Ruggieri}},\ }\bibfield  {title} {\bibinfo {title}
  {{Crystalline color superconductors}},\ }\href
  {https://doi.org/10.1103/RevModPhys.86.509} {\bibfield  {journal} {\bibinfo
  {journal} {Rev. Mod. Phys.}\ }\textbf {\bibinfo {volume} {86}},\ \bibinfo
  {pages} {509} (\bibinfo {year} {2014})},\ \Eprint
  {https://arxiv.org/abs/1302.4264} {arXiv:1302.4264 [hep-ph]} \BibitemShut
  {NoStop}%
\bibitem [{\citenamefont {Buballa}\ and\ \citenamefont
  {Carignano}(2015)}]{Buballa:2014tba}%
  \BibitemOpen
  \bibfield  {author} {\bibinfo {author} {\bibfnamefont {M.}~\bibnamefont
  {Buballa}}\ and\ \bibinfo {author} {\bibfnamefont {S.}~\bibnamefont
  {Carignano}},\ }\bibfield  {title} {\bibinfo {title} {{Inhomogeneous chiral
  condensates}},\ }\href {https://doi.org/10.1016/j.ppnp.2014.11.001}
  {\bibfield  {journal} {\bibinfo  {journal} {Prog. Part. Nucl. Phys.}\
  }\textbf {\bibinfo {volume} {81}},\ \bibinfo {pages} {39} (\bibinfo {year}
  {2015})},\ \Eprint {https://arxiv.org/abs/1406.1367} {arXiv:1406.1367
  [hep-ph]} \BibitemShut {NoStop}%
\bibitem [{\citenamefont {Braun}\ \emph {et~al.}(2019)\citenamefont {Braun},
  \citenamefont {Leonhardt},\ and\ \citenamefont {Pawlowski}}]{Braun:2018svj}%
  \BibitemOpen
  \bibfield  {author} {\bibinfo {author} {\bibfnamefont {J.}~\bibnamefont
  {Braun}}, \bibinfo {author} {\bibfnamefont {M.}~\bibnamefont {Leonhardt}},\
  and\ \bibinfo {author} {\bibfnamefont {J.~M.}\ \bibnamefont {Pawlowski}},\
  }\bibfield  {title} {\bibinfo {title} {{Renormalization group consistency and
  low-energy effective theories}},\ }\href
  {https://doi.org/10.21468/SciPostPhys.6.5.056} {\bibfield  {journal}
  {\bibinfo  {journal} {SciPost Phys.}\ }\textbf {\bibinfo {volume} {6}},\
  \bibinfo {pages} {056} (\bibinfo {year} {2019})},\ \Eprint
  {https://arxiv.org/abs/1806.04432} {arXiv:1806.04432 [hep-ph]} \BibitemShut
  {NoStop}%
\bibitem [{\citenamefont {Braun}\ and\ \citenamefont
  {Schallmo}(2022{\natexlab{b}})}]{Braun:2022olp}%
  \BibitemOpen
  \bibfield  {author} {\bibinfo {author} {\bibfnamefont {J.}~\bibnamefont
  {Braun}}\ and\ \bibinfo {author} {\bibfnamefont {B.}~\bibnamefont
  {Schallmo}},\ }\bibfield  {title} {\bibinfo {title} {{Zero-temperature
  thermodynamics of dense asymmetric strong-interaction matter}},\ }\href
  {https://doi.org/10.1103/PhysRevD.106.076010} {\bibfield  {journal} {\bibinfo
   {journal} {Phys. Rev. D}\ }\textbf {\bibinfo {volume} {106}},\ \bibinfo
  {pages} {076010} (\bibinfo {year} {2022}{\natexlab{b}})},\ \Eprint
  {https://arxiv.org/abs/2204.00358} {arXiv:2204.00358 [nucl-th]} \BibitemShut
  {NoStop}%
\bibitem [{\citenamefont {Andersen}\ and\ \citenamefont
  {N{\o}dtvedt}(2025{\natexlab{a}})}]{Andersen:2024qus}%
  \BibitemOpen
  \bibfield  {author} {\bibinfo {author} {\bibfnamefont {J.~O.}\ \bibnamefont
  {Andersen}}\ and\ \bibinfo {author} {\bibfnamefont {M.~P.}\ \bibnamefont
  {N{\o}dtvedt}},\ }\bibfield  {title} {\bibinfo {title} {{Color
  superconductivity and speed of sound in the two-flavor quark-meson diquark
  model}},\ }\href {https://doi.org/10.1103/PhysRevD.111.034031} {\bibfield
  {journal} {\bibinfo  {journal} {Phys. Rev. D}\ }\textbf {\bibinfo {volume}
  {111}},\ \bibinfo {pages} {034031} (\bibinfo {year} {2025}{\natexlab{a}})},\
  \Eprint {https://arxiv.org/abs/2408.12361} {arXiv:2408.12361 [hep-ph]}
  \BibitemShut {NoStop}%
\bibitem [{\citenamefont {Gholami}\ \emph
  {et~al.}(2025{\natexlab{a}})\citenamefont {Gholami}, \citenamefont
  {Hofmann},\ and\ \citenamefont {Buballa}}]{Gholami:2024diy}%
  \BibitemOpen
  \bibfield  {author} {\bibinfo {author} {\bibfnamefont {H.}~\bibnamefont
  {Gholami}}, \bibinfo {author} {\bibfnamefont {M.}~\bibnamefont {Hofmann}},\
  and\ \bibinfo {author} {\bibfnamefont {M.}~\bibnamefont {Buballa}},\
  }\bibfield  {title} {\bibinfo {title} {{Renormalization-group consistent
  treatment of color superconductivity in the NJL model}},\ }\href
  {https://doi.org/10.1103/PhysRevD.111.014006} {\bibfield  {journal} {\bibinfo
   {journal} {Phys. Rev. D}\ }\textbf {\bibinfo {volume} {111}},\ \bibinfo
  {pages} {014006} (\bibinfo {year} {2025}{\natexlab{a}})},\ \Eprint
  {https://arxiv.org/abs/2408.06704} {arXiv:2408.06704 [hep-ph]} \BibitemShut
  {NoStop}%
\bibitem [{\citenamefont {Gholami}\ \emph
  {et~al.}(2025{\natexlab{b}})\citenamefont {Gholami}, \citenamefont {Kurth},
  \citenamefont {Mire}, \citenamefont {Buballa},\ and\ \citenamefont
  {Schaefer}}]{Gholami:2025afm}%
  \BibitemOpen
  \bibfield  {author} {\bibinfo {author} {\bibfnamefont {H.}~\bibnamefont
  {Gholami}}, \bibinfo {author} {\bibfnamefont {L.}~\bibnamefont {Kurth}},
  \bibinfo {author} {\bibfnamefont {U.}~\bibnamefont {Mire}}, \bibinfo {author}
  {\bibfnamefont {M.}~\bibnamefont {Buballa}},\ and\ \bibinfo {author}
  {\bibfnamefont {B.-J.}\ \bibnamefont {Schaefer}},\ }\bibfield  {title}
  {\bibinfo {title} {{Renormalizing the Quark-Meson-Diquark Model}},\ }\Eprint
  {https://arxiv.org/abs/2505.22542} {arXiv:2505.22542 [hep-ph]}  (\bibinfo
  {year} {2025}{\natexlab{b}})\BibitemShut {NoStop}%
\bibitem [{\citenamefont {Andersen}\ and\ \citenamefont
  {N{\o}dtvedt}(2025{\natexlab{b}})}]{Andersen:2025uzh}%
  \BibitemOpen
  \bibfield  {author} {\bibinfo {author} {\bibfnamefont {J.~O.}\ \bibnamefont
  {Andersen}}\ and\ \bibinfo {author} {\bibfnamefont {M.~P.}\ \bibnamefont
  {N{\o}dtvedt}},\ }\bibfield  {title} {\bibinfo {title} {{Renormalization of
  the three-flavor quark-meson diquark model}},\ }\Eprint
  {https://arxiv.org/abs/2506.02941} {arXiv:2506.02941 [hep-ph]}  (\bibinfo
  {year} {2025}{\natexlab{b}})\BibitemShut {NoStop}%
\bibitem [{\citenamefont {Elitzur}(1975)}]{Elitzur:1975im}%
  \BibitemOpen
  \bibfield  {author} {\bibinfo {author} {\bibfnamefont {S.}~\bibnamefont
  {Elitzur}},\ }\bibfield  {title} {\bibinfo {title} {{Impossibility of
  Spontaneously Breaking Local Symmetries}},\ }\href
  {https://doi.org/10.1103/PhysRevD.12.3978} {\bibfield  {journal} {\bibinfo
  {journal} {Phys. Rev. D}\ }\textbf {\bibinfo {volume} {12}},\ \bibinfo
  {pages} {3978} (\bibinfo {year} {1975})}\BibitemShut {NoStop}%
\bibitem [{\citenamefont {Bardeen}\ \emph {et~al.}(1957)\citenamefont
  {Bardeen}, \citenamefont {Cooper},\ and\ \citenamefont
  {Schrieffer}}]{Bardeen:1957mv}%
  \BibitemOpen
  \bibfield  {author} {\bibinfo {author} {\bibfnamefont {J.}~\bibnamefont
  {Bardeen}}, \bibinfo {author} {\bibfnamefont {L.~N.}\ \bibnamefont
  {Cooper}},\ and\ \bibinfo {author} {\bibfnamefont {J.~R.}\ \bibnamefont
  {Schrieffer}},\ }\bibfield  {title} {\bibinfo {title} {{Theory of
  superconductivity}},\ }\href {https://doi.org/10.1103/PhysRev.108.1175}
  {\bibfield  {journal} {\bibinfo  {journal} {Phys. Rev.}\ }\textbf {\bibinfo
  {volume} {108}},\ \bibinfo {pages} {1175} (\bibinfo {year}
  {1957})}\BibitemShut {NoStop}%
\bibitem [{\citenamefont {Wetterich}(1993)}]{Wetterich:1992yh}%
  \BibitemOpen
  \bibfield  {author} {\bibinfo {author} {\bibfnamefont {C.}~\bibnamefont
  {Wetterich}},\ }\bibfield  {title} {\bibinfo {title} {{Exact evolution
  equation for the effective potential}},\ }\href
  {https://doi.org/10.1016/0370-2693(93)90726-X} {\bibfield  {journal}
  {\bibinfo  {journal} {Phys. Lett. B}\ }\textbf {\bibinfo {volume} {301}},\
  \bibinfo {pages} {90} (\bibinfo {year} {1993})},\ \Eprint
  {https://arxiv.org/abs/1710.05815} {arXiv:1710.05815 [hep-th]} \BibitemShut
  {NoStop}%
\bibitem [{\citenamefont {Koenigstein}\ \emph
  {et~al.}(2022{\natexlab{a}})\citenamefont {Koenigstein}, \citenamefont
  {Steil}, \citenamefont {Wink}, \citenamefont {Grossi}, \citenamefont {Braun},
  \citenamefont {Buballa},\ and\ \citenamefont
  {Rischke}}]{Koenigstein:2021syz}%
  \BibitemOpen
  \bibfield  {author} {\bibinfo {author} {\bibfnamefont {A.}~\bibnamefont
  {Koenigstein}}, \bibinfo {author} {\bibfnamefont {M.~J.}\ \bibnamefont
  {Steil}}, \bibinfo {author} {\bibfnamefont {N.}~\bibnamefont {Wink}},
  \bibinfo {author} {\bibfnamefont {E.}~\bibnamefont {Grossi}}, \bibinfo
  {author} {\bibfnamefont {J.}~\bibnamefont {Braun}}, \bibinfo {author}
  {\bibfnamefont {M.}~\bibnamefont {Buballa}},\ and\ \bibinfo {author}
  {\bibfnamefont {D.~H.}\ \bibnamefont {Rischke}},\ }\bibfield  {title}
  {\bibinfo {title} {{Numerical fluid dynamics for FRG flow equations:
  Zero-dimensional QFTs as numerical test cases. I. The O(N) model}},\ }\href
  {https://doi.org/10.1103/PhysRevD.106.065012} {\bibfield  {journal} {\bibinfo
   {journal} {Phys. Rev. D}\ }\textbf {\bibinfo {volume} {106}},\ \bibinfo
  {pages} {065012} (\bibinfo {year} {2022}{\natexlab{a}})},\ \Eprint
  {https://arxiv.org/abs/2108.02504} {arXiv:2108.02504 [cond-mat.stat-mech]}
  \BibitemShut {NoStop}%
\bibitem [{\citenamefont {Koenigstein}\ \emph
  {et~al.}(2022{\natexlab{b}})\citenamefont {Koenigstein}, \citenamefont
  {Steil}, \citenamefont {Wink}, \citenamefont {Grossi},\ and\ \citenamefont
  {Braun}}]{Koenigstein:2021rxj}%
  \BibitemOpen
  \bibfield  {author} {\bibinfo {author} {\bibfnamefont {A.}~\bibnamefont
  {Koenigstein}}, \bibinfo {author} {\bibfnamefont {M.~J.}\ \bibnamefont
  {Steil}}, \bibinfo {author} {\bibfnamefont {N.}~\bibnamefont {Wink}},
  \bibinfo {author} {\bibfnamefont {E.}~\bibnamefont {Grossi}},\ and\ \bibinfo
  {author} {\bibfnamefont {J.}~\bibnamefont {Braun}},\ }\bibfield  {title}
  {\bibinfo {title} {{Numerical fluid dynamics for FRG flow equations:
  Zero-dimensional QFTs as numerical test cases. II. Entropy production and
  irreversibility of RG flows}},\ }\href
  {https://doi.org/10.1103/PhysRevD.106.065013} {\bibfield  {journal} {\bibinfo
   {journal} {Phys. Rev. D}\ }\textbf {\bibinfo {volume} {106}},\ \bibinfo
  {pages} {065013} (\bibinfo {year} {2022}{\natexlab{b}})},\ \Eprint
  {https://arxiv.org/abs/2108.10085} {arXiv:2108.10085 [cond-mat.stat-mech]}
  \BibitemShut {NoStop}%
\bibitem [{\citenamefont {Steil}\ and\ \citenamefont
  {Koenigstein}(2022)}]{Steil:2021cbu}%
  \BibitemOpen
  \bibfield  {author} {\bibinfo {author} {\bibfnamefont {M.~J.}\ \bibnamefont
  {Steil}}\ and\ \bibinfo {author} {\bibfnamefont {A.}~\bibnamefont
  {Koenigstein}},\ }\bibfield  {title} {\bibinfo {title} {{Numerical fluid
  dynamics for FRG flow equations: Zero-dimensional QFTs as numerical test
  cases. III. Shock and rarefaction waves in RG flows reveal limitations of the
  N{\textrightarrow}{\ensuremath{\infty}} limit in O(N)-type models}},\ }\href
  {https://doi.org/10.1103/PhysRevD.106.065014} {\bibfield  {journal} {\bibinfo
   {journal} {Phys. Rev. D}\ }\textbf {\bibinfo {volume} {106}},\ \bibinfo
  {pages} {065014} (\bibinfo {year} {2022})},\ \Eprint
  {https://arxiv.org/abs/2108.04037} {arXiv:2108.04037 [cond-mat.stat-mech]}
  \BibitemShut {NoStop}%
\bibitem [{\citenamefont {Zorbach}\ \emph {et~al.}(2024)\citenamefont
  {Zorbach}, \citenamefont {Koenigstein},\ and\ \citenamefont
  {Braun}}]{Zorbach:2024rre}%
  \BibitemOpen
  \bibfield  {author} {\bibinfo {author} {\bibfnamefont {N.}~\bibnamefont
  {Zorbach}}, \bibinfo {author} {\bibfnamefont {A.}~\bibnamefont
  {Koenigstein}},\ and\ \bibinfo {author} {\bibfnamefont {J.}~\bibnamefont
  {Braun}},\ }\bibfield  {title} {\bibinfo {title} {{Functional Renormalization
  Group meets Computational Fluid Dynamics: RG flows in a multi-dimensional
  field space}},\ }\Eprint {https://arxiv.org/abs/2412.16053} {arXiv:2412.16053
  [cond-mat.stat-mech]}  (\bibinfo {year} {2024})\BibitemShut {NoStop}%
\bibitem [{\citenamefont {Grossi}\ and\ \citenamefont
  {Wink}(2023)}]{Grossi:2019urj}%
  \BibitemOpen
  \bibfield  {author} {\bibinfo {author} {\bibfnamefont {E.}~\bibnamefont
  {Grossi}}\ and\ \bibinfo {author} {\bibfnamefont {N.}~\bibnamefont {Wink}},\
  }\bibfield  {title} {\bibinfo {title} {{Resolving phase transitions with
  discontinuous Galerkin methods}},\ }\href
  {https://doi.org/10.21468/SciPostPhysCore.6.4.071} {\bibfield  {journal}
  {\bibinfo  {journal} {SciPost Phys. Core}\ }\textbf {\bibinfo {volume} {6}},\
  \bibinfo {pages} {071} (\bibinfo {year} {2023})},\ \Eprint
  {https://arxiv.org/abs/1903.09503} {arXiv:1903.09503 [hep-th]} \BibitemShut
  {NoStop}%
\bibitem [{\citenamefont {Dupuis}\ \emph {et~al.}(2021)\citenamefont {Dupuis},
  \citenamefont {Canet}, \citenamefont {Eichhorn}, \citenamefont {Metzner},
  \citenamefont {Pawlowski}, \citenamefont {Tissier},\ and\ \citenamefont
  {Wschebor}}]{Dupuis:2020fhh}%
  \BibitemOpen
  \bibfield  {author} {\bibinfo {author} {\bibfnamefont {N.}~\bibnamefont
  {Dupuis}}, \bibinfo {author} {\bibfnamefont {L.}~\bibnamefont {Canet}},
  \bibinfo {author} {\bibfnamefont {A.}~\bibnamefont {Eichhorn}}, \bibinfo
  {author} {\bibfnamefont {W.}~\bibnamefont {Metzner}}, \bibinfo {author}
  {\bibfnamefont {J.~M.}\ \bibnamefont {Pawlowski}}, \bibinfo {author}
  {\bibfnamefont {M.}~\bibnamefont {Tissier}},\ and\ \bibinfo {author}
  {\bibfnamefont {N.}~\bibnamefont {Wschebor}},\ }\bibfield  {title} {\bibinfo
  {title} {{The nonperturbative functional renormalization group and its
  applications}},\ }\href {https://doi.org/10.1016/j.physrep.2021.01.001}
  {\bibfield  {journal} {\bibinfo  {journal} {Phys. Rept.}\ }\textbf {\bibinfo
  {volume} {910}},\ \bibinfo {pages} {1} (\bibinfo {year} {2021})},\ \Eprint
  {https://arxiv.org/abs/2006.04853} {arXiv:2006.04853 [cond-mat.stat-mech]}
  \BibitemShut {NoStop}%
\bibitem [{\citenamefont {Rapp}\ \emph {et~al.}(1998)\citenamefont {Rapp},
  \citenamefont {Sch\"afer}, \citenamefont {Shuryak},\ and\ \citenamefont
  {Velkovsky}}]{Rapp:1997zu}%
  \BibitemOpen
  \bibfield  {author} {\bibinfo {author} {\bibfnamefont {R.}~\bibnamefont
  {Rapp}}, \bibinfo {author} {\bibfnamefont {T.}~\bibnamefont {Sch\"afer}},
  \bibinfo {author} {\bibfnamefont {E.~V.}\ \bibnamefont {Shuryak}},\ and\
  \bibinfo {author} {\bibfnamefont {M.}~\bibnamefont {Velkovsky}},\ }\bibfield
  {title} {\bibinfo {title} {{Diquark Bose condensates in high density matter
  and instantons}},\ }\href {https://doi.org/10.1103/PhysRevLett.81.53}
  {\bibfield  {journal} {\bibinfo  {journal} {Phys. Rev. Lett.}\ }\textbf
  {\bibinfo {volume} {81}},\ \bibinfo {pages} {53} (\bibinfo {year} {1998})},\
  \Eprint {https://arxiv.org/abs/hep-ph/9711396} {arXiv:hep-ph/9711396}
  \BibitemShut {NoStop}%
\bibitem [{\citenamefont {Alford}\ \emph {et~al.}(1998)\citenamefont {Alford},
  \citenamefont {Rajagopal},\ and\ \citenamefont {Wilczek}}]{Alford:1997zt}%
  \BibitemOpen
  \bibfield  {author} {\bibinfo {author} {\bibfnamefont {M.~G.}\ \bibnamefont
  {Alford}}, \bibinfo {author} {\bibfnamefont {K.}~\bibnamefont {Rajagopal}},\
  and\ \bibinfo {author} {\bibfnamefont {F.}~\bibnamefont {Wilczek}},\
  }\bibfield  {title} {\bibinfo {title} {{QCD at finite baryon density: Nucleon
  droplets and color superconductivity}},\ }\href
  {https://doi.org/10.1016/S0370-2693(98)00051-3} {\bibfield  {journal}
  {\bibinfo  {journal} {Phys. Lett. B}\ }\textbf {\bibinfo {volume} {422}},\
  \bibinfo {pages} {247} (\bibinfo {year} {1998})},\ \Eprint
  {https://arxiv.org/abs/hep-ph/9711395} {arXiv:hep-ph/9711395} \BibitemShut
  {NoStop}%
\bibitem [{\citenamefont {Berges}\ and\ \citenamefont
  {Rajagopal}(1999)}]{Berges:1998rc}%
  \BibitemOpen
  \bibfield  {author} {\bibinfo {author} {\bibfnamefont {J.}~\bibnamefont
  {Berges}}\ and\ \bibinfo {author} {\bibfnamefont {K.}~\bibnamefont
  {Rajagopal}},\ }\bibfield  {title} {\bibinfo {title} {{Color
  superconductivity and chiral symmetry restoration at nonzero baryon density
  and temperature}},\ }\href {https://doi.org/10.1016/S0550-3213(98)00620-8}
  {\bibfield  {journal} {\bibinfo  {journal} {Nucl. Phys. B}\ }\textbf
  {\bibinfo {volume} {538}},\ \bibinfo {pages} {215} (\bibinfo {year}
  {1999})},\ \Eprint {https://arxiv.org/abs/hep-ph/9804233}
  {arXiv:hep-ph/9804233} \BibitemShut {NoStop}%
\bibitem [{\citenamefont {Haensch}\ \emph {et~al.}(2024)\citenamefont
  {Haensch}, \citenamefont {Rennecke},\ and\ \citenamefont {von
  Smekal}}]{Haensch:2023sig}%
  \BibitemOpen
  \bibfield  {author} {\bibinfo {author} {\bibfnamefont {M.}~\bibnamefont
  {Haensch}}, \bibinfo {author} {\bibfnamefont {F.}~\bibnamefont {Rennecke}},\
  and\ \bibinfo {author} {\bibfnamefont {L.}~\bibnamefont {von Smekal}},\
  }\bibfield  {title} {\bibinfo {title} {{Medium induced mixing, spatial
  modulations, and critical modes in QCD}},\ }\href
  {https://doi.org/10.1103/PhysRevD.110.036018} {\bibfield  {journal} {\bibinfo
   {journal} {Phys. Rev. D}\ }\textbf {\bibinfo {volume} {110}},\ \bibinfo
  {pages} {036018} (\bibinfo {year} {2024})},\ \Eprint
  {https://arxiv.org/abs/2308.16244} {arXiv:2308.16244 [hep-ph]} \BibitemShut
  {NoStop}%
\bibitem [{\citenamefont {Mark\'o}\ \emph {et~al.}(2014)\citenamefont
  {Mark\'o}, \citenamefont {Reinosa},\ and\ \citenamefont
  {Sz\'ep}}]{Marko:2014hea}%
  \BibitemOpen
  \bibfield  {author} {\bibinfo {author} {\bibfnamefont {G.}~\bibnamefont
  {Mark\'o}}, \bibinfo {author} {\bibfnamefont {U.}~\bibnamefont {Reinosa}},\
  and\ \bibinfo {author} {\bibfnamefont {Z.}~\bibnamefont {Sz\'ep}},\
  }\bibfield  {title} {\bibinfo {title} {{Bose-Einstein condensation and Silver
  Blaze property from the two-loop $\Phi$-derivable approximation}},\ }\href
  {https://doi.org/10.1103/PhysRevD.90.125021} {\bibfield  {journal} {\bibinfo
  {journal} {Phys. Rev. D}\ }\textbf {\bibinfo {volume} {90}},\ \bibinfo
  {pages} {125021} (\bibinfo {year} {2014})},\ \Eprint
  {https://arxiv.org/abs/1410.6998} {arXiv:1410.6998 [hep-ph]} \BibitemShut
  {NoStop}%
\bibitem [{\citenamefont {Khan}\ \emph {et~al.}(2015)\citenamefont {Khan},
  \citenamefont {Pawlowski}, \citenamefont {Rennecke},\ and\ \citenamefont
  {Scherer}}]{Khan:2015puu}%
  \BibitemOpen
  \bibfield  {author} {\bibinfo {author} {\bibfnamefont {N.}~\bibnamefont
  {Khan}}, \bibinfo {author} {\bibfnamefont {J.~M.}\ \bibnamefont {Pawlowski}},
  \bibinfo {author} {\bibfnamefont {F.}~\bibnamefont {Rennecke}},\ and\
  \bibinfo {author} {\bibfnamefont {M.~M.}\ \bibnamefont {Scherer}},\
  }\bibfield  {title} {\bibinfo {title} {{The Phase Diagram of QC2D from
  Functional Methods}},\ }\Eprint {https://arxiv.org/abs/1512.03673}
  {arXiv:1512.03673 [hep-ph]}  (\bibinfo {year} {2015})\BibitemShut {NoStop}%
\bibitem [{\citenamefont {Braun}\ \emph {et~al.}(2021)\citenamefont {Braun},
  \citenamefont {D\"ornfeld}, \citenamefont {Schallmo},\ and\ \citenamefont
  {T\"opfel}}]{Braun:2020bhy}%
  \BibitemOpen
  \bibfield  {author} {\bibinfo {author} {\bibfnamefont {J.}~\bibnamefont
  {Braun}}, \bibinfo {author} {\bibfnamefont {T.}~\bibnamefont {D\"ornfeld}},
  \bibinfo {author} {\bibfnamefont {B.}~\bibnamefont {Schallmo}},\ and\
  \bibinfo {author} {\bibfnamefont {S.}~\bibnamefont {T\"opfel}},\ }\bibfield
  {title} {\bibinfo {title} {{Renormalization group studies of dense
  relativistic systems}},\ }\href {https://doi.org/10.1103/PhysRevD.104.096002}
  {\bibfield  {journal} {\bibinfo  {journal} {Phys. Rev. D}\ }\textbf {\bibinfo
  {volume} {104}},\ \bibinfo {pages} {096002} (\bibinfo {year} {2021})},\
  \Eprint {https://arxiv.org/abs/2008.05978} {arXiv:2008.05978 [hep-ph]}
  \BibitemShut {NoStop}%
\bibitem [{\citenamefont {Cohen}(2003)}]{Cohen:2003kd}%
  \BibitemOpen
  \bibfield  {author} {\bibinfo {author} {\bibfnamefont {T.~D.}\ \bibnamefont
  {Cohen}},\ }\bibfield  {title} {\bibinfo {title} {{Functional integrals for
  QCD at nonzero chemical potential and zero density}},\ }\href
  {https://doi.org/10.1103/PhysRevLett.91.222001} {\bibfield  {journal}
  {\bibinfo  {journal} {Phys. Rev. Lett.}\ }\textbf {\bibinfo {volume} {91}},\
  \bibinfo {pages} {222001} (\bibinfo {year} {2003})},\ \Eprint
  {https://arxiv.org/abs/hep-ph/0307089} {arXiv:hep-ph/0307089} \BibitemShut
  {NoStop}%
\bibitem [{\citenamefont {Litim}\ and\ \citenamefont
  {Pawlowski}(2006)}]{Litim:2006ag}%
  \BibitemOpen
  \bibfield  {author} {\bibinfo {author} {\bibfnamefont {D.~F.}\ \bibnamefont
  {Litim}}\ and\ \bibinfo {author} {\bibfnamefont {J.~M.}\ \bibnamefont
  {Pawlowski}},\ }\bibfield  {title} {\bibinfo {title} {{Non-perturbative
  thermal flows and resummations}},\ }\href
  {https://doi.org/10.1088/1126-6708/2006/11/026} {\bibfield  {journal}
  {\bibinfo  {journal} {JHEP}\ }\textbf {\bibinfo {volume} {11}},\ \bibinfo
  {pages} {026}},\ \Eprint {https://arxiv.org/abs/hep-th/0609122}
  {arXiv:hep-th/0609122} \BibitemShut {NoStop}%
\bibitem [{\citenamefont {Blaizot}\ \emph {et~al.}(2007)\citenamefont
  {Blaizot}, \citenamefont {Ipp}, \citenamefont {Mendez-Galain},\ and\
  \citenamefont {Wschebor}}]{Blaizot:2006rj}%
  \BibitemOpen
  \bibfield  {author} {\bibinfo {author} {\bibfnamefont {J.-P.}\ \bibnamefont
  {Blaizot}}, \bibinfo {author} {\bibfnamefont {A.}~\bibnamefont {Ipp}},
  \bibinfo {author} {\bibfnamefont {R.}~\bibnamefont {Mendez-Galain}},\ and\
  \bibinfo {author} {\bibfnamefont {N.}~\bibnamefont {Wschebor}},\ }\bibfield
  {title} {\bibinfo {title} {{Perturbation theory and non-perturbative
  renormalization flow in scalar field theory at finite temperature}},\ }\href
  {https://doi.org/10.1016/j.nuclphysa.2006.11.139} {\bibfield  {journal}
  {\bibinfo  {journal} {Nucl. Phys. A}\ }\textbf {\bibinfo {volume} {784}},\
  \bibinfo {pages} {376} (\bibinfo {year} {2007})},\ \Eprint
  {https://arxiv.org/abs/hep-ph/0610004} {arXiv:hep-ph/0610004} \BibitemShut
  {NoStop}%
\bibitem [{\citenamefont {Litim}(2000)}]{Litim:2000ci}%
  \BibitemOpen
  \bibfield  {author} {\bibinfo {author} {\bibfnamefont {D.~F.}\ \bibnamefont
  {Litim}},\ }\bibfield  {title} {\bibinfo {title} {{Optimization of the exact
  renormalization group}},\ }\href
  {https://doi.org/10.1016/S0370-2693(00)00748-6} {\bibfield  {journal}
  {\bibinfo  {journal} {Phys. Lett. B}\ }\textbf {\bibinfo {volume} {486}},\
  \bibinfo {pages} {92} (\bibinfo {year} {2000})},\ \Eprint
  {https://arxiv.org/abs/hep-th/0005245} {arXiv:hep-th/0005245} \BibitemShut
  {NoStop}%
\bibitem [{\citenamefont {Litim}(2001)}]{Litim:2001up}%
  \BibitemOpen
  \bibfield  {author} {\bibinfo {author} {\bibfnamefont {D.~F.}\ \bibnamefont
  {Litim}},\ }\bibfield  {title} {\bibinfo {title} {{Optimized renormalization
  group flows}},\ }\href {https://doi.org/10.1103/PhysRevD.64.105007}
  {\bibfield  {journal} {\bibinfo  {journal} {Phys. Rev. D}\ }\textbf {\bibinfo
  {volume} {64}},\ \bibinfo {pages} {105007} (\bibinfo {year} {2001})},\
  \Eprint {https://arxiv.org/abs/hep-th/0103195} {arXiv:hep-th/0103195}
  \BibitemShut {NoStop}%
\bibitem [{\citenamefont {Fu}\ \emph {et~al.}(2016)\citenamefont {Fu},
  \citenamefont {Pawlowski}, \citenamefont {Rennecke},\ and\ \citenamefont
  {Schaefer}}]{Fu:2016tey}%
  \BibitemOpen
  \bibfield  {author} {\bibinfo {author} {\bibfnamefont {W.-j.}\ \bibnamefont
  {Fu}}, \bibinfo {author} {\bibfnamefont {J.~M.}\ \bibnamefont {Pawlowski}},
  \bibinfo {author} {\bibfnamefont {F.}~\bibnamefont {Rennecke}},\ and\
  \bibinfo {author} {\bibfnamefont {B.-J.}\ \bibnamefont {Schaefer}},\
  }\bibfield  {title} {\bibinfo {title} {{Baryon number fluctuations at finite
  temperature and density}},\ }\href
  {https://doi.org/10.1103/PhysRevD.94.116020} {\bibfield  {journal} {\bibinfo
  {journal} {Phys. Rev. D}\ }\textbf {\bibinfo {volume} {94}},\ \bibinfo
  {pages} {116020} (\bibinfo {year} {2016})},\ \Eprint
  {https://arxiv.org/abs/1608.04302} {arXiv:1608.04302 [hep-ph]} \BibitemShut
  {NoStop}%
\bibitem [{\citenamefont {Braun}\ \emph {et~al.}(2017)\citenamefont {Braun},
  \citenamefont {Leonhardt},\ and\ \citenamefont {Pospiech}}]{Braun:2017srn}%
  \BibitemOpen
  \bibfield  {author} {\bibinfo {author} {\bibfnamefont {J.}~\bibnamefont
  {Braun}}, \bibinfo {author} {\bibfnamefont {M.}~\bibnamefont {Leonhardt}},\
  and\ \bibinfo {author} {\bibfnamefont {M.}~\bibnamefont {Pospiech}},\
  }\bibfield  {title} {\bibinfo {title} {{Fierz-complete NJL model study: Fixed
  points and phase structure at finite temperature and density}},\ }\href
  {https://doi.org/10.1103/PhysRevD.96.076003} {\bibfield  {journal} {\bibinfo
  {journal} {Phys. Rev. D}\ }\textbf {\bibinfo {volume} {96}},\ \bibinfo
  {pages} {076003} (\bibinfo {year} {2017})},\ \Eprint
  {https://arxiv.org/abs/1705.00074} {arXiv:1705.00074 [hep-ph]} \BibitemShut
  {NoStop}%
\bibitem [{\citenamefont {Braun}\ \emph {et~al.}(2023)\citenamefont {Braun}
  \emph {et~al.}}]{Braun:2022mgx}%
  \BibitemOpen
  \bibfield  {author} {\bibinfo {author} {\bibfnamefont {J.}~\bibnamefont
  {Braun}} \emph {et~al.},\ }\bibfield  {title} {\bibinfo {title}
  {{Renormalised spectral flows}},\ }\href
  {https://doi.org/10.21468/SciPostPhysCore.6.3.061} {\bibfield  {journal}
  {\bibinfo  {journal} {SciPost Phys. Core}\ }\textbf {\bibinfo {volume} {6}},\
  \bibinfo {pages} {061} (\bibinfo {year} {2023})},\ \Eprint
  {https://arxiv.org/abs/2206.10232} {arXiv:2206.10232 [hep-th]} \BibitemShut
  {NoStop}%
\bibitem [{\citenamefont {Rajagopal}\ and\ \citenamefont
  {Wilczek}(2000)}]{Rajagopal:2000wf}%
  \BibitemOpen
  \bibfield  {author} {\bibinfo {author} {\bibfnamefont {K.}~\bibnamefont
  {Rajagopal}}\ and\ \bibinfo {author} {\bibfnamefont {F.}~\bibnamefont
  {Wilczek}},\ }\bibfield  {title} {\bibinfo {title} {The condensed matter
  physics of qcd},\ }\Eprint {https://arxiv.org/abs/hep-ph/0011333}
  {arXiv:hep-ph/0011333}  (\bibinfo {year} {2000}),\ \bibinfo {note} {in *At
  the Frontier of Particle Physics: Handbook of QCD*, Vol. 3, edited by
  M.~Shifman and B.~Ioffe, pp.~2061--2151}\BibitemShut {NoStop}%
\bibitem [{\citenamefont {T{\"o}pfel}\ \emph {et~al.}(2025)\citenamefont
  {T{\"o}pfel}, \citenamefont {Gei{\ss}el},\ and\ \citenamefont
  {Braun}}]{Topfel:2024cll}%
  \BibitemOpen
  \bibfield  {author} {\bibinfo {author} {\bibfnamefont {S.}~\bibnamefont
  {T{\"o}pfel}}, \bibinfo {author} {\bibfnamefont {A.}~\bibnamefont
  {Gei{\ss}el}},\ and\ \bibinfo {author} {\bibfnamefont {J.}~\bibnamefont
  {Braun}},\ }\bibfield  {title} {\bibinfo {title} {{Subtleties in the
  calculation of correlation functions for hot and dense systems}},\ }\href
  {https://doi.org/10.1103/PhysRevD.111.016023} {\bibfield  {journal} {\bibinfo
   {journal} {Phys. Rev. D}\ }\textbf {\bibinfo {volume} {111}},\ \bibinfo
  {pages} {016023} (\bibinfo {year} {2025})},\ \Eprint
  {https://arxiv.org/abs/2410.06674} {arXiv:2410.06674 [nucl-th]} \BibitemShut
  {NoStop}%
\bibitem [{\citenamefont {Nielsen}\ and\ \citenamefont
  {Chadha}(1976)}]{Nielsen:1975hm}%
  \BibitemOpen
  \bibfield  {author} {\bibinfo {author} {\bibfnamefont {H.~B.}\ \bibnamefont
  {Nielsen}}\ and\ \bibinfo {author} {\bibfnamefont {S.}~\bibnamefont
  {Chadha}},\ }\bibfield  {title} {\bibinfo {title} {{On How to Count Goldstone
  Bosons}},\ }\href {https://doi.org/10.1016/0550-3213(76)90025-0} {\bibfield
  {journal} {\bibinfo  {journal} {Nucl. Phys. B}\ }\textbf {\bibinfo {volume}
  {105}},\ \bibinfo {pages} {445} (\bibinfo {year} {1976})}\BibitemShut
  {NoStop}%
\bibitem [{\citenamefont {Andersen}(2005)}]{Andersen:2005yk}%
  \BibitemOpen
  \bibfield  {author} {\bibinfo {author} {\bibfnamefont {J.~O.}\ \bibnamefont
  {Andersen}},\ }\bibfield  {title} {\bibinfo {title} {{Relativistic Bose gases
  at finite density}},\ }\Eprint {https://arxiv.org/abs/hep-ph/0501094}
  {arXiv:hep-ph/0501094}  (\bibinfo {year} {2005})\BibitemShut {NoStop}%
\bibitem [{\citenamefont {Watanabe}(2020)}]{Watanabe:2019xul}%
  \BibitemOpen
  \bibfield  {author} {\bibinfo {author} {\bibfnamefont {H.}~\bibnamefont
  {Watanabe}},\ }\bibfield  {title} {\bibinfo {title} {{Counting Rules of
  Nambu{\textendash}Goldstone Modes}},\ }\href
  {https://doi.org/10.1146/annurev-conmatphys-031119-050644} {\bibfield
  {journal} {\bibinfo  {journal} {Ann. Rev. Condensed Matter Phys.}\ }\textbf
  {\bibinfo {volume} {11}},\ \bibinfo {pages} {169} (\bibinfo {year} {2020})},\
  \Eprint {https://arxiv.org/abs/1904.00569} {arXiv:1904.00569
  [cond-mat.other]} \BibitemShut {NoStop}%
\bibitem [{\citenamefont {Litim}\ \emph {et~al.}(2006)\citenamefont {Litim},
  \citenamefont {Pawlowski},\ and\ \citenamefont {Vergara}}]{Litim:2006nn}%
  \BibitemOpen
  \bibfield  {author} {\bibinfo {author} {\bibfnamefont {D.~F.}\ \bibnamefont
  {Litim}}, \bibinfo {author} {\bibfnamefont {J.~M.}\ \bibnamefont
  {Pawlowski}},\ and\ \bibinfo {author} {\bibfnamefont {L.}~\bibnamefont
  {Vergara}},\ }\bibfield  {title} {\bibinfo {title} {{Convexity of the
  effective action from functional flows}},\ }\Eprint
  {https://arxiv.org/abs/hep-th/0602140} {arXiv:hep-th/0602140}  (\bibinfo
  {year} {2006})\BibitemShut {NoStop}%
\bibitem [{\citenamefont {Zorbach}\ \emph {et~al.}(2025)\citenamefont
  {Zorbach}, \citenamefont {Stoll},\ and\ \citenamefont
  {Braun}}]{Zorbach:2024zjx}%
  \BibitemOpen
  \bibfield  {author} {\bibinfo {author} {\bibfnamefont {N.}~\bibnamefont
  {Zorbach}}, \bibinfo {author} {\bibfnamefont {J.}~\bibnamefont {Stoll}},\
  and\ \bibinfo {author} {\bibfnamefont {J.}~\bibnamefont {Braun}},\ }\bibfield
   {title} {\bibinfo {title} {{Optimization and stabilization of functional
  renormalization group flows}},\ }\href
  {https://doi.org/10.1103/PhysRevD.111.096022} {\bibfield  {journal} {\bibinfo
   {journal} {Phys. Rev. D}\ }\textbf {\bibinfo {volume} {111}},\ \bibinfo
  {pages} {096022} (\bibinfo {year} {2025})},\ \Eprint
  {https://arxiv.org/abs/2401.12854} {arXiv:2401.12854 [hep-ph]} \BibitemShut
  {NoStop}%
\bibitem [{\citenamefont {Morones-Ibarra}\ \emph {et~al.}(2017)\citenamefont
  {Morones-Ibarra}, \citenamefont {Enriquez-Perez-Gavilan}, \citenamefont
  {Rodriguez}, \citenamefont {Flores-Baez}, \citenamefont {Mata-Carrizalez},\
  and\ \citenamefont {Ordo\~nez}}]{Morones-Ibarra:2017avu}%
  \BibitemOpen
  \bibfield  {author} {\bibinfo {author} {\bibfnamefont {J.~R.}\ \bibnamefont
  {Morones-Ibarra}}, \bibinfo {author} {\bibfnamefont {A.}~\bibnamefont
  {Enriquez-Perez-Gavilan}}, \bibinfo {author} {\bibfnamefont {A.~I.~H.}\
  \bibnamefont {Rodriguez}}, \bibinfo {author} {\bibfnamefont {F.~V.}\
  \bibnamefont {Flores-Baez}}, \bibinfo {author} {\bibfnamefont {N.~B.}\
  \bibnamefont {Mata-Carrizalez}},\ and\ \bibinfo {author} {\bibfnamefont
  {E.~V.}\ \bibnamefont {Ordo\~nez}},\ }\bibfield  {title} {\bibinfo {title}
  {{Chiral symmetry restoration and the critical end point in QCD}},\ }\href
  {https://doi.org/10.1515/phys-2017-0130} {\bibfield  {journal} {\bibinfo
  {journal} {Open Phys.}\ }\textbf {\bibinfo {volume} {15}},\ \bibinfo {pages}
  {1039} (\bibinfo {year} {2017})}\BibitemShut {NoStop}%
\bibitem [{\citenamefont {Terazaki}\ \emph {et~al.}(2024)\citenamefont
  {Terazaki}, \citenamefont {Mameda},\ and\ \citenamefont
  {Suzuki}}]{Terazaki:2024evv}%
  \BibitemOpen
  \bibfield  {author} {\bibinfo {author} {\bibfnamefont {F.}~\bibnamefont
  {Terazaki}}, \bibinfo {author} {\bibfnamefont {K.}~\bibnamefont {Mameda}},\
  and\ \bibinfo {author} {\bibfnamefont {K.}~\bibnamefont {Suzuki}},\
  }\bibfield  {title} {\bibinfo {title} {{Relativistic BEC extracted from a
  complex FRG flow equation}},\ }\href {https://doi.org/10.1093/ptep/ptae166}
  {\bibfield  {journal} {\bibinfo  {journal} {PTEP}\ }\textbf {\bibinfo
  {volume} {2024}},\ \bibinfo {pages} {123B02} (\bibinfo {year} {2024})},\
  \Eprint {https://arxiv.org/abs/2409.04361} {arXiv:2409.04361 [hep-ph]}
  \BibitemShut {NoStop}%
\bibitem [{\citenamefont {Gattringer}\ and\ \citenamefont
  {Kloiber}(2013)}]{Gattringer:2012df}%
  \BibitemOpen
  \bibfield  {author} {\bibinfo {author} {\bibfnamefont {C.}~\bibnamefont
  {Gattringer}}\ and\ \bibinfo {author} {\bibfnamefont {T.}~\bibnamefont
  {Kloiber}},\ }\bibfield  {title} {\bibinfo {title} {{Lattice study of the
  Silver Blaze phenomenon for a charged scalar $\phi^4$ field}},\ }\href
  {https://doi.org/10.1016/j.nuclphysb.2012.12.005} {\bibfield  {journal}
  {\bibinfo  {journal} {Nucl. Phys. B}\ }\textbf {\bibinfo {volume} {869}},\
  \bibinfo {pages} {56} (\bibinfo {year} {2013})},\ \Eprint
  {https://arxiv.org/abs/1206.2954} {arXiv:1206.2954 [hep-lat]} \BibitemShut
  {NoStop}%
\bibitem [{\citenamefont {Svanes}\ and\ \citenamefont
  {Andersen}(2011)}]{Svanes:2010we}%
  \BibitemOpen
  \bibfield  {author} {\bibinfo {author} {\bibfnamefont {E.~E.}\ \bibnamefont
  {Svanes}}\ and\ \bibinfo {author} {\bibfnamefont {J.~O.}\ \bibnamefont
  {Andersen}},\ }\bibfield  {title} {\bibinfo {title} {{Functional
  renormalization group at finite density and Bose condensation}},\ }\href
  {https://doi.org/10.1016/j.nuclphysa.2011.03.007} {\bibfield  {journal}
  {\bibinfo  {journal} {Nucl. Phys. A}\ }\textbf {\bibinfo {volume} {857}},\
  \bibinfo {pages} {16} (\bibinfo {year} {2011})},\ \Eprint
  {https://arxiv.org/abs/1009.0430} {arXiv:1009.0430 [hep-ph]} \BibitemShut
  {NoStop}%
\bibitem [{\citenamefont {Palhares}(2012)}]{Palhares:2012fv}%
  \BibitemOpen
  \bibfield  {author} {\bibinfo {author} {\bibfnamefont {L.~F.}\ \bibnamefont
  {Palhares}},\ }\bibfield  {title} {\bibinfo {title} {Exploring the different
  phase diagrams of strong interactions},\ }\Eprint
  {https://arxiv.org/abs/1208.0574} {arXiv:1208.0574 [hep-ph]}  (\bibinfo
  {year} {2012}),\ \bibinfo {note} {phD thesis, Rio de Janeiro Federal
  University}\BibitemShut {NoStop}%
\bibitem [{\citenamefont {Kapusta}(1981)}]{Kapusta:1981aa}%
  \BibitemOpen
  \bibfield  {author} {\bibinfo {author} {\bibfnamefont {J.~I.}\ \bibnamefont
  {Kapusta}},\ }\bibfield  {title} {\bibinfo {title} {{Bose-Einstein
  Condensation, Spontaneous Symmetry Breaking, and Gauge Theories}},\ }\href
  {https://doi.org/10.1103/PhysRevD.24.426} {\bibfield  {journal} {\bibinfo
  {journal} {Phys. Rev. D}\ }\textbf {\bibinfo {volume} {24}},\ \bibinfo
  {pages} {426} (\bibinfo {year} {1981})}\BibitemShut {NoStop}%
\bibitem [{\citenamefont {Strodthoff}\ \emph {et~al.}(2012)\citenamefont
  {Strodthoff}, \citenamefont {Schaefer},\ and\ \citenamefont {von
  Smekal}}]{Strodthoff:2011tz}%
  \BibitemOpen
  \bibfield  {author} {\bibinfo {author} {\bibfnamefont {N.}~\bibnamefont
  {Strodthoff}}, \bibinfo {author} {\bibfnamefont {B.-J.}\ \bibnamefont
  {Schaefer}},\ and\ \bibinfo {author} {\bibfnamefont {L.}~\bibnamefont {von
  Smekal}},\ }\bibfield  {title} {\bibinfo {title} {{Quark-meson-diquark model
  for two-color QCD}},\ }\href {https://doi.org/10.1103/PhysRevD.85.074007}
  {\bibfield  {journal} {\bibinfo  {journal} {Phys. Rev. D}\ }\textbf {\bibinfo
  {volume} {85}},\ \bibinfo {pages} {074007} (\bibinfo {year} {2012})},\
  \Eprint {https://arxiv.org/abs/1112.5401} {arXiv:1112.5401 [hep-ph]}
  \BibitemShut {NoStop}%
\bibitem [{\citenamefont {Cichutek}\ \emph {et~al.}(2020)\citenamefont
  {Cichutek}, \citenamefont {Divotgey},\ and\ \citenamefont
  {Eser}}]{Cichutek:2020bli}%
  \BibitemOpen
  \bibfield  {author} {\bibinfo {author} {\bibfnamefont {N.}~\bibnamefont
  {Cichutek}}, \bibinfo {author} {\bibfnamefont {F.}~\bibnamefont {Divotgey}},\
  and\ \bibinfo {author} {\bibfnamefont {J.}~\bibnamefont {Eser}},\ }\bibfield
  {title} {\bibinfo {title} {{Fluctuation-induced higher-derivative couplings
  and infrared dynamics of the Quark-Meson-Diquark Model}},\ }\href
  {https://doi.org/10.1103/PhysRevD.102.034030} {\bibfield  {journal} {\bibinfo
   {journal} {Phys. Rev. D}\ }\textbf {\bibinfo {volume} {102}},\ \bibinfo
  {pages} {034030} (\bibinfo {year} {2020})},\ \Eprint
  {https://arxiv.org/abs/2006.12473} {arXiv:2006.12473 [hep-ph]} \BibitemShut
  {NoStop}%
\bibitem [{\citenamefont {Khan}(2015)}]{Khan:2015etz}%
  \BibitemOpen
  \bibfield  {author} {\bibinfo {author} {\bibfnamefont {N.}~\bibnamefont
  {Khan}},\ }\bibfield  {title} {\bibinfo {title} {Interplay of mesonic and
  baryonic degrees of freedom in quark matter}} (\bibinfo {year} {2015}),\
  \bibinfo {note} {phD thesis, Heidelberg University}\BibitemShut {NoStop}%
\bibitem [{\citenamefont {Lakaschus}\ \emph {et~al.}(2021)\citenamefont
  {Lakaschus}, \citenamefont {Buballa},\ and\ \citenamefont
  {Rischke}}]{Lakaschus:2020caq}%
  \BibitemOpen
  \bibfield  {author} {\bibinfo {author} {\bibfnamefont {P.}~\bibnamefont
  {Lakaschus}}, \bibinfo {author} {\bibfnamefont {M.}~\bibnamefont {Buballa}},\
  and\ \bibinfo {author} {\bibfnamefont {D.~H.}\ \bibnamefont {Rischke}},\
  }\bibfield  {title} {\bibinfo {title} {{Competition of inhomogeneous chiral
  phases and two-flavor color superconductivity in the NJL model}},\ }\href
  {https://doi.org/10.1103/PhysRevD.103.034030} {\bibfield  {journal} {\bibinfo
   {journal} {Phys. Rev. D}\ }\textbf {\bibinfo {volume} {103}},\ \bibinfo
  {pages} {034030} (\bibinfo {year} {2021})},\ \Eprint
  {https://arxiv.org/abs/2012.07520} {arXiv:2012.07520 [hep-ph]} \BibitemShut
  {NoStop}%
\bibitem [{\citenamefont {Lakaschus}(2021)}]{Lakaschus:2021ewd}%
  \BibitemOpen
  \bibfield  {author} {\bibinfo {author} {\bibfnamefont {P.}~\bibnamefont
  {Lakaschus}},\ }\bibfield  {title} {\bibinfo {title} {Inhomogeneous chiral
  condensates in low-energy color-superconductivity models of qcd}} (\bibinfo
  {year} {2021}),\ \bibinfo {note} {phD thesis, Goethe University, Frankfurt
  (Main)}\BibitemShut {NoStop}%
\bibitem [{\citenamefont {Yuan}\ \emph {et~al.}(2023)\citenamefont {Yuan},
  \citenamefont {Chao},\ and\ \citenamefont {Li}}]{Yuan:2023dco}%
  \BibitemOpen
  \bibfield  {author} {\bibinfo {author} {\bibfnamefont {W.-L.}\ \bibnamefont
  {Yuan}}, \bibinfo {author} {\bibfnamefont {J.}~\bibnamefont {Chao}},\ and\
  \bibinfo {author} {\bibfnamefont {A.}~\bibnamefont {Li}},\ }\bibfield
  {title} {\bibinfo {title} {{Diquark and chiral condensates in a
  self-consistent NJL-type model}},\ }\href
  {https://doi.org/10.1103/PhysRevD.108.043008} {\bibfield  {journal} {\bibinfo
   {journal} {Phys. Rev. D}\ }\textbf {\bibinfo {volume} {108}},\ \bibinfo
  {pages} {043008} (\bibinfo {year} {2023})},\ \Eprint
  {https://arxiv.org/abs/2304.12050} {arXiv:2304.12050 [hep-ph]} \BibitemShut
  {NoStop}%
\bibitem [{fQC()}]{fQCD}%
  \BibitemOpen
  \href@noop {} {}\bibinfo {note} {{\it fQCD Collaboration:} Braun, Jens and
  Chen, Yong-rui and Fu, Wei-jie and Gao, Fei and Ihssen, Friederike and
  Geißel, Andreas and Huang, Chuang and Pawlowski, Jan M. and Rennecke, Fabian
  and Sattler, Franz R. and Schallmo, Benedikt and Stoll, Jonas and Tan,
  Yang-yang and T{\"o}pfel, Sebastian and Turnwald, Jonas and Wen, Rui and
  Wessely, Jonas and Wink, Nicolas and Yin, Shi and Zorbach, Niklas (September
  2025)}\BibitemShut {NoStop}%
\bibitem [{\citenamefont {Stoll}\ \emph {et~al.}(2021)\citenamefont {Stoll},
  \citenamefont {Zorbach}, \citenamefont {Koenigstein}, \citenamefont {Steil},\
  and\ \citenamefont {Rechenberger}}]{Stoll:2021ori}%
  \BibitemOpen
  \bibfield  {author} {\bibinfo {author} {\bibfnamefont {J.}~\bibnamefont
  {Stoll}}, \bibinfo {author} {\bibfnamefont {N.}~\bibnamefont {Zorbach}},
  \bibinfo {author} {\bibfnamefont {A.}~\bibnamefont {Koenigstein}}, \bibinfo
  {author} {\bibfnamefont {M.~J.}\ \bibnamefont {Steil}},\ and\ \bibinfo
  {author} {\bibfnamefont {S.}~\bibnamefont {Rechenberger}},\ }\bibfield
  {title} {\bibinfo {title} {{Bosonic fluctuations in the $( 1 + 1
  )$-dimensional Gross-Neveu(-Yukawa) model at varying $\mu$ and $T$ and finite
  $N$}},\ }\Eprint {https://arxiv.org/abs/2108.10616} {arXiv:2108.10616
  [hep-ph]}  (\bibinfo {year} {2021})\BibitemShut {NoStop}%
\bibitem [{\citenamefont {Kurganov}\ and\ \citenamefont
  {Tadmor}(2000)}]{Kurganov:2000ovy}%
  \BibitemOpen
  \bibfield  {author} {\bibinfo {author} {\bibfnamefont {A.}~\bibnamefont
  {Kurganov}}\ and\ \bibinfo {author} {\bibfnamefont {E.}~\bibnamefont
  {Tadmor}},\ }\bibfield  {title} {\bibinfo {title} {{New High-Resolution
  Central Schemes for Nonlinear Conservation Laws and
  Convection\textendash{}Diffusion Equations}},\ }\href
  {https://doi.org/10.1006/jcph.2000.6459} {\bibfield  {journal} {\bibinfo
  {journal} {J. Comput. Phys.}\ }\textbf {\bibinfo {volume} {160}},\ \bibinfo
  {pages} {241} (\bibinfo {year} {2000})}\BibitemShut {NoStop}%
\bibitem [{\citenamefont {Virtanen}\ \emph {et~al.}(2020)\citenamefont
  {Virtanen}, \citenamefont {Gommers}, \citenamefont {Oliphant}, \citenamefont
  {Haberland}, \citenamefont {Reddy}, \citenamefont {Cournapeau}, \citenamefont
  {Burovski}, \citenamefont {Peterson}, \citenamefont {Weckesser},
  \citenamefont {Bright}, \citenamefont {{van der Walt}}, \citenamefont
  {Brett}, \citenamefont {Wilson}, \citenamefont {Millman}, \citenamefont
  {Mayorov}, \citenamefont {Nelson}, \citenamefont {Jones}, \citenamefont
  {Kern}, \citenamefont {Larson}, \citenamefont {Carey}, \citenamefont {Polat},
  \citenamefont {Feng}, \citenamefont {Moore}, \citenamefont {{VanderPlas}},
  \citenamefont {Laxalde}, \citenamefont {Perktold}, \citenamefont {Cimrman},
  \citenamefont {Henriksen}, \citenamefont {Quintero}, \citenamefont {Harris},
  \citenamefont {Archibald}, \citenamefont {Ribeiro}, \citenamefont
  {Pedregosa}, \citenamefont {{van Mulbregt}},\ and\ \citenamefont {{SciPy 1.0
  Contributors}}}]{2020SciPy-NMeth}%
  \BibitemOpen
  \bibfield  {author} {\bibinfo {author} {\bibfnamefont {P.}~\bibnamefont
  {Virtanen}}, \bibinfo {author} {\bibfnamefont {R.}~\bibnamefont {Gommers}},
  \bibinfo {author} {\bibfnamefont {T.~E.}\ \bibnamefont {Oliphant}}, \bibinfo
  {author} {\bibfnamefont {M.}~\bibnamefont {Haberland}}, \bibinfo {author}
  {\bibfnamefont {T.}~\bibnamefont {Reddy}}, \bibinfo {author} {\bibfnamefont
  {D.}~\bibnamefont {Cournapeau}}, \bibinfo {author} {\bibfnamefont
  {E.}~\bibnamefont {Burovski}}, \bibinfo {author} {\bibfnamefont
  {P.}~\bibnamefont {Peterson}}, \bibinfo {author} {\bibfnamefont
  {W.}~\bibnamefont {Weckesser}}, \bibinfo {author} {\bibfnamefont
  {J.}~\bibnamefont {Bright}}, \bibinfo {author} {\bibfnamefont {S.~J.}\
  \bibnamefont {{van der Walt}}}, \bibinfo {author} {\bibfnamefont
  {M.}~\bibnamefont {Brett}}, \bibinfo {author} {\bibfnamefont
  {J.}~\bibnamefont {Wilson}}, \bibinfo {author} {\bibfnamefont {K.~J.}\
  \bibnamefont {Millman}}, \bibinfo {author} {\bibfnamefont {N.}~\bibnamefont
  {Mayorov}}, \bibinfo {author} {\bibfnamefont {A.~R.~J.}\ \bibnamefont
  {Nelson}}, \bibinfo {author} {\bibfnamefont {E.}~\bibnamefont {Jones}},
  \bibinfo {author} {\bibfnamefont {R.}~\bibnamefont {Kern}}, \bibinfo {author}
  {\bibfnamefont {E.}~\bibnamefont {Larson}}, \bibinfo {author} {\bibfnamefont
  {C.~J.}\ \bibnamefont {Carey}}, \bibinfo {author} {\bibfnamefont
  {{\.I}.}~\bibnamefont {Polat}}, \bibinfo {author} {\bibfnamefont
  {Y.}~\bibnamefont {Feng}}, \bibinfo {author} {\bibfnamefont {E.~W.}\
  \bibnamefont {Moore}}, \bibinfo {author} {\bibfnamefont {J.}~\bibnamefont
  {{VanderPlas}}}, \bibinfo {author} {\bibfnamefont {D.}~\bibnamefont
  {Laxalde}}, \bibinfo {author} {\bibfnamefont {J.}~\bibnamefont {Perktold}},
  \bibinfo {author} {\bibfnamefont {R.}~\bibnamefont {Cimrman}}, \bibinfo
  {author} {\bibfnamefont {I.}~\bibnamefont {Henriksen}}, \bibinfo {author}
  {\bibfnamefont {E.~A.}\ \bibnamefont {Quintero}}, \bibinfo {author}
  {\bibfnamefont {C.~R.}\ \bibnamefont {Harris}}, \bibinfo {author}
  {\bibfnamefont {A.~M.}\ \bibnamefont {Archibald}}, \bibinfo {author}
  {\bibfnamefont {A.~H.}\ \bibnamefont {Ribeiro}}, \bibinfo {author}
  {\bibfnamefont {F.}~\bibnamefont {Pedregosa}}, \bibinfo {author}
  {\bibfnamefont {P.}~\bibnamefont {{van Mulbregt}}},\ and\ \bibinfo {author}
  {\bibnamefont {{SciPy 1.0 Contributors}}},\ }\bibfield  {title} {\bibinfo
  {title} {{{SciPy} 1.0: Fundamental Algorithms for Scientific Computing in
  Python}},\ }\href {https://doi.org/10.1038/s41592-019-0686-2} {\bibfield
  {journal} {\bibinfo  {journal} {Nature Methods}\ }\textbf {\bibinfo {volume}
  {17}},\ \bibinfo {pages} {261} (\bibinfo {year} {2020})}\BibitemShut
  {NoStop}%
\bibitem [{\citenamefont {Van~Rossum}\ and\ \citenamefont
  {Drake}(2009)}]{10.5555/1593511}%
  \BibitemOpen
  \bibfield  {author} {\bibinfo {author} {\bibfnamefont {G.}~\bibnamefont
  {Van~Rossum}}\ and\ \bibinfo {author} {\bibfnamefont {F.~L.}\ \bibnamefont
  {Drake}},\ }\href@noop {} {\emph {\bibinfo {title} {Python 3 Reference
  Manual}}}\ (\bibinfo  {publisher} {CreateSpace},\ \bibinfo {address} {Scotts
  Valley, CA},\ \bibinfo {year} {2009})\BibitemShut {NoStop}%
\bibitem [{\citenamefont {Hunter}(2007)}]{Hunter:2007}%
  \BibitemOpen
  \bibfield  {author} {\bibinfo {author} {\bibfnamefont {J.~D.}\ \bibnamefont
  {Hunter}},\ }\bibfield  {title} {\bibinfo {title} {Matplotlib: A 2d graphics
  environment},\ }\href {https://doi.org/10.1109/MCSE.2007.55} {\bibfield
  {journal} {\bibinfo  {journal} {Computing in Science \& Engineering}\
  }\textbf {\bibinfo {volume} {9}},\ \bibinfo {pages} {90} (\bibinfo {year}
  {2007})}\BibitemShut {NoStop}%
\bibitem [{\citenamefont {Harris}\ \emph {et~al.}(2020)\citenamefont {Harris},
  \citenamefont {Millman}, \citenamefont {van~der Walt}, \citenamefont
  {Gommers}, \citenamefont {Virtanen}, \citenamefont {Cournapeau},
  \citenamefont {Wieser}, \citenamefont {Taylor}, \citenamefont {Berg},
  \citenamefont {Smith}, \citenamefont {Kern}, \citenamefont {Picus},
  \citenamefont {Hoyer}, \citenamefont {van Kerkwijk}, \citenamefont {Brett},
  \citenamefont {Haldane}, \citenamefont {del R{\'{i}}o}, \citenamefont
  {Wiebe}, \citenamefont {Peterson}, \citenamefont {G{\'{e}}rard-Marchant},
  \citenamefont {Sheppard}, \citenamefont {Reddy}, \citenamefont {Weckesser},
  \citenamefont {Abbasi}, \citenamefont {Gohlke},\ and\ \citenamefont
  {Oliphant}}]{harris2020array}%
  \BibitemOpen
  \bibfield  {author} {\bibinfo {author} {\bibfnamefont {C.~R.}\ \bibnamefont
  {Harris}}, \bibinfo {author} {\bibfnamefont {K.~J.}\ \bibnamefont {Millman}},
  \bibinfo {author} {\bibfnamefont {S.~J.}\ \bibnamefont {van~der Walt}},
  \bibinfo {author} {\bibfnamefont {R.}~\bibnamefont {Gommers}}, \bibinfo
  {author} {\bibfnamefont {P.}~\bibnamefont {Virtanen}}, \bibinfo {author}
  {\bibfnamefont {D.}~\bibnamefont {Cournapeau}}, \bibinfo {author}
  {\bibfnamefont {E.}~\bibnamefont {Wieser}}, \bibinfo {author} {\bibfnamefont
  {J.}~\bibnamefont {Taylor}}, \bibinfo {author} {\bibfnamefont
  {S.}~\bibnamefont {Berg}}, \bibinfo {author} {\bibfnamefont {N.~J.}\
  \bibnamefont {Smith}}, \bibinfo {author} {\bibfnamefont {R.}~\bibnamefont
  {Kern}}, \bibinfo {author} {\bibfnamefont {M.}~\bibnamefont {Picus}},
  \bibinfo {author} {\bibfnamefont {S.}~\bibnamefont {Hoyer}}, \bibinfo
  {author} {\bibfnamefont {M.~H.}\ \bibnamefont {van Kerkwijk}}, \bibinfo
  {author} {\bibfnamefont {M.}~\bibnamefont {Brett}}, \bibinfo {author}
  {\bibfnamefont {A.}~\bibnamefont {Haldane}}, \bibinfo {author} {\bibfnamefont
  {J.~F.}\ \bibnamefont {del R{\'{i}}o}}, \bibinfo {author} {\bibfnamefont
  {M.}~\bibnamefont {Wiebe}}, \bibinfo {author} {\bibfnamefont
  {P.}~\bibnamefont {Peterson}}, \bibinfo {author} {\bibfnamefont
  {P.}~\bibnamefont {G{\'{e}}rard-Marchant}}, \bibinfo {author} {\bibfnamefont
  {K.}~\bibnamefont {Sheppard}}, \bibinfo {author} {\bibfnamefont
  {T.}~\bibnamefont {Reddy}}, \bibinfo {author} {\bibfnamefont
  {W.}~\bibnamefont {Weckesser}}, \bibinfo {author} {\bibfnamefont
  {H.}~\bibnamefont {Abbasi}}, \bibinfo {author} {\bibfnamefont
  {C.}~\bibnamefont {Gohlke}},\ and\ \bibinfo {author} {\bibfnamefont {T.~E.}\
  \bibnamefont {Oliphant}},\ }\bibfield  {title} {\bibinfo {title} {Array
  programming with {NumPy}},\ }\href
  {https://doi.org/10.1038/s41586-020-2649-2} {\bibfield  {journal} {\bibinfo
  {journal} {Nature}\ }\textbf {\bibinfo {volume} {585}},\ \bibinfo {pages}
  {357} (\bibinfo {year} {2020})}\BibitemShut {NoStop}%
\bibitem [{\citenamefont {Inc.}()}]{Mathematica:12.1}%
  \BibitemOpen
  \bibfield  {author} {\bibinfo {author} {\bibfnamefont {W.~R.}\ \bibnamefont
  {Inc.}},\ }\href {https://www.wolfram.com/mathematica} {\bibinfo {title}
  {Mathematica, {V}ersion 12.1}},\ \bibinfo {note} {champaign, IL,
  2023}\BibitemShut {NoStop}%
\end{thebibliography}%

\end{document}